\documentclass{article}
\usepackage{graphicx}  
\usepackage{amsmath}   
\usepackage{amssymb}   
\usepackage{bm} 
\usepackage{dcolumn}
\usepackage{color}
\usepackage{mathrsfs}
\usepackage{amsfonts}
\usepackage{varioref}
\usepackage[margin=1.6in]{geometry}
\RequirePackage[colorlinks,citecolor=blue,urlcolor=magenta,linkcolor=blue]{hyperref}
\def\p{\partial}
\def\e{\epsilon}

\def\be{\begin{equation}}
\def\ee{\end{equation}}

\labelformat{section}{Section #1} 
\labelformat{subsection}{Section #1} 
\labelformat{subsubsection}{Section #1}
\labelformat{subsubsubsection}{Section #1}
\labelformat{equation}{Eq.~(#1)} 
\labelformat{figure}{Fig.~#1} 
\labelformat{subfigure}{Fig.~\thefigure#1} 
\labelformat{table}{Tab.~#1} 
\labelformat{appendix}{Appendix #1}

\title{\bf Non-perturbative analysis for a massless minimal quantum scalar with ${\footnotesize  V(\phi)=\lambda \phi^4/4!+\beta \phi^3/3!}$  in the inflationary de Sitter spacetime }
\author{$^{1,2}$Sourav Bhattacharya\footnote{sbhatta.physics@jadavpuruniversity.in; On lien from IIT Ropar, Punjab, India.},\,\, and $^2$Nitin Joshi\footnote{2018phz0014@iitrpr.ac.in}\\
\small{$^1$Relativity and Cosmology Research Centre, Department of Physics, Jadavpur University, Kolkata 700 032, India}\\
\small{$^2$Department of Physics, Indian Institute of Technology Ropar, Rupnagar, Punjab 140 001, India}\\}
\begin{document}
\maketitle
\begin{abstract}
\noindent
We consider a massless, minimally coupled quantum scalar field theory with an asymmetric self interaction, $V (\phi) = \lambda\phi^4/4!+\beta\phi^3/3!$ ($\lambda >0$) in the inflationary de Sitter spacetime. The potential is bounded from below. While the $\beta=0$ case has been much well studied, the motivation behind taking such a hybrid potential corresponds to the fact that it might generate finite negative vacuum expectation values of $V(\phi)$ as well of $\phi$, leading to  some dynamical screening of the inflationary cosmological constant $\Lambda$, at late times, with the initial conditions, $\langle \phi \rangle=0=\langle V(\phi) \rangle $.  In this work we first compute   the vacuum expectation values of $\phi,\, \phi^2$ and $V(\phi)$, using the late time, non-perturbative and infrared effective stochastic formalism. The backreactions to the inflationary  $\Lambda$  are estimated. We also compute the dynamically generated mass of the scalar field using  $\langle \phi^2 \rangle$. We next compute  $\langle\phi^2\rangle$ using quantum field theory with respect to  the initial Bunch-Davies vacuum at one and two loops, using the Schwinger-Keldysh formalism. These  results show non-perturbative secular logarithms, growing with the cosmological time.   Using next a recently proposed renormalisation group inspired formalism, we attempt to find out a resummed $\langle\phi^2\rangle$.  We have been able to resum some part of the same which contains contributions only from the local self energy. The corresponding dynamically generated mass is  computed. Comparison of the stochastic and the quantum field theory  results shows that they differ numerically, although they have  similar qualitative behaviour. Possible reasons for such quantitative mismatch is discussed. The manifestation of strong non-classical effects in the results found via both the formalisms has  been emphasised. 
\end{abstract}
\vskip .5cm

\newpage

\tableofcontents

\section{Introduction}\label{S1}

The Hot Big Bang cosmological model  is very successful in describing the redshifts of galaxies, the origin of cosmological microwave background radiation, abundances of light elements and the formation of large scale structures of our universe. However, several issues such as the spatial flatness, horizon and the rarity of relics like the magnetic monopoles cannot be explained by this model~\cite{Rindler, Dicke, Wein, Mukhanov:2005sc}. The  primordial cosmic inflation is a phase of  rapid, near exponential expansion of our very early universe, i.e. an initial condition,  introduced to provide a possible solution to such problems. Apart from being a solution to them, inflation also provides elegant mechanism  to generate primordial cosmological density perturbations in the very early universe, as a seed of the large scale cosmic structures we observe today in the sky, see~\cite{Wein, Mukhanov:2005sc} and references therein for theoretical as well as observational aspects. 

The  inflationary phase was expected to be dominated by dark energy/positive cosmological constant $\Lambda$, whose density was supposed to be much larger compared to the one we observe today.  Thus it is an interesting task to understand how the inflationary $\Lambda$-value got diminished to reach the current one. Since we are essentially talking about a time dependent, expanding background here, we must not ignore the  quantum effects, which one may reasonably expect to be large. The problem then essentially boils down to estimating the backreaction of quantum fields, to see whether or to what extent they can screen the inflationary $\Lambda$ or break the de Sitter invariance, so that the inflation reaches a graceful exit, see~\cite{Tsamis, Ringeval, Miao:2021gic} and references therein. We also refer our reader to  e.g.~\cite{Dadhich, Padmanabhan, Alberte, Appleby, Khan:2022bxs, Evnin:2018zeo} for alternative proposals to the solution of the aforementioned cosmological constant or the cosmic coincidence problem. In any case, analysing matter field's backreaction in the inflationary background seems to be an important task.  

Massless but conformally non-invariant quantum fields such as gravitons and massless minimally coupled scalars break the de Sitter invariance and can generate large non-perturbative effects growing with time, popularly known as the {\it secular effect}, at late cosmological times~\cite{Floratos}. In some instances, such super-Hubble or infrared contributions can be resummed, to produce a finite answer. The problem of resumming the secular effect for infrared gravitons however, remains as an open challenge~\cite{Miao:2021gic} (also references therein). 

Quantisation of the free massless or massive  scalar fields and the corresponding Unruh-DeWitt detector responses with or without non-minimal coupling in the de Sitter spacetime can be seen in~\cite{Chernikov:1968zm, Bunch:1978yq, Linde:1982uu, Starobinsky:1982ee, Allen:1985ux, Allen, Karakaya:2017evp, Ali:2020gij}. For a massless minimal scalar, there exists no de Sitter invariant vacuum states, reflected in the Wightman functions as the appearance of $\ln a $, where $a\equiv a(t)$  is the de Sitter scale factor. As a consequence, each internal line in a given Feynman diagram may yield one such secular logarithm at late times, eventually the corresponding amplitude growing non-perturbatively large  and hence leading to the breakdown of the perturbation theory. One- and two loop such computations in interacting scalar field theories can be seen in~\cite{Ali:2020gij, Onemli:2002hr, Brunier:2004sb, Kahya:2009sz, Boyanovsky:2012qs, Onemli:2015pma, Prokopec:2003tm, Miao:2006pn, Prokopec:2007ak, Liao:2018sci, Miao:2020zeh, Glavan:2019uni, Karakaya:2019vwg, Cabrer:2007xm, Prokopec:2003qd, Boran:2017fsx} and references therein.  Efforts to resum such non-perturbative infrared effects and the differences of such resummation with that of the standard dynamical renormalisation group can be seen in e.g.~\cite{ Miao:2021gic, Moreau:2018ena, Moreau:2018lmz, Gautier:2015pca, Serreau:2013eoa, Serreau:2013koa, Serreau:2013psa, Ferreira:2017ogo, Burgess:2009bs, Burgess:2015ajz, Youssef:2013by, Baumgart:2019clc, Kitamoto:2018dek}. In particular, a renormalisation  group inspired autonomous method to resum secular logarithms was proposed recently in~\cite{Kamenshchik:2020yyn, Kamenshchik:2021tjh}, which will be useful for our purpose. This method involves  constructing a first order differential equation out of the perturbation series containing secular logarithms and then increasing its domain to the non-perturbative regime and finally, integrating it. 

However, perhaps the most efficient formalism so far to resum these secular logarithms for scalar field theories without derivative interactions  is the stochastic method, proposed in the pioneering works~\cite{Starobinsky:1986fx, Starobinsky:1994bd}.  We further refer our reader to~\cite{Cho:2015pwa, Prokopec:2015owa, Garbrecht:2013coa, Vennin:2015hra, Cruces:2022imf, Finelli:2008zg, Markkanen:2019kpv, Markkanen:2020bfc, Tsamis:2005hd} and references therein for recent developments. This is basically a  non-perturbative method for calculating expectation values associated with the long wavelength or infrared part of the scalar field. These super-Hubble modes receive  quantum-kicks (similar to that of the Brownian motion) from the stochastic forces rendered by sub-Hubble modes. One uses here a probability distribution function that satisfies the Fokker-Planck equation. The stochastic formalism is highly successful because  it changes a problem of quantum field theory into a statistical one. In any case, all these non-perturbative methods suggest that despite the appearances of secular logarithms at perturbative orders indicating breakdown of the de Sitter invariance, the resummed non-perturbative results are not only finite, but also are consistent with the de Sitter symmetry, owing to the dynamical generation of mass of the scalar field. A massive scalar field, no matter how tiny its mass is, does not break de Sitter invariance. However, as we have emphasised earlier, estimating the precise backreaction for any given model is important, in order to understand how much screening of the inflationary $\Lambda$ is possible.

\begin{figure}[h!]
\begin{center}
  \includegraphics[width=7.0cm]{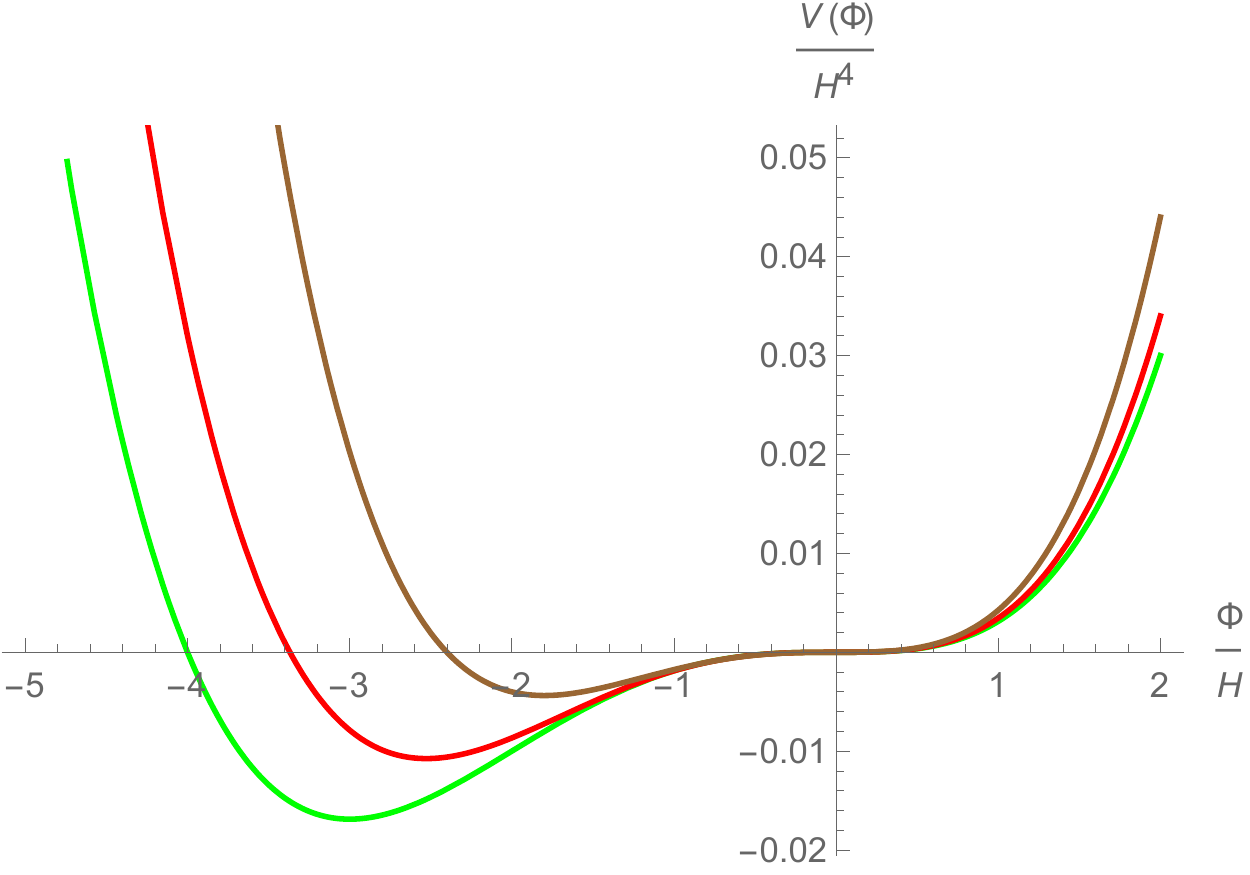}
  \caption{\small Variation of $V (\phi)=\lambda \phi^4/4!+\beta \phi^3/3!$  with respect to $\phi$. We have made things dimensionless by scaling with respect to the Hubble rate, $H$.  The system is assumed to be located at the around the flat plateau at initial times, $t \to 0$. The green, red and brown curves respectively correspond to $\lambda$ and $\beta/H$ values : $(0.015,\, 0.015)$, $(0.019,\, 0.016)$ and $(0.03,\, 0.018)$. For $\beta<0$, the minimum just shifts to the opposite side.}
  \label{poten}
\end{center}
\end{figure}

In this paper, we  wish to consider the backreaction problem of a massless minimally coupled quantum scalar field in the early inflationary de Sitter universe, with an asymmetric self interaction  potential 
\be
V (\phi) = \frac{\lambda\phi^4}{4!}+\frac{\beta\phi^3}{3!}\qquad (\lambda >0)
\label{poten0}
\ee
depicted in \ref{poten} with different parameter values. Vacuum decay in flat spacetime with the above $V(\phi)$  was addressed long time ago  in~\cite{Veneziano} for a massless or massive scalar and the mass generation as well as the condition for the initial vacuum state to be stable was derived. In~\cite{Tsamis}, the same model was investigated using the in-in formalism and a modified version of it. The temporal growth of the vacuum expectation value of the scalar field up to two loop corresponding to the cubic self interaction was explicitly demonstrated. This temporal growth seems to be different from the pure logarithmic secular growth in the inflationary background, owing to the absence of any natural length scale associated  with the Minkowski spacetime. Subsequently, the critical  time from which perturbation theory might fail was estimated.   We further refer our reader to~\cite{Cirigliano:2004yh}, for discussion on the utility of such potential with a non-vanishing scalar mass in a wide range of inflationary paradigms, inspired by the WMAP data. 

To the best of our knowledge, the hybrid potential of \ref{poten0} for a {\it massless and minimally coupled} scalar in the primordial inflationary scenario was first addressed recently in~\cite{Bhattacharya:2022aqi}. The motivation behind this choice is as follows. First of all, being bounded from below irrespective of the sign and magnitude of $\beta$, the potential is expected to yield a late time equilibrium state, so that the  eternal roll down disaster  is avoided. Let us now assume that the system is located around $\phi \sim 0 $ in \ref{poten}, {\it initially}. As time goes on, the system will roll down towards the minima of the potential to settle down there at late times. However, in this due course, we expect there will be strong non-perturbative radiative processes as well, owing to the secular effects generated by the loops of the massless and minimal scalar. Clearly,  these strong  quantum effects must be taken into account in order obtain any sensible answer about the final state of the system. Flat spacetime results may not give us correct indication  here, due to these effects which can only be present in the inflationary scenario. For example, for a quartic self interaction for a massless and minimal scalar in de Sitter, it is an established result that there is a dynamical generation of mass at late times, e.g.~\cite{Youssef:2013by} and references therein, even though there is no analogous phenomenon  in the flat spacetime. 

Now, from \ref{poten}, we expect that the late time vacuum expectation values of $ V(\phi)$ and $\phi$ (owing to the tadpoles, non-vanishing due to the appearances of secular logarithms), may be negative and hence they  will act as an effective {\it negative} cosmological constant. In other words, they may at least partially screen the inflationary  $\Lambda$. The quartic self interaction case ($\beta=0$) does not possess such screening feature. Such non-perturbative backreaction due to $\langle \phi \rangle$  was computed  using quantum field theory in~\cite{Bhattacharya:2022aqi}. 

In this work, we shall compute $\langle \phi \rangle $, $\langle \phi^2 \rangle $ and $\langle V(\phi) \rangle$ using the stochastic formalism of~\cite{Starobinsky:1986fx, Starobinsky:1994bd}, in \ref{stocastic}. The subsequent backreaction on inflationary $\Lambda$  will be estimated in \ref{backreaction}.  We shall also compute   $\langle \phi^2 \rangle $ non-perturbatively, using the Schwinger-Keldysh or in-in formalism of quantum field theory and the renormalisation group inspired method of~\cite{Kamenshchik:2020yyn, Kamenshchik:2021tjh}, in \ref{phiresum} and \ref{resum}. As we shall discuss below, we have been able to resum some diagrams containing local contributions from the self energy. We compute the dynamical  masses generated by the non-vanishing and non-perturbative $\langle \phi^2 \rangle $ obtained via the stochastic method as well as  quantum field theory and point out that even though their behaviour are similar with respect to the variation of the parameters, they differ numerically, and the difference grows with the increasing strength of the cubic coupling parameter.   We discuss some possible reason behind such mismatch in \ref{discussion}, where we also put concluding remarks along with some future directions. In any case, we have emphasised the strong non-classical effects, manifested by both the approaches. Also, we have added  \ref{A}-\ref{nlo3}, containing some detail for the computations. 

 We shall work with the mostly positive signature of the metric in $d = 4 - \epsilon$ ($\epsilon$ = $0^{+}$) dimensions and will set $c = 1 = \hbar$ throughout. The vacuum expectation value of any operator $O$ will be denoted as $\langle O \rangle$. Also for the sake of brevity, we shall denote for powers of propagators and logarithms respectively as, $(i\Delta)^n\equiv i\Delta^n$ and $(\ln x )^n \equiv \ln^n x$.

\section{ Vacuum expectation values via the stochastic formalism}\label{stocastic}

\subsection{The general setup}\label{basics}
%
The metric for the inflationary de Sitter spacetime we are interested in reads
\be
ds^2 = -dt^2 + e^{2Ht}d\vec{x}\cdot  d\vec{x}
\label{l0-}
\ee
where  $H=\sqrt{\Lambda/3}$ is the Hubble rate, and $\Lambda$ is the positive cosmological constant. The above metric can also be written in the conformally flat form
\begin{eqnarray}
ds^2 = a^2(\eta) \left[-d\eta^2 + d\vec{x}\cdot  d\vec{x}\right],
\label{l0}
\end{eqnarray}
where $a(\eta)= - 1/H\eta $ is the scale factor and the conformal time, $\eta = - e^{-H t}/H$. We have the range $0\leq t <\infty$, so that $-H^{-1}\leq \eta \leq 0^-$.

The bare Lagrangian density is given by,
\begin{eqnarray}
\sqrt{-g}{\cal L}= \sqrt{-g}\left[-\frac12 (\nabla_{\mu} \psi)(\nabla_{\nu} \psi) g^{\mu \nu} -\frac12 m_0^2 \psi^2  - \frac{\lambda_0}{4!}\psi^4 - \frac{\beta_0}{3!} \psi^3 - \tau_0 \psi \right] 
\label{l1}
\end{eqnarray}
 In terms of the field strength  renormalisation, $\phi = \psi/{\sqrt Z} $,
\begin{eqnarray}
\sqrt{-g}{\cal L}= -\frac{Z}{2}(\p_{\mu} \phi)(\p_{\nu} \phi) a^{d-2}\, \eta^{\mu \nu} -\frac12 \,Z m_0^2\, \phi^2 a^{d} - \frac{Z^2 \lambda_0}{4!}\, \phi^4 a^{d} - \frac{Z^{3/2} \beta_0}{3!}\, \phi^3 a^{d} - \tau_0 \sqrt{Z} \,\phi a^{d}
\label{l2}
\end{eqnarray}
We decompose the bare Lagrangian density in terms of the renormalised quantities and the counterterms. We write for a massless field,
\begin{eqnarray}
Z = 1+ \delta Z \qquad Z m_0^2 = 0 + \delta m^2 \qquad Z^2 \lambda_0 = \lambda + \delta \lambda \qquad Z^{3/2} \beta_0 = \beta +\delta \beta \qquad \tau_0 \sqrt{Z} = 0 + \delta \tau
\label{l3}
\end{eqnarray}
Thus \ref{l2} is rewritten as,
\begin{eqnarray}
\sqrt{-g}{\cal L}=&& -\frac{1}{2}(\p_{\mu} \phi)(\p_{\nu} \phi) a^{d-2}\, \eta^{\mu \nu}  - \frac{\lambda}{4!} \phi^4 a^d     -\frac{\beta}{3!} \phi^3 a^d  \nonumber\\ && -\frac{\delta Z}{2}(\p_{\mu} \phi)(\p_{\nu} \phi) a^{d-2}\, \eta^{\mu \nu} -  \frac12 \delta m^2  \phi^2 a^{d} - \frac{\delta \lambda}{4!}\, \phi^4 a^{d} - \frac{\delta \beta}{3!}\, \phi^3 a^{d} -  \delta \tau \phi a^{d}
\label{l4}
\end{eqnarray}
where the second line in the above equation corresponds to the counterterm Lagrangian density.
However, since renormalisation is essentially an ultraviolet phenomenon, the above counterterms will not be necessary for the infrared (IR) effective stochastic formalism or IR field theory. However, we shall require some counterterms in \ref{resum}, while  computing the local part of the self energy.

\subsection{Computation of $\langle \phi \rangle$, $\langle \phi^2 \rangle$ and $\langle V(\phi) \rangle$ using stochastic formalism}\label{stoch}

We wish to compute below some correlation functions using the late time IR effective stochastic formalism, describing the long wavelength or super Hubble dynamics of the scalar field~\cite{Starobinsky:1986fx, Starobinsky:1994bd}, which we first wish to review briefly. At sufficiently late cosmological times, we split the scalar field into two parts : $\phi(\vec{x}, t)=\tilde{\phi}(\vec{x}, t)+\phi(\vec{x}, t)_{\rm short}$, where $\tilde{\phi}$ is the long wavelength part of the field, coarse grained over a volume assumed to be larger than what is enclosed within the Hubble radius, whereas $\phi(\vec{x}, t)_{\rm short}$ is the short wavelength fluctuations over it. We have
\begin{eqnarray}\label{sto}
\phi(\vec{x}, t)=\tilde{\phi}(\vec{x}, t)+\int \frac{\mathrm{d}^{3} \vec{k}}{(2 \pi)^{3/2}} \theta(k-\varepsilon a(t) H)\left[\hat{a}_{\vec{k}} u(\vec{k},t)_{\rm short} e^{-i \vec{k} \cdot \vec{x}}+\hat{a}_{\vec{k}}^{\dagger} u^*(\vec{k},t)_{\rm short} e^{i \vec{k} \cdot \vec{x}}\right]
\end{eqnarray}
The Heaviside step  function appearing in the second term on the right hand side ensures that only the short wavelength part with wavenumber $k>\varepsilon a(t)H$ ($k=|\vec{k}|$) is quantised, whereas the long wavelength part  with the condition $k<\varepsilon a(t) H$ is included in $\tilde{\phi}$. $\varepsilon$ is a small positive parameter. The coarse graining suggests that  any two points closer than the length scale $1/(\varepsilon H)$ are considered to be the same point. The equation of motion for $\phi$ with a potential $V(\phi)$ is given by 
\begin{eqnarray}
\ddot{\phi}(\vec{x}, t)+3 H \dot{\phi}(\vec{x}, t)-\frac{\vec{\p}^2}{a^2} \phi(\vec{x}, t)+V^{\prime}(\phi(\vec{x}, t))=0
\label{gen}
\end{eqnarray}
where a dot denotes derivative once with respect to the cosmological time and prime denotes the same with respect to the field $\phi$. Substituting  now $\phi(\vec{x}, t)=\tilde{\phi}(\vec{x}, t) +\phi(\vec{x}, t)_{\rm short}$ into the above and ignoring the spatial variation of the course grained part of the field, we have 
\begin{eqnarray}\label{split}
\begin{aligned}
&\ddot{\tilde \phi}(t)+3 H \dot{\tilde\phi}(t)+\ddot\phi(\vec{x}, t)_{\text {short }}+3 H \dot \phi(\vec{x}, t)_{\text {short }}-\frac{\vec{\p}^2}{a^{2}}\phi(\vec{x}, t)_{\text {short }} 
+V^{\prime}(\tilde{\phi})+ V^{\prime \prime}({\tilde{\phi}})\phi(\vec{x}, t)_{\text {short }}+O\left( \phi^2(\vec{x}, t)_{\text {short }}\right)=0,
\end{aligned}
\end{eqnarray}
where we have discarded higher order short wavelength corrections. We can cast the above equation into two separate parts for $k>\varepsilon a(t) H$ and $k<\varepsilon a(t) H$, so that in the three-momentum space, we have for $k>\varepsilon a(t) H$
\be
\ddot{\phi}(t,\vec{k})_{\rm short}+3 H \dot{\phi}(t,\vec{k})_{\rm short}+\frac{{k}^{2}}{a^{2}} {\phi}(t,\vec{k})_{\rm short}+V^{\prime \prime}(\tilde{\phi}) \phi(t,\vec{k})_{\rm short}=0
\label{primeprime}
\ee
For the inflation to last for a sufficiently long time,  $V(\tilde{\phi})$ must be sufficiently flat.  Thus \ref{primeprime} can further be simplified to the free equation of motion at the leading approximation
\be
\ddot{\phi}(t,\vec{k})_{\rm short}+3 H \dot{\phi}(t,\vec{k})_{\rm short}+\frac{k^{2}}{a^{2}} {\phi}(t,\vec{k})_{\rm short}=0
\label{shorty}
\ee
The asymptotically positive frequency solution in the past of the above equation is given by~\cite{Allen:1985ux}
\be\label{hankel}
{u}(\vec{k},\eta)_{\rm short}=\frac{H \eta}{\sqrt{2 k}}\left(1-\frac{i}{k \eta}\right) e^{-i k \eta},
\ee
and the corresponding vacuum state is known as the Bunch-Davies state. We shall require the late time limit of the  above expression later for computing various two point functions. Now, after ignoring some of the terms in \ref{split} by the virtue of the discussion above, we have
\begin{eqnarray}\label{lit}
\ddot{\tilde \phi}+3 H \dot{\tilde \phi}+\ddot\phi(\vec{x}, t)_{\text {short }}+3 H \dot \phi(\vec{x}, t)_{\text {short }}+\frac{k^{2}}{a^{2}} {\phi}(t,\vec{k})_{\rm short}+V^{\prime}(\tilde{\phi})=0
\end{eqnarray}
  Owing to the slow role condition, we now ignore the double temporal derivatives to obtain
\begin{eqnarray}\label{lit}
3 H \dot{\tilde\phi}+V^{\prime}(\tilde{\phi})+3 H \dot \phi(\vec{x}, t)_{\text{short }}+\frac{k^{2}}{a^{2}} {\phi}(t,\vec{k})_{\rm short} \approx 0
\end{eqnarray}
Using \ref{sto}  into the above equation, we get
\begin{eqnarray}\label{pre}
\begin{aligned}
3 H \dot{\tilde\phi}+V^{\prime}(\tilde{\phi})&+\int \frac{\mathrm{d}^{3} \vec{k}}{(2 \pi)^{3/2}} \theta(k-\varepsilon a(t) H)\left[\hat{a}_{\vec{k}} \left(3H\dot{u}(\vec{k},t)_{\rm short}+\frac{k^2}{a^2} {u}(\vec{k},t)_{\rm short} \right) e^{-i \vec{k} \cdot \vec{x}}+ {\rm h.c.}\right]\\
&-3H\varepsilon a(t) H^2 \int \frac{\mathrm{d}^{3} \vec{k}}{(2 \pi)^{3/2}} \delta(k-\varepsilon a(t) H)\left[\hat{a}_{\vec{k}} u(\vec{k},t)_{\rm short} e^{-i \vec{k} \cdot \vec{x}}+ {\rm h.c.} \right]=0
\end{aligned}
\end{eqnarray}
The third term in the above equation can be ignored via \ref{shorty}, since we have discarded  the double derivative term subject to the slow role condition. Thus we finally have the {\it Langevin equation } satisfied by the long wavelength, super-Hubble or infrared part of the scalar field
\begin{equation}
\dot{\tilde \phi}=-\frac{1}{3 H} V^{\prime}(\tilde{\phi})+f(\vec{x}, t)
\label{lg}
\end{equation}
where $f(\vec{x}, t)$ is regarded as the noise given by
\begin{eqnarray}\label{noise}
f(\vec{x}, t)=\dot{\phi}(\vec{x}, t)_{\rm short}=\varepsilon a(t) H^2 \int \frac{\mathrm{d}^{3} \vec{k}}{(2 \pi)^{3/2}} \delta(k-\varepsilon a(t) H)\left[\hat{a}_{\vec{k}} u(\vec{k},t)_{\rm short} e^{-i \vec{k} \cdot \vec{x}}+\hat{a}_{\vec{k}}^{\dagger} u^{*}(\vec{k},t)_{\rm short} e^{i \vec{k} \cdot \vec{x}}\right]
\end{eqnarray}
In \ref{lg}, the derivative of the potential term can be thought of as a classical drift, whereas $f(\vec{x}, t)$ can be thought of as the random kick.

Let us now compute the correlation function of \ref{noise}. After dropping  off the suffix `short' without any loss of generality, we have
\begin{eqnarray}
\left\langle f\left(\vec{x}_{1}, t_{1}\right) f\left(\vec{x}_{2}, t_{2}\right)\right\rangle 
= H^{4}\int \frac{\mathrm{d}^{3} \vec{k}_{1}}{(2 \pi)^{3}}\left[\varepsilon^{2} a(t_{1}) a(t_{2})  \delta(k_{1}-\varepsilon a(t_{1}) H) \delta(k_{1}-\varepsilon a(t_{2}) H) e^{-i \vec{k}_{1} \cdot\left(\vec{x}_{1}-\vec{x}_{2}\right)} u(\vec{k}_1,t_1) u^{*}(\vec{k}_2,t_2)\right]\hskip .3cm
\label{ad1}
\end{eqnarray}
where we have used the canonical commutation relation $[\hat{a}_{\vec{k}_{1}}, \hat{a}_{\vec{k}_{2}}^{\dagger}]=\delta^{3}(\vec{k}_{1}-\vec{k}_{2})$.  Converting now into the polar coordinates of momentum space, using
$\delta(f(x)-f(y))=\delta(x-y)/|f^{\prime}(x)|$,  and substituting  \ref{hankel} into the above, we have
\be
\begin{aligned}
\left\langle f\left(\vec{x}_{1}, t_{1}\right) f\left(\vec{x}_{2}, t_{2}\right)\right\rangle
=& \frac{H^{3} }{4 \pi^{2}} \delta(t_{1}-t_{2}) \frac{\sin (\varepsilon a(t_{1}) H\left|\vec{x}_{1}-\vec{x}_{2}\right|)}{ \varepsilon a(t_{1}) H\left|\vec{x}_{1}-\vec{x}_{2}\right|}+{\cal O}(\varepsilon^2),
\end{aligned}
\ee
which in the spatial coincidence limit, $\vec{x}_1 \to \vec{x}_2$, reduces to
\begin{eqnarray}
\langle f(t, \vec{x}) f(t', \vec{x}) \rangle= \frac{H^3}{4\pi^2} \delta(t-t')
\label{es}
\end{eqnarray}

Note that the correlators are the chief observables in cosmology. However, the Langevin equation~\ref{lg}, may often be not very useful to compute them.  Instead, the {\it Fokker-Planck equation} \cite{Risken, pathria} is much more illuminating for this purpose, which, corresponding to \ref{lg} reads,   
\begin{equation}
{\partial\rho \over\partial t} 
= {H^3\over8\pi^2}{\partial^2\rho\over\partial{\tilde \phi}^2} 
+ {1\over3H}{\partial\over\partial\tilde{\phi}}
\biggl({\partial V\over\partial{\tilde{\phi}}} \,\rho \biggr)
\label{F-P}
\end{equation}
where $\rho  \equiv \rho(\tilde{\phi},t)$ is the one-point probability density distribution function. The general solution of \ref{F-P} is given by~\cite{Starobinsky:1994bd}, 
\begin{equation}
\rho(\tilde{\phi}, t)=e^{-\nu(\tilde{\phi})} \sum_{n=0}^{\infty} a_{n} \Phi_{n}(\tilde{\phi}) e^{-\Lambda_{n}\left(t-t_{0}\right)}
\label{i1}
\end{equation}
where $t_0$ is some initial time, $\nu(\tilde{\phi})= 4 \pi^{2} V(\tilde{\phi})/3 H^{4}$, and $a_n$'s are coefficients independent of time,  whereas $\Phi_{n}(\tilde{\phi})$ and $\Lambda_{n}$ are respectively the eigenfunctions and eigenvalues corresponding to the Schr\"{o}dinger-like equation
\begin{equation}
-\frac{1}{2} \frac{d^2 \Phi_{n}(\tilde{\phi})}{d \tilde{\phi}^{2}}+\frac{1}{2}\left[(\nu^{\prime}(\tilde{\phi}))^{2}-\nu^{\prime \prime}(\tilde{\phi})\right] \Phi_{n}(\tilde{\phi})=\frac{4 \pi^{2} \Lambda_{n}}{H^{3}} \Phi_{n}(\tilde{\phi})
\label{i2}
\end{equation}
The eigenfunctions $\Phi_{n}(\tilde{\phi})$'s are orthonormal,
\begin{eqnarray}
\int \mathrm{d} \tilde{\phi}\, \Phi_{n}(\tilde{\phi}) \Phi_{m}(\tilde{\phi})=\delta_{n m}
\label{i8}
\end{eqnarray}
using which one can find out the coefficients $a_n$ in \ref{i1}. Note also that \ref{i2} can be rewritten in the form
\begin{eqnarray}
\frac{1}{2}\left(-\frac{\partial}{\partial \tilde{\phi}}+\nu^{\prime}(\tilde{\phi})\right)\left(\frac{\partial}{\partial \tilde{\phi}}+\nu^{\prime}(\tilde{\phi})\right) \Phi_{n}(\tilde{\phi})=\frac{4 \pi^{2} \Lambda_{n}}{H^{3}} \Phi_{n}(\tilde{\phi})
\label{i10}
\end{eqnarray}
Since $\tilde{\phi}$ is real, we have $(\partial_{\tilde{\phi}})^{\dagger}=-\partial_{\tilde{\phi}}$. Hence  $\Lambda_{n}$'s are eigenvalues of a positive operator of the form $A^{\dagger}A$, implying $\Lambda_{n}\geq 0$, where $\Lambda_0=0$ is the ground state. The corresponding wave function is
\begin{eqnarray}\label{i10'}
\Phi_{0}(\tilde{\phi})=N^{-1/2}e^{-\frac{4\pi^2 V(\tilde{\phi})}{3H^4}}
\end{eqnarray}
where $N$ is the normalisation,  exists only if $V(\tilde{\phi})$ is bounded from below, in which case late time equilibrium state is possible. The equilibrium probability distribution is given from \ref{i1}, \ref{i10'} by
\begin{equation}
\rho_{\rm eq}(\tilde{\phi}) = N^{-1}e^{-\frac{8\pi^2 V(\tilde{\phi})}{3H^4}}
\label{stat}
\end{equation}
%
%
The equilibrium probability distribution function allows us to compute the expectation value of any operator $\chi(\tilde{\phi})$, at late times    as
\begin{eqnarray}\label{ran}
\langle \chi(\tilde{\phi}) \rangle&:=&\int_{-\infty}^{\infty}d\tilde{\phi}\,\chi(\tilde{\phi}) \rho_{\rm eq}(\tilde{\phi}), 
\end{eqnarray}

Being equipped with all these, we are now ready to work with the cubic plus quartic self interaction potential. For our convenience, from now on we shall write $\tilde{\phi}$ simply as $\phi$, without any loss of generality. Let us first compute the normalisation corresponding to \ref{stat}. Using Mathematica, we find
\begin{eqnarray}
N&=&\int_{-\infty}^{\infty}e^{-{8\pi^2\over3H^4}\left( \frac{\lambda \phi^4}{4!}       +\frac{\beta \phi^3}{3!}   \right)}d\phi\nonumber\ \\
&=&\frac{3^{1/4}8  \lambda^{3/2} \Gamma \left(\frac{1}{12}\right) \Gamma \left(\frac{5}{12}\right) \Gamma \left(\frac{7}{4}\right) \,
   _2F_2\left(\frac{1}{12},\frac{5}{12};\frac{1}{4},\frac{1}{2};\frac{3 \pi ^2 \bar{\beta }^4}{\lambda ^3}\right)-\ 3^{3/4}9 \pi  \Gamma \left(-\frac{3}{4}\right) \Gamma
   \left(\frac{7}{12}\right) \Gamma \left(\frac{11}{12}\right) \bar{\beta }^2 \, _2F_2\left(\frac{7}{12},\frac{11}{12};\frac{3}{4},\frac{3}{2};\frac{3 \pi ^2 \bar{\beta
   }^4}{\lambda ^3}\right)}{24 \pi ^{3/2} \lambda ^{7/4}}\nonumber\\
\label{norm}
\end{eqnarray}
where $\overline{\beta} = \beta/H$ is dimensionless and $_2F_2$ is the hypergeometric function. For $\bar\beta\to 0$ above expresion simplifies to,
\begin{eqnarray}
N= \frac{\Gamma \left(\frac{1}{12}\right) \Gamma \left(\frac{5}{12}\right) \Gamma \left(\frac{7}{4}\right)}{3^{3/4} \pi ^{3/2} \lambda^{1/4}}-\frac{
   3^{3/4} 3\Gamma \left(-\frac{3}{4}\right) \Gamma \left(\frac{7}{12}\right) \Gamma \left(\frac{11}{12}\right) \bar{\beta }^2}{8 \sqrt{\pi
   } \lambda ^{7/4}}+O\left(\bar{\beta }^3\right)
\label{normalization}   
\end{eqnarray}

On the other hand, for $\bar{\beta}^4/\lambda^3 \gg 1$, by expanding the hypergeometric functions for large arguments\footnote{See e.g., Heck : https://stars.library.ucf.edu/etd/193},
\begin{equation}
    _2F_2(a, b; c, d; z) \to \frac{\Gamma(c)\Gamma(d)\Gamma(b-a)}{\Gamma(b)\Gamma(c-a)\Gamma(d-a)}\times z^{-a}+ \frac{\Gamma(c)\Gamma(d)\Gamma(a-b)}{\Gamma(a)\Gamma(c-b)\Gamma(d-b)}\times z^{-b},
    \label{hypergeom}
\end{equation}
\ref{norm} takes the  form
\begin{eqnarray}
N= \frac{\bar{\beta }^2 \left(0.244545 \sqrt[3]{\frac{\bar{\beta }^4}{\lambda ^3}}+0.0469689\right)+0.912655 \lambda ^{3/2} \left( \left(\frac{\bar{\beta }^4}{\lambda ^3}\right)^{5/6}+0.0137897
   \sqrt{\frac{\bar{\beta }^4}{\lambda ^3}}\right)}{\lambda ^{7/4} \left(\frac{\bar{\beta }^4}{\lambda ^3}\right)^{11/12}}    
\end{eqnarray}

Let us now compute the  expectation values of $\phi$, $\phi^2$ and $V(\phi)$ using \ref{ran} and \ref{norm}. 
For the first, we have
\begin{eqnarray}
\resizebox{1.0\hsize}{!}{$
\langle \bar\phi \rangle=-\frac{48 \left[\sqrt{3} \pi  \Gamma \left(\frac{1}{4}\right) \Gamma \left(\frac{11}{12}\right) \Gamma \left(\frac{19}{12}\right) \bar{\beta }^3 \,
   _2F_2\left(\frac{11}{12},\frac{19}{12};\frac{3}{2},\frac{7}{4};\frac{3 \pi ^2 \bar{\beta }^4}{\lambda ^3}\right)+2  \lambda ^{3/2} \Gamma
   \left(\frac{5}{12}\right) \Gamma \left(\frac{13}{12}\right) \Gamma \left(\frac{7}{4}\right) \bar{\beta } \,
   _2F_2\left(\frac{5}{12},\frac{13}{12};\frac{1}{2},\frac{5}{4};\frac{3 \pi ^2 \bar{\beta }^4}{\lambda ^3}\right)\right]}{\lambda  \left[
  8 \lambda^{3/2} \Gamma \left(\frac{1}{12}\right) \Gamma \left(\frac{5}{12}\right) \Gamma \left(\frac{7}{4}\right) \,
   _2F_2\left(\frac{1}{12},\frac{5}{12};\frac{1}{4},\frac{1}{2};\frac{3 \pi ^2 \bar{\beta }^4}{\lambda ^3}\right)-\ 9\sqrt{3} \pi  \Gamma
   \left(-\frac{3}{4}\right) \Gamma \left(\frac{7}{12}\right) \Gamma \left(\frac{11}{12}\right) \bar{\beta }^2 \,
   _2F_2\left(\frac{7}{12},\frac{11}{12};\frac{3}{4},\frac{3}{2};\frac{3 \pi ^2 \bar{\beta }^4}{\lambda ^3}\right)\right]}$}
\label{ssa}
\end{eqnarray}
 where $\bar{\phi}=\phi/H$ is dimensionless. Note that $\langle \bar\phi \rangle$ changes sign as $\bar \beta \to -\bar \beta$, as expected from the asymmetric behaviour of \ref{poten}. 
 
\begin{figure}[h!]
\begin{center}
  \includegraphics[width=6.5cm]{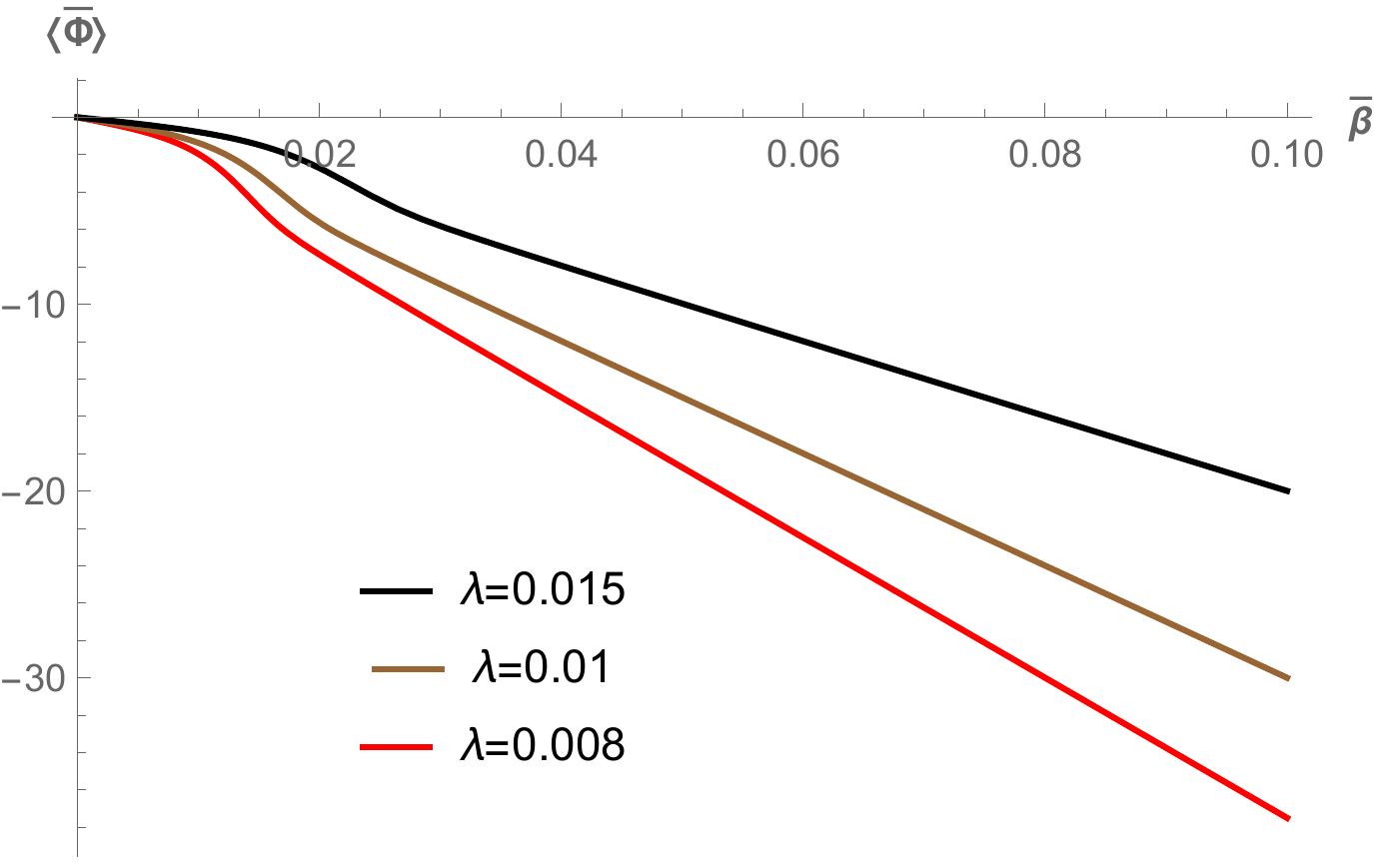}
  \includegraphics[width=7.0cm]{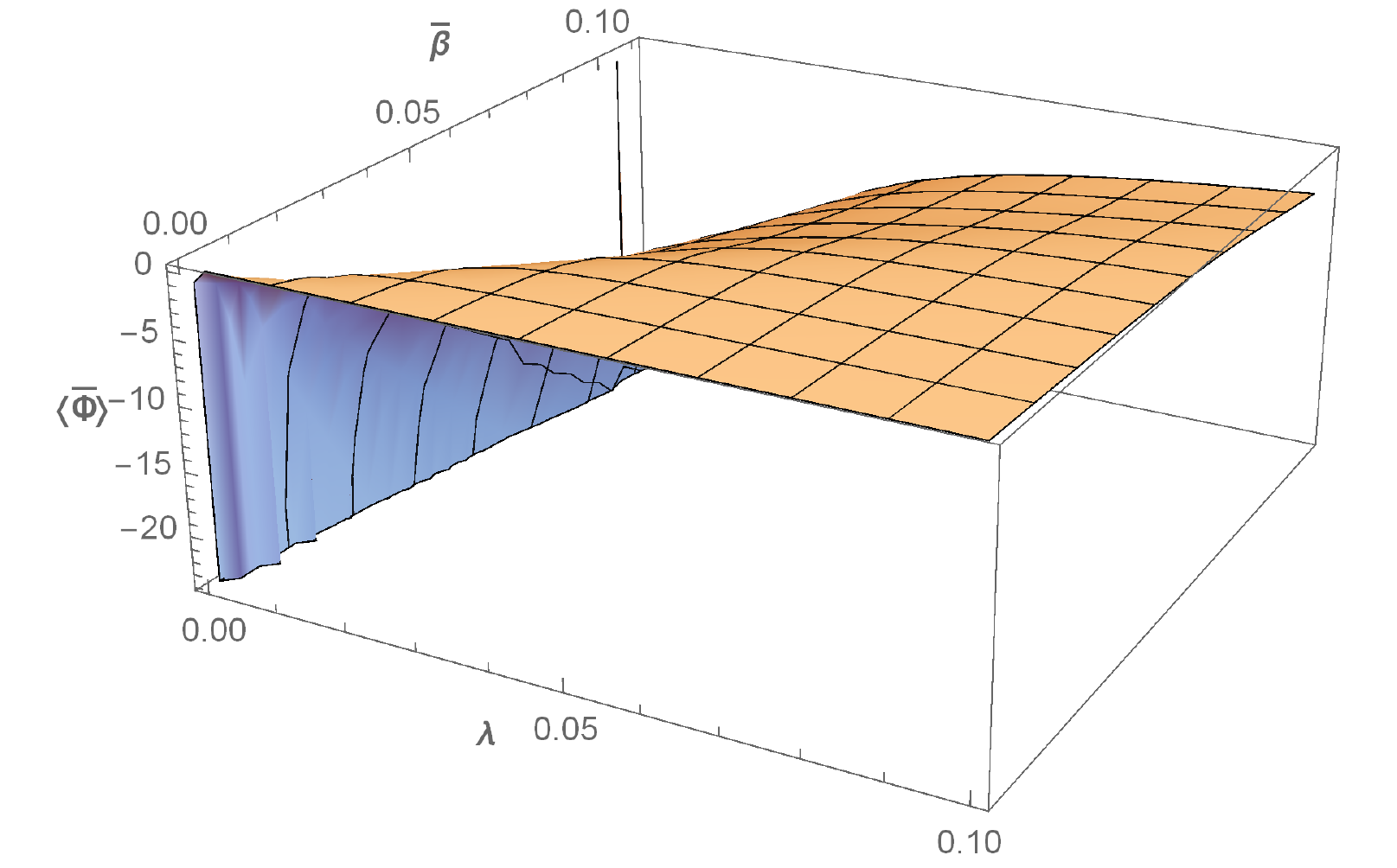}
  \caption{\small Plots for $\langle\bar\phi\rangle$ vs. $\lambda$ and $\bar \beta$, \ref{ssa}. For $\bar \beta>0$, the expectation value is negative. Note that $\langle\bar\phi\rangle \to 0$ for $\overline \beta \to 0$ and  $\langle\bar\phi\rangle \to -\infty$ for $\lambda \to 0$. See main text for discussion.}
  \label{phi}
\end{center}
\end{figure}

Let us now examine the two special cases, i.e. respectively small and large cubic couplings. For $\bar\beta\to 0$, \ref{ssa} becomes
 \begin{eqnarray}
\langle \bar\phi \rangle= -\frac{\bar{\beta }}{\lambda}-\frac{3 \left[\sqrt{3} \pi  \Gamma
   \left(\frac{11}{12}\right) \left(9 \Gamma \left(-\frac{3}{4}\right) \Gamma \left(\frac{7}{12}\right) \Gamma \left(\frac{13}{12}\right)+4 \Gamma
   \left(\frac{1}{12}\right) \Gamma \left(\frac{1}{4}\right) \Gamma \left(\frac{19}{12}\right)\right)\right] \bar{\beta }^3}{2 \lambda ^{5/2} \Gamma^2
   \left(\frac{1}{12}\right) \Gamma \left(\frac{5}{12}\right) \Gamma \left(\frac{7}{4}\right)}+O\left(\bar{\beta }^4\right)
   \label{ssa1}
 \end{eqnarray}
which as expected, is vanishing for $\bar{\beta}=0$.  For large $\bar{\beta }^4/\lambda^3$ on the other hand, we get using \ref{hypergeom},
\begin{eqnarray}
 \langle \bar\phi \rangle = -\frac{0.39152\times  \bar{\beta}^{1/3}\left( \bar{\beta }^2 \sqrt[6]{\frac{\bar{\beta }^4}{\lambda ^3}}+0.0239261 \lambda ^{3/2}\right)}{\lambda^{1/2} \left[\bar{\beta }^2 \left(0.267949 \sqrt[3]{\frac{\bar{\beta }^4}{\lambda ^3}}+0.051464\right)+ \lambda ^{3/2}
   \left( \left(\frac{\bar{\beta }^4}{\lambda ^3}\right)^{5/6}+0.0137897 \sqrt{\frac{\bar{\beta }^4}{\lambda ^3}}\right)\right]}  
   \label{ssa2} 
\end{eqnarray}
The variation of  $\langle \bar\phi \rangle$ with respect to the dimensionless coupling parameters $\lambda$ and $\bar \beta$ has been depicted in \ref{phi}. As one expects from  \ref{poten}, $\langle \bar\phi \rangle$ is negative for all $\overline\beta >0$. We also note that $\langle \bar\phi \rangle$  decreases unboundedly as $\lambda \to 0$, a manifestation of the fact that the cubic potential is unbounded from below and hence in that case no equilibrium state is possible. Also, we note from \ref{ssa} that $\langle \bar\phi \rangle$ in general  may not equal to $-3\bar{\beta}/\lambda$, the position of the classical minima of $V(\phi)$. This is also manifest for the small cubic coupling, \ref{ssa1}. However, for a strong cubic coupling, \ref{ssa2}, it is easy to see from \ref{phi} that $\langle \bar{\phi}\rangle$ matches well with the classical value. We shall compare this result with the one found using quantum field theory in \ref{resum}.

\begin{figure}[h!]
\begin{center}
  \includegraphics[width=6.5cm]{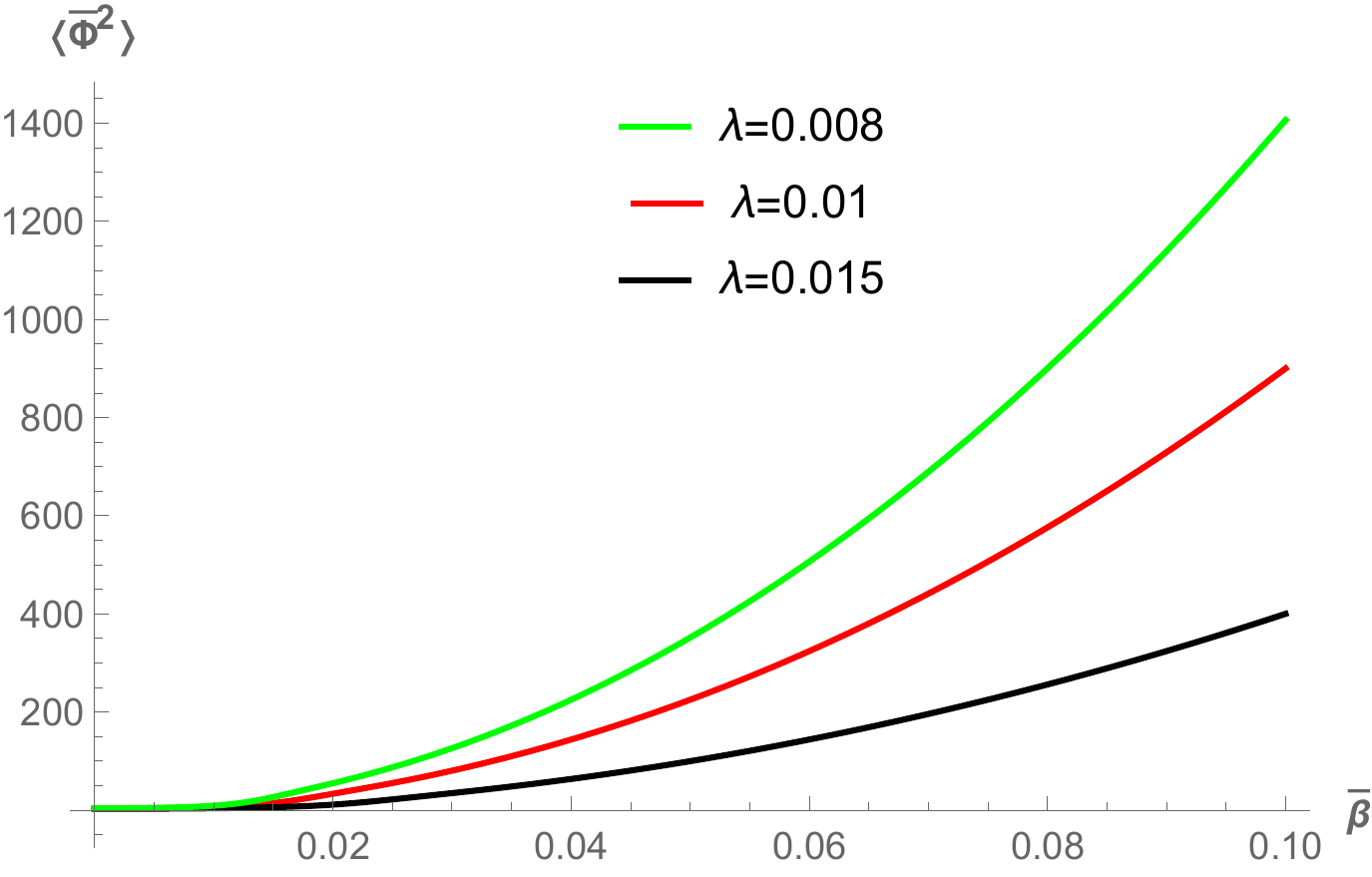}
  \includegraphics[width=7.5cm]{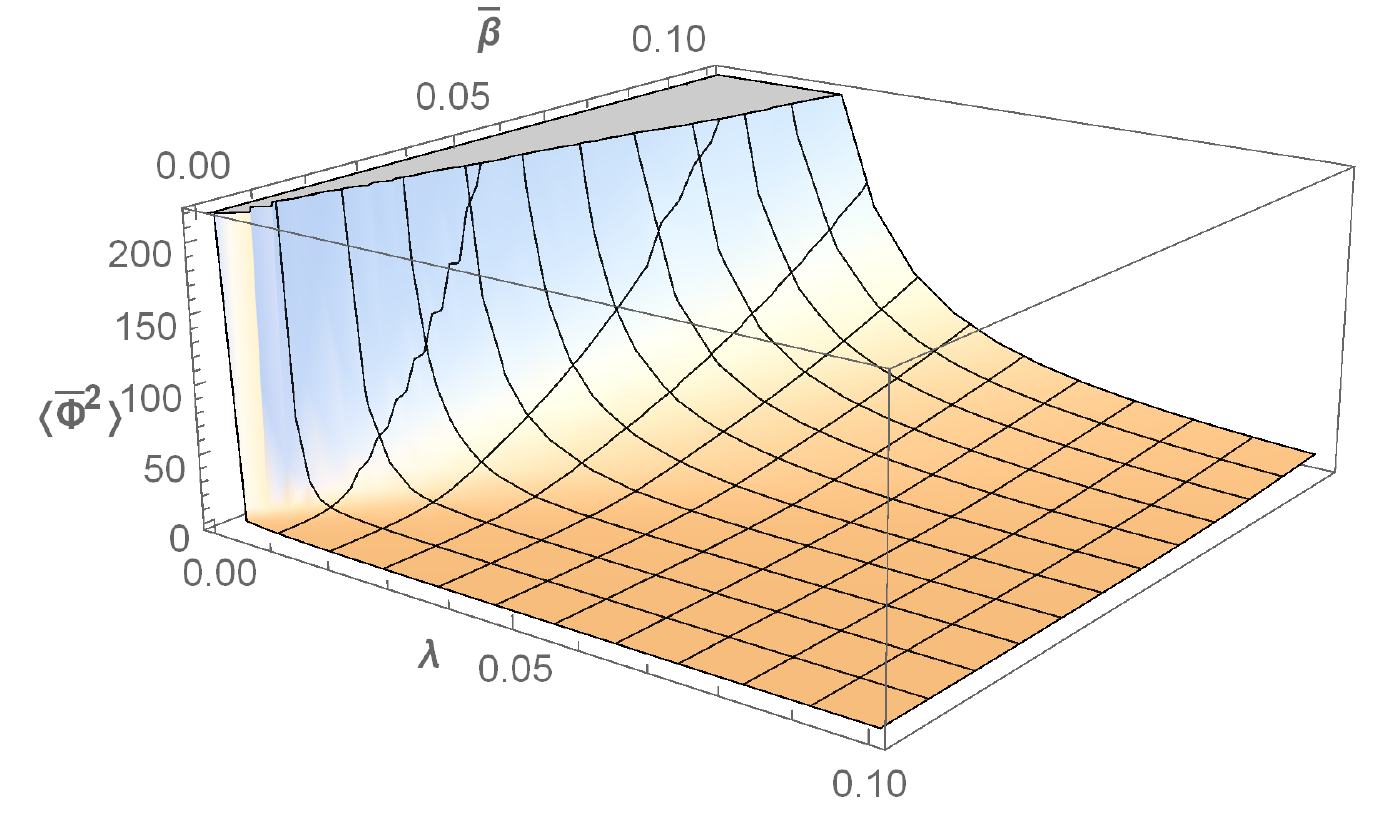}
  \caption{\small Plots for $\langle\bar\phi^2\rangle$ vs. $\lambda$ and $\bar \beta$, \ref{av}. We notice as $\lambda$ decreases, $\langle\bar\phi^2\rangle$ increases rapidly and tends to $+\infty$.   See the main text for discussion.}
  \label{phis}
\end{center}
\end{figure}

Using \ref{ran}, we next compute  
%
%
%
\begin{eqnarray}
\resizebox{1.0\hsize}{!}{$
\langle\bar{\phi}^2\rangle =-\frac{9 \left[\pi  \Gamma \left(-\frac{1}{4}\right) \Gamma \left(\frac{1}{12}\right) \Gamma \left(\frac{17}{12}\right) \bar{\beta }^2 \,
   _2F_2\left(\frac{13}{12},\frac{17}{12};\frac{5}{4},\frac{3}{2};\frac{3 \pi ^2 \bar{\beta }^4}{\lambda ^3}\right)-8\sqrt{3} \lambda ^{3/2} \Gamma \left(\frac{7}{12}\right)
   \Gamma \left(\frac{11}{12}\right) \Gamma \left(\frac{5}{4}\right) \, _2F_2\left(\frac{7}{12},\frac{11}{12};\frac{1}{2},\frac{3}{4};\frac{3 \pi ^2 \bar{\beta }^4}{\lambda
   ^3}\right)\right]}{\pi  \sqrt{\lambda } \left[8  \lambda ^{3/2} \Gamma \left(\frac{1}{12}\right) \Gamma \left(\frac{5}{12}\right) \Gamma \left(\frac{7}{4}\right)
   \, _2F_2\left(\frac{1}{12},\frac{5}{12};\frac{1}{4},\frac{1}{2};\frac{3 \pi ^2 \bar{\beta }^4}{\lambda ^3}\right)-9\sqrt{3}\pi  \Gamma \left(-\frac{3}{4}\right) \Gamma
   \left(\frac{7}{12}\right) \Gamma \left(\frac{11}{12}\right) \bar{\beta }^2 \, _2F_2\left(\frac{7}{12},\frac{11}{12};\frac{3}{4},\frac{3}{2};\frac{3 \pi ^2 \bar{\beta
   }^4}{\lambda ^3}\right)\right]}$}
\label{av}
\end{eqnarray}
We have plotted the variation of the  above  in \ref{phis}. We see that as   $\lambda \to 0$, $\langle\bar{\phi}^2\rangle$ tends to diverge, once again a manifestation of the fact that the cubic potential alone must lead to an unstable system. 
 Note also  that as $\overline{\beta}\to 0$, we reproduce the result for the quartic theory~\cite{Starobinsky:1994bd} 
\begin{eqnarray}
\langle\bar{\phi}^2\rangle =\frac{0.322756}{\sqrt{\lambda}} +{\cal O}\left(\frac{{\bar \beta}^2}{\lambda^2}\right)
\label{value}
\end{eqnarray}
and for  large  $ \bar{\beta }^4/\lambda ^3 \gg 1$, we get
\begin{eqnarray}
\langle\bar{\phi}^2\rangle = \frac{\lambda \left(0.094534 \left(\frac{\bar{\beta }^4}{\lambda^3}\right)^{1/3}+0.00651799\right)-0.0253303 \lambda ^{5/2} \left( \left(\frac{\bar{\beta }^4}{\lambda ^3}\right)^{5/6}+0.960331
\sqrt{\frac{\bar{\beta }^4}{\lambda ^3}}\right)}{\bar{\beta }^2 \left(3.73205 \left(\frac{\bar{\beta }^4}{\lambda ^3}\right)^{1/3}+0.051464\right)+ \lambda ^{3/2} \left(\left(\frac{\bar{\beta}^4}{\lambda^3}\right)^{5/6}+0.192066 \sqrt{\frac{\bar{\beta }^4}{\lambda ^3}}\right)}
\end{eqnarray}

Finally we compute $\langle V(\phi)\rangle$, given by 
\begin{eqnarray}
\resizebox{1.0\hsize}{!}{$
\left\langle \frac{V(\phi)}{H^4} \right\rangle=\frac{3 \sqrt[4]{3} \left(9 \lambda ^3 \Gamma \left(\frac{5}{12}\right) \Gamma \left(\frac{3}{4}\right) \Gamma \left(\frac{13}{12}\right) \,
_2F_2\left(\frac{5}{12},\frac{13}{12};\frac{1}{4},\frac{1}{2};\frac{3 \pi ^2 \bar{\beta }^4}{\lambda ^3}\right)-2 \pi  \bar{\beta }^2 \left(80 \pi  \Gamma
\left(\frac{5}{12}\right) \Gamma \left(\frac{13}{12}\right) \Gamma \left(\frac{7}{4}\right) \bar{\beta }^2 \,
_2F_2\left(\frac{13}{12},\frac{17}{12};\frac{5}{4},\frac{3}{2};\frac{3 \pi ^2 \bar{\beta }^4}{\lambda ^3}\right)+24 \sqrt{3} \lambda ^{3/2} \Gamma \left(\frac{7}{12}\right)
\Gamma \left(\frac{11}{12}\right) \Gamma \left(\frac{5}{4}\right) \, _2F_2\left(\frac{7}{12},\frac{11}{12};\frac{1}{2},\frac{3}{4};\frac{3 \pi ^2 \bar{\beta }^4}{\lambda^3}\right)-21 \sqrt{3} \lambda ^{3/2}
\Gamma \left(\frac{7}{12}\right) \Gamma \left(\frac{11}{12}\right) \Gamma \left(\frac{5}{4}\right) \,
_2F_2\left(\frac{11}{12},\frac{19}{12};\frac{3}{4},\frac{3}{2};\frac{3 \pi ^2 \bar{\beta }^4}{\lambda ^3}\right)\right)\right)}{4 \pi ^2 \lambda ^{3/2} \left(8 \sqrt[4]{3}
\lambda ^{3/2} \Gamma \left(\frac{1}{12}\right) \Gamma \left(\frac{5}{12}\right) \Gamma \left(\frac{7}{4}\right) \,
_2F_2\left(\frac{1}{12},\frac{5}{12};\frac{1}{4},\frac{1}{2};\frac{3 \pi ^2 \bar{\beta }^4}{\lambda ^3}\right)-9\ 3^{3/4} \pi  \Gamma \left(-\frac{3}{4}\right) \Gamma
\left(\frac{7}{12}\right) \Gamma \left(\frac{11}{12}\right) \bar{\beta }^2 \, _2F_2\left(\frac{7}{12},\frac{11}{12};\frac{3}{4},\frac{3}{2};\frac{3 \pi ^2 \bar{\beta
}^4}{\lambda ^3}\right)\right)}$}
\label{pot1}
\end{eqnarray}
In particular, for the special case $\bar\beta\to0$, we have
\begin{eqnarray}
\left\langle \frac{V(\phi)}{H^4} \right\rangle= \frac{3}{32\pi^2}+\frac{9 \sqrt{3} \Gamma \left(\frac{7}{12}\right) \Gamma \left(\frac{11}{12}\right) \left[27 \Gamma \left(-\frac{3}{4}\right)
   \Gamma \left(\frac{3}{4}\right) \Gamma \left(\frac{13}{12}\right)-16 \Gamma \left(\frac{1}{12}\right) \Gamma \left(\frac{5}{4}\right) \Gamma
   \left(\frac{7}{4}\right)\right] \bar{\beta }^2}{256 \pi  \lambda^{3/2} \Gamma^2 \left(\frac{1}{12}\right) \Gamma \left(\frac{5}{12}\right) \Gamma^2
   \left(\frac{7}{4}\right)}+O\left(\bar{\beta }^3\right)
   \label{small}
\end{eqnarray}
For large $ \bar{\beta }^4/\lambda ^3$, we have
\begin{eqnarray}
\resizebox{1.0\hsize}{!}{$
 \left\langle \frac{V(\phi)}{H^4} \right\rangle=\frac{\lambda ^3 \bar{\beta }^2 \left(0.0013036 \left(\frac{\bar{\beta }^4}{\lambda ^3}\right)^{2/3}-0.0000626801\right)-0.0126651 \lambda ^{3/2} \bar{\beta }^4 \left( \sqrt{\frac{\bar{\beta }^4}{\lambda
   ^3}}+0.110318 \sqrt[6]{\frac{\bar{\beta }^4}{\lambda ^3}}\right)-0.000519354 \lambda ^{9/2} \sqrt{\frac{\bar{\beta }^4}{\lambda ^3}}+0.00339362 \bar{\beta }^6}{\frac{0.051464 \bar{\beta
   }^6}{\sqrt[3]{\frac{\bar{\beta }^4}{\lambda ^3}}}+ \lambda ^{9/2} \left(\frac{\bar{\beta }^4}{\lambda ^3}\right)^{3/2}+0.0137897 \lambda ^{9/2} \left(\frac{\bar{\beta }^4}{\lambda ^3}\right)^{7/6}+0.267949
   \bar{\beta }^6}$}   
\end{eqnarray}

\begin{figure}[h!]
\begin{center}
  \includegraphics[width=7.0cm]{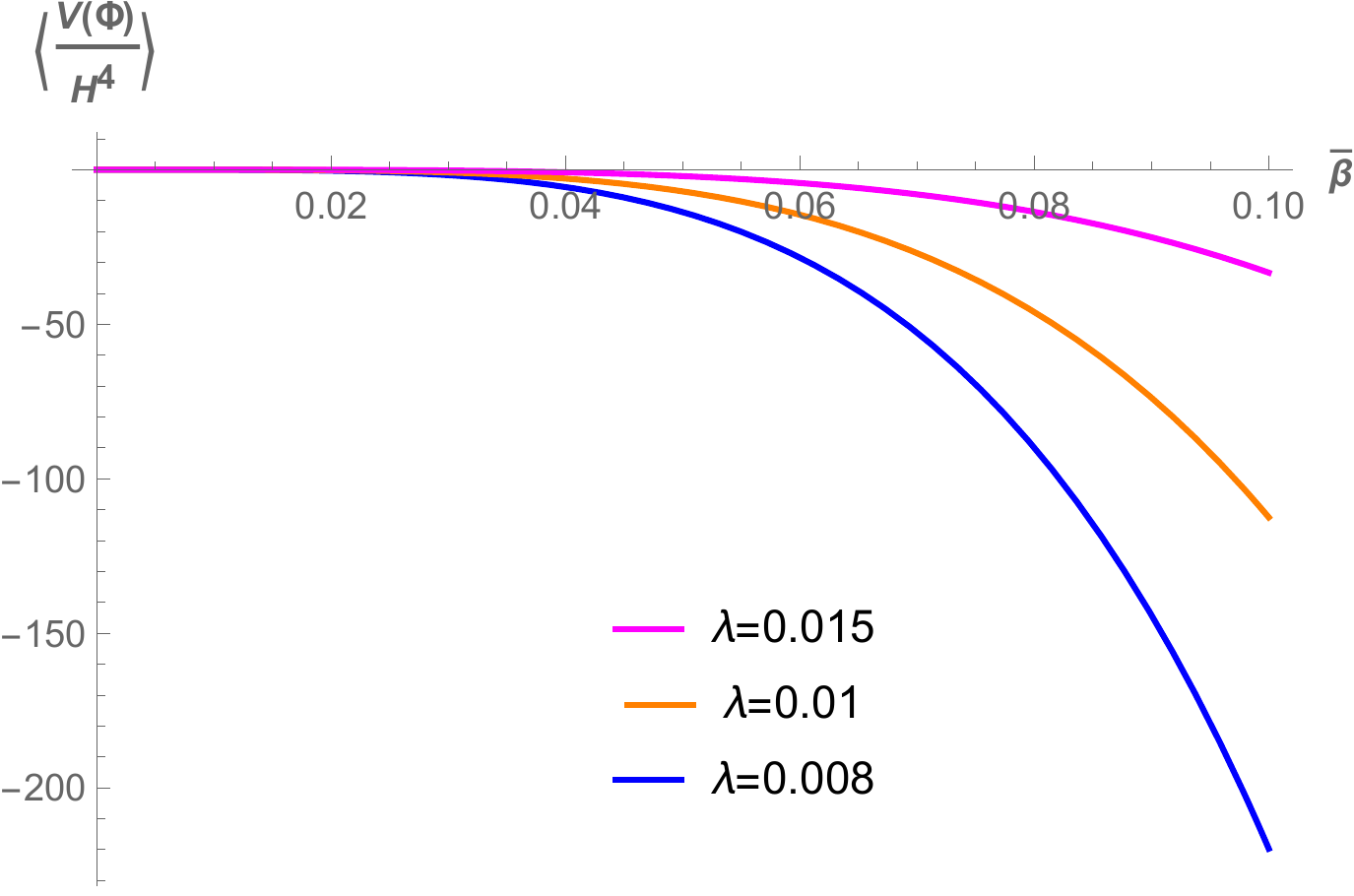}
 \includegraphics[width=7.5cm]{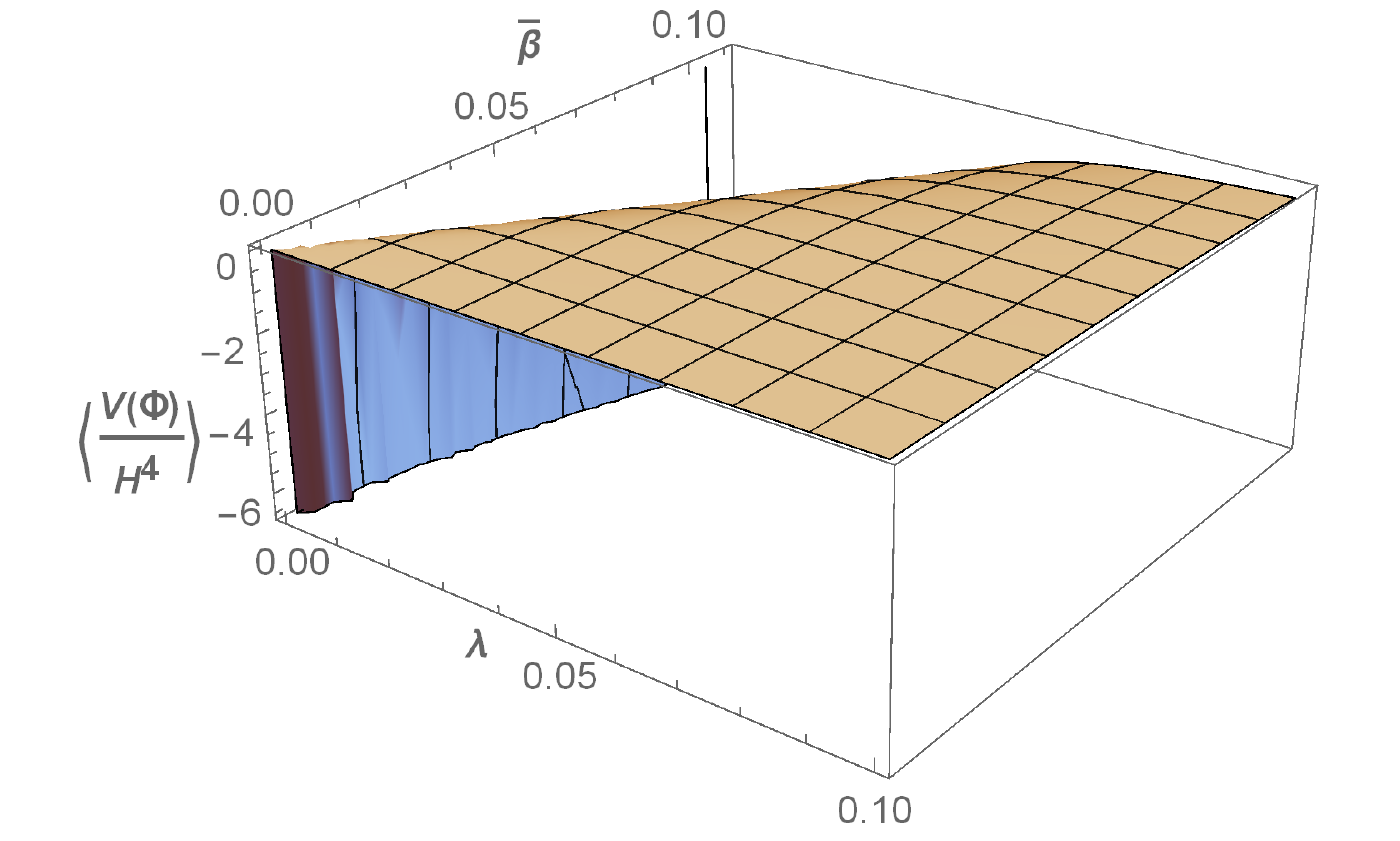}
  \caption{\small The variation of $\langle V(\phi)/H^4\rangle$  with respect to $\lambda$ and $\bar \beta$.  Note that as $\lambda$ decreases, the expectation value   decreases rapidly and unboundedly.  See main text for discussion.}
  \label{V}
\end{center}
\end{figure}

We have depicted the behaviour of \ref{pot1} in  \ref{V}.  As was expected from \ref{poten},  $\langle V(\phi) \rangle$ can be negative, and reducing $\lambda$ (equivalently, increasing $\bar \beta$) increases the negativity.  

We may compare  the $\langle V(\phi)\rangle$ obtained via the stochastic formalism, with the $V(\phi)$  estimated at the classical minimum, $-3\beta/\lambda$, given by $V_{\rm cl, min}=-9\beta^4/(8\lambda^3)$. While it is easy to see from \ref{small} that they do not match for small cubic coupling,   \ref{V} shows  that they match well when it gets strong, similar to the case of $\langle \bar{\phi}\rangle$.

We shall estimate the backreaction of $\langle \bar \phi\rangle$ and $\langle V(\phi)/H^4\rangle$ on the inflationary $\Lambda$ in \ref{backreaction}. However before we end this section, we wish to also compute the dynamically  generated mass in our potential. Such dynamical mass generation occurs due to the finite $\langle \phi^2 \rangle $ appearing at late times, due to the non-perturbative radiative corrections arising from self interaction. One compares  this non-perturbative expression of $\langle \phi^2 \rangle $  with that of a free massive minimally coupled scalar. Comparison of these two leads to the dynamically generated mass, which can have interesting footprints on the cosmological correlation functions~\cite{ Youssef:2013by, davis, Beneke:2012kn}. For a free massive scalar, $V(\phi)=m^2\phi^2/2$ and we have from \ref{ran}
\begin{eqnarray}
\langle\phi^2\rangle=\frac{3 H^4}{8 \pi ^2 m^2}
\end{eqnarray}
One can obtain the above result using quantum field theory as well. We conclude for our case from the above equation

\begin{eqnarray}\label{dynamical}
\frac{m^2_{\rm dyn}}{H^2}=\frac{3}{8\pi^2\langle\bar{\phi}^2\rangle} 
\end{eqnarray}
\begin{figure}[h!]
\begin{center}
  \includegraphics[width=6.5cm]{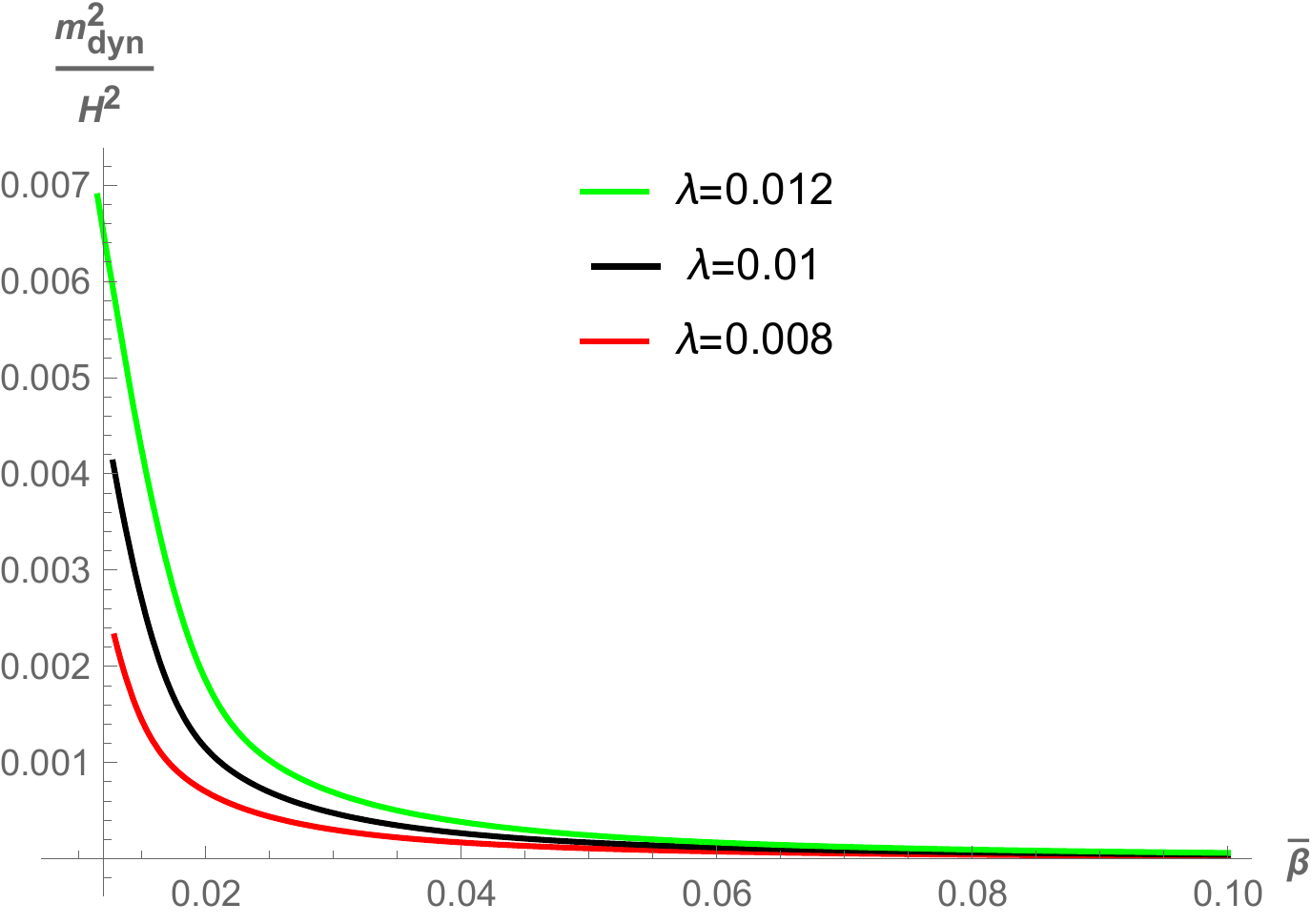}
  \includegraphics[width=7.5cm]{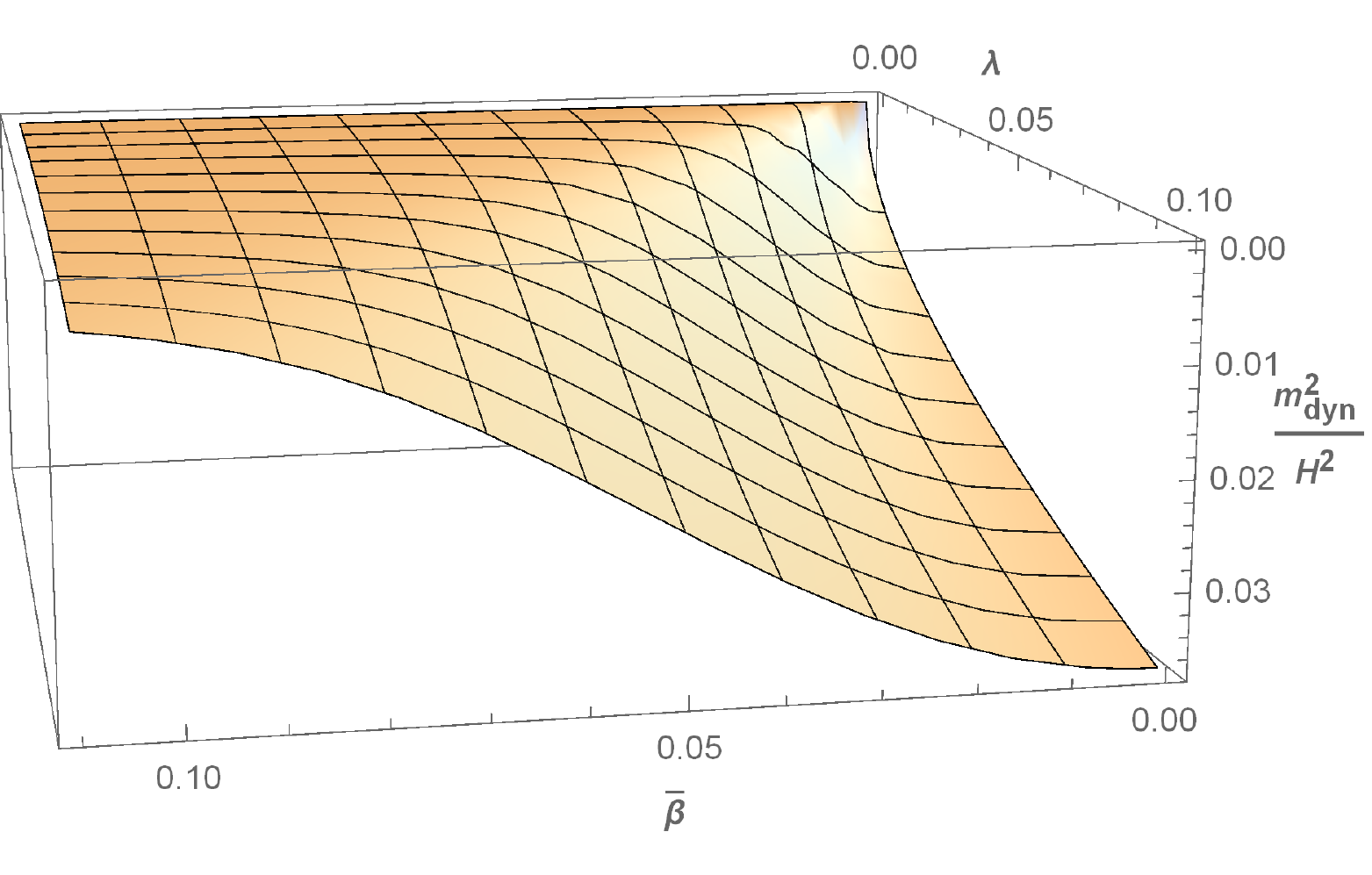}
  \caption{\small Variation of the dynamically generated mass $m^2_{\rm dyn}/H^2$ found via the stochastic formalism, \ref{dynamical}, with respect to $\lambda$ and $\overline{\beta}$. See  main text for discussion.}
  \label{dy'}
\end{center}
\end{figure}
where $\langle\bar{\phi}^2\rangle$ is given by \ref{av}.  Since a massive scalar field's two point correlations, no matter how tiny the mass is, obeys de Sitter invariance, the generation of the dynamical mass ensures that the de Sitter symmetry is after all not broken at late times.
We have plotted  $m^2_{\rm dyn}$  with respect to the parameters $\lambda$ and $\overline\beta$ in \ref{dy'}. Note that as we increase $\lambda$, $m^2_{\rm dyn}$ increases. On the other hand, $m^2_{\rm dyn} \to 0$ as $\lambda \to 0$, owing to the fact that the cubic potential is unbounded from below and hence unstable, as manifested by \ref{av}, which diverges for pure cubic self interaction.

\section{Estimation of late time backreactions from $\langle \phi \rangle$ and $\langle V(\phi)\rangle$}\label{backreaction}

In this Section, we estimate the backreactions of  $\langle\phi\rangle$ and $\langle V(\phi)\rangle$  computed stochastically in the preceding Section, to the
inflationary cosmological constant $\Lambda$. In~\cite{Bhattacharya:2022aqi}, a non-perturbative $\langle\phi\rangle$  was computed using quantum  field theory, and  it was shown that its contribution to the late time energy-momentum tensor is of the form $-2 \gamma \Lambda\langle\phi\rangle g_{\mu \nu}$, where $\gamma$ is a constant of length dimension minus one. Accordingly, the backreaction to the inflationary  $\Lambda$ in the Einstein equation will be,
\be
\Lambda \rightarrow \Lambda(1+16 \pi G \gamma\langle\bar\phi\rangle H) = \Lambda\left(1 +16 \pi \langle\bar\phi\rangle \times \frac{L_{P}^{2}}{L_{C} L_{\gamma}}\right) = \Lambda(1+\delta_1)\,\,\,\,({\rm say})
\label{L1}
\ee
where $H^{-1}=L_{C}$ is the length scale of the cosmological event horizon, $L_{\gamma}=\gamma^{-1}$ is the length scale associated with $\gamma$  and $\sqrt{G}=L_P = 1.61\times10^{-35}{\rm m}$ is the Planck length. The Hubble rate during the inflation is estimated as  $ H=8 \times 10^{13} \sqrt{r / 0.1} \mathrm{GeV}$~\cite{Enqvist:2017kzh}, where $r<0.03$ is the primordial tensor to scalar ratio. This yields, $L_{C}=H^{-1} \gtrsim 10^{-30} \mathrm{~m}$,  giving, $L_{P} / L_{C} \lesssim \mathcal{O}\left(10^{-5}\right)$. We do not possibly have any way to determine $L_{\gamma}$ uniquely from any physical principle. We have made some estimations of $\delta_1$ in \ref{tab:coeffslambda} using \ref{ssa} and \ref{phi},  by taking $L_{\gamma}$ to be as large as $L_C$ or as small as $L_P$. Any significant shift in the inflationary $\Lambda$ seems to be possible only if we consider the latter case. For this, and with $\bar\beta/ \lambda$ as large as $\mathcal{O}(10^{2})$ or even larger, there seems to be a significant shift in $\Lambda$ of $\mathcal{O}(1)$. However, we would like to emphasise that we are not sure whether such a large $\bar\beta/ \lambda$ is physically viable. It  needs to be determined by computing the observable cosmological correlation functions only, which we reserve  for  a future work.

Let us now estimate the backreaction due to  $\langle V(\phi) \rangle$ in the Einstein equation, given by
\be
\Lambda \rightarrow \Lambda\left(1+ \frac{8 \pi G}{3H^2}\langle V(\phi)\rangle\right) = \Lambda\left(1+ \frac{8 \pi \langle V(\phi)\rangle}{3H^4}\frac{L_{P}^{2}}{L^2_{C} }\right) = \Lambda(1+\delta_2)\,\,\,\,({\rm say})
\label{L2}
\ee
\begin{table}[]
    \centering
    \renewcommand{\arraystretch}{1.8}
    \begin{tabular}{|c|c|c|c|c|c|c|}
\hline
${\rm S.No.}$ & $\lambda$ & $\bar\beta$ & $\delta_1$ with $L_{\gamma} \sim L_{C}$ & $\delta_1$ with $L_{\gamma} \sim L_{P}$ & $\delta_2$ \\
        \hline
         $1$ & $0.0001\,$ & $0.1$ & $-0.15\times 10^{-4}$ & $-0.15\times 10^{1}$ & $-10^{-1}$  \\
         \hline
         $2$ & $0.001$ & $0.1$ & $-0.15\times 10^{-5}$ & $-0.15\times 10^{0}$ & $-10^{-4}$ \\
         \hline
         $3$ & $0.01\,$ & $0.1$ & $-0.15\times 10^{-6}$  & $-0.15\times 10^{-1}$ & $-10^{-7}$
          \\
         \hline
         $4$ & $0.1$ & $0.1$ & $-0.13\times 10^{-7}$ & $-0.13\times 10^{-2}$ &  $-0.7\times 10^{-10}$ 
         \\ 
         \hline
         $5$ & $0.1$ & $0.01$ & $-0.5\times 10^{-9}$ & $-0.5\times 10^{-4}$ & $0.077\times 10^{-10}$ 
         \\
          \hline
         $6$ & $0.1$ & $0.001$ & $-0.5\times  10^{-10}$ & $-0.5\times 10^{-5}$ & $0.078\times 10^{-10}$ 
         \\
       \hline
    \end{tabular}
    \caption{ Estimation of $\delta_1$, \ref{L1}, for $L_{\gamma} \sim L_{C}$ and $L_{\gamma} \sim L_{P}$, and $\delta_2$, \ref{L2}, for different values of $\lambda$ and $\bar\beta$. See main text for discussion.}
    \label{tab:coeffslambda}
\end{table}
Using \ref{V}, we estimate $\delta_2$ in \ref{tab:coeffslambda} for different values of the coupling parameters.  Note  in \ref{tab:coeffslambda} that when $\bar\beta/\lambda$ becomes small, $\delta_2$ becomes positive. This correspond to the fact that for $\bar\beta = 0$, $\langle V(\phi)\rangle$ becomes a positive operator, whose expectation value must be positive.

In the next section we shall do the non-perturbative analysis for $\langle \phi^2 \rangle$  via quantum field theory.

\section{Non-perturbative $\langle \phi^2 \rangle$  via quantum field theory}\label{phiresum}

The non-perturbative expectation values we have found via the stochastic method in \ref{stoch}, at least in principle,  can also be computed using quantum field theory. However  as we have discussed in \ref{S1}, the perturbative results generated by loops are plagued by the monotonically growing secular logarithms of the scale factor at late times, clearly indicating breakdown of the perturbation theory.  Obtaining any physically sensible result then strictly  necessitates some kind of resummation. However, the standard dynamical renormalisation group techniques may not be very useful for this purpose, chiefly due to the fact that the scale factor $a(t)$ does not lead to simple scaling as that of the flat spacetime quantum field theory~\cite{Miao:2021gic} (also references therein). Recently, a dynamical renormalisation group inspired autonomous method to resum the secular logarithms for the quartic self interaction was proposed  in~\cite{Kamenshchik:2020yyn, Kamenshchik:2021tjh}. This method involves constructing a first order differential equation out of the perturbation series, then promoting various quantities to non-perturbative level to increase the domain of the equation and finally integrating it. This  was used to resum the two point correlators for the quartic self interaction and the resultant $m_{\rm dyn}^2$ was in good agreement with that of the stochastic formalism~\cite{Starobinsky:1994bd}.  In~\cite{Bhattacharya:2022aqi}, this formalism was used to compute $\langle \phi \rangle$ for our quartic plus cubic potential. It was shown that the result differs from that of the stochastic formalism by approximately order unity multiplicative factors, although they have similar qualitative features. We wish to apply the same resummation technique below      to compute $m_{\rm dyn}^2$ non-perturbatively. In this case we shall also see that although it has  qualitative behaviour similar to that of the stochastic formalism, the numerical values differ.  We shall discuss some possible reasons behind such disagreement in~\ref{discussion}.

We assume that the initial ($t\to 0$) vacuum  is the Bunch-Davies state. Since the Bunch Davies vacuum is defined with respect to the free part of \ref{l4}, this means that the field must be located around the flat plateau around $\phi\sim 0$ in \ref{poten}, {\it initially}. As time goes on, the field would tend to move towards the minimum of $V(\phi)$. However, we expect $V(\phi)$ to receive corrections due to strong quantum effects and accordingly, the position of its minima should change. Indeed, we shall emphasise the  manifestation of strong quantum/stochastic effects towards  the end of \ref{resum}.

We would first like to show below that at least for our $V(\phi)$,  we may not be able to resum a perturbation series containing contributions from both    local and non-local parts of the self energies very efficiently for $\langle \phi^2 \rangle$, using the formalism of~\cite{Kamenshchik:2020yyn, Kamenshchik:2021tjh}. The {\it local part} corresponds to the scenario when a self energy loop entirely shrinks to a {\it single} point, and hence contains a $\delta$-function which identifies the two external vertices of the corresponding Feynman diagram. Clearly, such local parts must be related to renormalisation, after which, the reminiscent finite part will yield the desired local contribution to the self energy.
The self energy contributions correspond to the amputated versions of the one particle irreducible (1PI) diagrams for the two point correlator. The one and two loop Feynman diagrams for $\langle \phi^2 (x)\rangle$  are given by \ref{f1}, \ref{f2} and \ref{f3}. Among these, only the ${\cal O}(\lambda)$ contribution is purely local, \ref{lo1}, whereas the contribution at ${\cal O}(\beta^2)$,  ${\cal O}(\lambda^2)$ and ${\cal O}(\lambda \beta^2)$ are partially local and non-local. These local contributions to $\langle \phi^2(x)\rangle$ have been obtained below in \ref{lo1}, \ref{lo3},  \ref{lambdasquare}, and \ref{lo4}. Whereas the derivation of the non-local contributions, since they will not be directly useful  to the main results of this paper,  have been kept in \ref{nlo1}, \ref{nlo2} and \ref{nlo3}.  
Although excellently detailed self energy computation at one and two loop for the quartic self interaction can be seen in~\cite{Brunier:2004sb}, we have retained some calculations for the same  here,  first because we are interested to compute the two point correlators. And second, we shall  be using in many instances the infrared (IR) effective propagators outlined in~\ref{corr1},  instead of the full ones. The late time IR effective computations for the quartic self interaction can also be found in~\cite{Onemli:2015pma}, but the perturbative expansion of the IR field seems to be explicitly used there instead of the  in-in or diagrammatic  techniques. Although both methods should be equivalent, we shall keep below the derivations on the quartic sector briefly, for the sake of completeness and clarity.  

Collecting now the results of \ref{nlo1}, \ref{nlo2} and \ref{nlo3}, we have the leading late time secular contribution
\begin{eqnarray}
&&\langle \phi^2(x)\rangle_{\rm free}=\frac{H^2}{4\pi^2} \ln a,\qquad   \langle \phi^2(x)\rangle_{\lambda}=   -\frac{\lambda H^2}{2^4\times 9\pi^4} \ln^3 a,\qquad \langle \phi^2(x)\rangle_{\beta^2}=\frac{\beta^2}{2^5\times 9\pi^4}\ln^4 a\nonumber\\
&&\langle \phi^2(x)\rangle_{\lambda^2} = \frac{\lambda^2 H^2}{2^6  \times 45 \pi^6} \ln^5 a, \qquad \langle \phi^2(x)\rangle_{\lambda \beta^2}= -\frac{\lambda \beta^2}{2^{10}\times 9 \pi^6} \frac{29 \ln^6 a}{27}
\label{cc17}
\end{eqnarray}
where the free correlator in the coincidence limit has been computed in \ref{corr1} (\ref{free}). Adding now the contributions of \ref{cc17}, abbreviating $\langle {\phi}^2(x)\rangle/H^2$ as $f$, and differentiating it once with respect to the number of e-foldings, ${\cal N}=\ln a= Ht$, we have
 \begin{eqnarray}
 \frac{d f}{d {\cal N}}=c_1 -3 c_2 \lambda {\cal N}^2 + 4 c_3 \overline\beta^2{\cal N}^3 +5 c_4 \lambda^2{\cal N}^4 -6 c_5 \lambda \overline\beta^2 {\cal N}^5
\label{kk26}
 \end{eqnarray}
 where ${\bar \beta}=\beta/H$ as earlier, and we have made abbreviations for the sake of brevity
\begin{eqnarray}
c_1=\frac{1}{4\pi^2},\qquad  c_2=\frac{1}{2^4\times 9 \pi^4},\qquad c_3=\frac{1}{2^5\times 9 \pi^4},\qquad c_4=\frac{1}{2^6\times 45 \pi^6},\qquad c_5=\frac{29 }{2^{10}\times 9 \pi^6\times 27}
\label{kk3}
\end{eqnarray}
Following~\cite{Kamenshchik:2020yyn, Kamenshchik:2021tjh}, we now replace ${\cal N}$ by $f$ in order to promote it to non-perturbative level. In order to be consistent with the perturbative expansion, we need to replace ${\cal N}$ simply by $f$ in for the ${\cal O}(\lambda^2)$ and ${\cal O}(\lambda \beta^2)$ terms in \ref{kk26}, whereas  for the ${\cal O}(\lambda)$ term,  ${\cal N}$ should contain further ${\cal O}(\lambda)$ and ${\cal O}(\beta^2)$ terms. Similar thing should be kept in mind for the ${\cal O}(\beta^2)$ term. Putting these in together, we have 
\begin{eqnarray}
 \frac{d f}{d {\cal N}}=c_1 - \frac{3 c_2 \lambda f^2}{c_1^2}+\frac{4 c_3 \overline{\beta}^2 f^3}{c_1^3}  + \lambda^2 \left(\frac{5c_4}{c_1^4}-  \frac{6c_2^2}{c_1^5} \right) f^4+6\lambda \overline\beta^2 \left(\frac{3c_2 c_3}{c_1^6}-  \frac{c_5}{c_1^5} \right) f^5
\label{kk88}
\end{eqnarray}
The next task is to integrate the above equation and find $f$ at late times, ${\cal N}\gg 1$.

\ref{kk88} can be solved numerically using Mathematica. However, it is easy to see that for $|\bar{\beta}| \gtrsim \lambda$, it yields complex $f=\langle \phi^2(x) \rangle$. This is unacceptable, as $\phi$ is Hermitian. Similar feature holds if we ignore the ${\cal O}(\lambda \beta^2)$ term in \ref{kk26}. Perhaps adding further higher order corrections might solve this issue. However, it seems reasonable to conclude that at least for the $V(\phi)$ with two coupling parameters we are interested in, the formalism proposed  
in~\cite{Kamenshchik:2020yyn, Kamenshchik:2021tjh} may not be very efficient to resum simultaneously the local and non-local contributions of self energies to  $\langle \phi^2(x) \rangle$. Hence let us now attempt to resum the part of  $\langle \phi^2(x) \rangle$ that contains only the local part of the self energy. Such non-perturbative $\langle \phi^2(x) \rangle_{\rm loc}$ will also be related to the dynamical generation of mass, as has been discussed in e.g.~\cite{Youssef:2013by} for the quartic self interaction using the Schwinger-Dyson resummation technique. The computation of such local contributions requires evaluation of diagrams \ref{f1}, the first, second and fourth of \ref{f2} and the first and second of \ref{f3}. It can be explicitly seen that none of the other diagrams, being  1-particle reducible (1PR), contains any local contribution. 

\subsection{One loop ${\cal O}(\lambda)$ local contribution to $\langle \phi^2 (x) \rangle$}\label{lo1}
\begin{figure}[h!]
		\includegraphics[height=3.0cm]{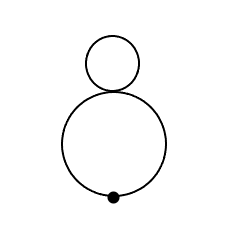}\centering \hskip 2cm
		\caption{ \footnotesize 1-loop ${\cal O}(\lambda)$ correction  to the two-point correlation function in the coincident limit, $\langle \phi^2(x)\rangle$. The dot denotes the spacetime point $x$, where the two external propagators are identified.}
		\label{f1}
	\end{figure}

\begin{figure}[h!]
		\includegraphics[height=3.5cm]{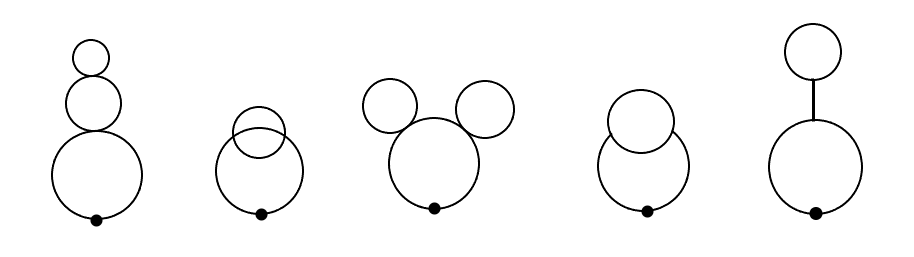}\centering \hskip 2cm
		\caption{ \footnotesize 2-loop ${\cal O}(\lambda^2)$ correction  to  $\langle \phi^2(x)\rangle$. The last two represent the same for one loop at ${\cal O}(\beta^2)$. The dot denotes the spacetime point $x$, where the two external propagators are identified.}
		\label{f2}
	\end{figure}

\begin{figure}[h!]
		\includegraphics[height=7cm]{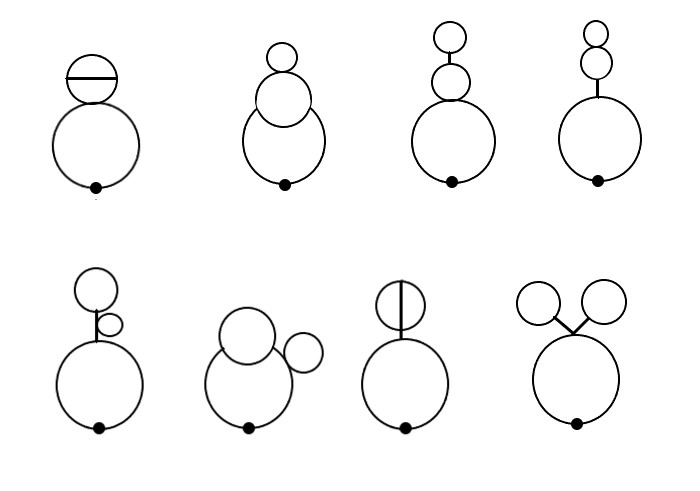}\centering \hskip 2cm
		\caption{ \footnotesize 2-loop ${\cal O}(\lambda \beta^2)$ correction to $\langle \phi^2(x)\rangle$. The dot denotes the spacetime point $x$, where the two external propagators are identified.}
		\label{f3}
	\end{figure}

In order to save space and for the sake of tidiness, we shall abbreviate in the following for the powers of the propagators and logarithms, $(i\Delta(x,x'))^n\equiv i\Delta^n(x,x')$, and $\ln^n a \equiv (\ln a)^n$. Also, for the diagrams which entirely contain local contribution to the self energy, we shall employ the IR effective propagators of \ref{corr1}, for the sake of simplicity in computation. The diagrams which make partial local contributions to the self energy, needs to be evaluated using the full propagators to know the local contribution. It is easy to see by amputating the external legs of the  diagrams  that only the one loop bubble (\ref{f1}) contains purely local contribution from the self energy. 
The bubble diagram contribution to the two point correlator at ${\cal O}(\lambda)$ in the in-in formalism outlined in \ref{A} is given by,
\begin{eqnarray}
\langle \phi^2(x)\rangle_{\lambda}=-\frac{i\lambda}{2} \int d \eta' a'^4 d^3{\vec x'} \left(i\Delta^2_{++}(x,x')-i\Delta^2_{+-}(x,x')\right) i\Delta_{+-}(x',x')
\label{nc5}
\end{eqnarray}
where the field species in the external line are taken to be `$+$' type, as they must evolve forward in time.

Using now the first of \ref{nc1}, we note that
$$i\Delta_{++}^2(x,x')-i\Delta_{+-}^2(x,x')= \theta(\eta -\eta')i\Delta_{-+}^2(x,x') + \theta(\eta' -\eta)i\Delta_{+-}^2(x,x')- i\Delta_{+-}^2(x,x')$$
where we have used the fact that $\theta^2 \equiv \theta$ and $\theta(\eta-\eta')\theta(\eta'-\eta)=0$. Since $\eta$ is the final observation time, we have  $\eta \gtrsim \eta'$. Thus
$$i\Delta^2_{++}(x,x')-i\Delta^2_{+-}(x,x') \equiv  i \Delta_{-+}^2(x,x') - i\Delta_{+-}^2(x,x')= (i \Delta_{-+}(x,x') + i\Delta_{+-}(x,x'))(i \Delta_{-+}(x,x') - i\Delta_{+-}(x,x'))$$
Using these into \ref{nc5}, and using the 3-momentum description described in \ref{corr1}, we have
\begin{eqnarray}
&&\langle \phi^2(x)\rangle_{\lambda}=\frac{i\lambda}{2} \int d \eta' a'^4 d^3{\vec x'} \int \frac{d^3{\vec k_1} d^3{\vec k_2}}{(2\pi)^6} e^{i({\vec k_1}+{\vec k_2})\cdot \vec{x}} e^{-i({\vec k_1}+{\vec k_2})\cdot \vec{x'}} \nonumber\\ &&\times \left(i\Delta_{+-}(k_1,\eta,\eta')+i\Delta_{-+}(k_1,\eta,\eta')\right)\left(i\Delta_{+-}(k_2,\eta,\eta')-i\Delta_{-+}(k_2,\eta,\eta')\right) \int \frac{d^3 {\vec k_3}}{(2\pi)^3} i\Delta(k_3, \eta',\eta') \nonumber\\
&&=\frac{i\lambda}{2} \int d \eta' a'^4 \int \frac{d^3{\vec k_1} }{(2\pi)^3}\left(i\Delta_{+-}(k_1,\eta,\eta')+i\Delta_{-+}(k_1,\eta,\eta')\right)\left(i\Delta_{+-}(k_1,\eta,\eta')-i\Delta_{-+}(k_1,\eta,\eta')\right)\int \frac{d^3 {\vec k_3}}{(2\pi)^3} i\Delta(k_3, \eta',\eta') \nonumber\\
\label{nc6}
\end{eqnarray}

Using now \ref{nc4}, the fact that $\eta' \lesssim \eta$ and recalling the lower limit of the momentum integral is the Hubble rate, $H$, in our IR effective super-Hubble framework~\cite{Onemli:2015pma}, we have at the leading order, 
\begin{eqnarray}
&&\langle \phi^2(x)\rangle_{\lambda}=\frac{i\lambda}{2} \int d\eta' a'^4 \int \frac{d^3 {\vec k_1}}{(2\pi)^3} \frac{H^2}{k_1^3} \frac{i}{3 H a'^3} \theta(Ha'-k_1)\theta(Ha-k_1) \int \frac{d^3 {\vec k_2}}{(2\pi)^3 k_2^3} \theta(Ha-k_2)\theta(Ha'-k_2)\nonumber\\
&&= -\frac{\lambda H^2}{2^4\times 3\pi^4} \int_{1}^{a} da' \frac{\ln^2a'}{a'} =  -\frac{\lambda H^2}{2^4\times 9\pi^4} \ln^3 a
\label{nc7}
\end{eqnarray}
%

\subsection{Local contribution to the self energy at ${\cal O}(\beta^2)$}\label{lo3}
We have mentioned earlier that only the fourth diagram contributes locally to $\langle\phi^2\rangle$ at $\mathcal{O}(\beta^2)$ of \ref{f2}. We have
\begin{eqnarray}
\langle \phi^2(x)\rangle_{\beta^2}=-\frac{\beta^2}{2} \int (a'a'')^d d^d x' d^d x'' \left[ i\Delta_{-+}(x,x') i\Delta_{-+}(x,x'')i\Delta^2_{++}(x',x'')+i\Delta_{+-}(x,x') i\Delta_{+-}(x,x'')i\Delta^2_{--}(x',x'')  \right. \nonumber\\ \left.
- i\Delta_{-+}(x,x') i\Delta_{+-}(x,x'')i\Delta^2_{+-}(x',x'')-i\Delta_{+-}(x,x') i\Delta_{-+}(x,x'')i\Delta^2_{-+}(x',x'')  \right]
\label{lo52}
\end{eqnarray}
Only $i\Delta^2_{++}(x',x'')$ and $i\Delta^2_{--}(x',x'')$ will yield $\delta$-functions. We have using the full propagator~\cite{Brunier:2004sb}
\be
i\Delta^2_{++}(x',x'')= \frac{(a'a'')^{-2+\e} \Gamma^2(1-\e/2)}{2^4\pi^{4-\e}} \frac{1}{\Delta x^{4-2\e}_{++}}+ \frac{H^4}{2^6\pi^4}\ln^2 \frac{\sqrt{e}H^2 \Delta x_{++}^2}{4} - \frac{H^2(a'a'')^{-1}}{2^4 \pi^4} \frac{\ln \frac{\sqrt{e}H^2 \Delta x_{++}^2}{4}}{\Delta x_{++}^2}
\label{lo53}
\ee
Using next
\be
\frac{1}{\Delta_{++}^{4-2\e}}= - \frac{2i \mu^{-\e} \pi^{2-\e/2}}{\e(1-\e) \Gamma(1-\e/2)} \delta^d(x'-x'')
\label{lo54}
\ee
we have
\be
i\Delta^2_{++}(x',x'')= -\frac{i \mu^{-\e}(a')^{-4+2\e} \Gamma(1-\e/2)}{2^3\pi^{2-\e/2}\e(1-\e)} \delta^d(x'-x'')+ \frac{H^4}{2^6\pi^4}\ln^2 \frac{\sqrt{e}H^2 \Delta x_{++}^2}{4} - \frac{H^2(a'a'')^{-1}}{2^4 \pi^4} \frac{\ln \frac{\sqrt{e}H^2 \Delta x_{++}^2}{4}}{\Delta x_{++}^2}
\label{lo55}
\ee
Thus the local part of the self energy  will correspond to 
\be
\frac{i\mu^{-\e}\beta^2 a'^{4-\e+\e} \Gamma(1-\e/2)}{2^4 \pi^{2-\e/2} \e (1-\e)}\delta^d(x'-x'')= \left[\frac{i\mu^{-\e}\beta^2 a'^{d} \Gamma(1-\e/2)}{2^4 \pi^{2-\e/2} \e (1-\e)}+\frac{i\beta^2}{2^4\pi^2} a'^{4} \ln a'\right]\delta^d(x'-x'')
\label{lo56}
\ee
The divergence can be absorbed in a mass renormalisation counterterm,
$$\delta m^2 =  \frac{\mu^{-\e}\beta^2\Gamma(1-\e/2)}{2^4 \pi^{2-\e/2} \e (1-\e)} $$
whereas the finite part is the local contribution to the self energy. The contribution of $i\Delta_{--}^2(x',x'')$ is given by the complex conjugation of \ref{lo55}.  Thus we have the renormalised expression
\be
\langle \phi^2(x)\rangle_{\beta^2,\,{\rm loc}}= \frac{i\beta^2}{2^4\pi^2} \int d^4 x' a'^4 \ln a'  \left(i\Delta_{-+}^2(x,x')- i\Delta_{+-}^2(x,x')\right)
\label{lo57}
\ee
The above integral, being ultraviolet finite, can be evaluated as of \ref{lo1} to obtain at the leading order
\be
\langle \phi^2(x)\rangle_{\beta^2,\,{\rm loc}}= \frac{\beta^2}{2^5\times 9 \pi^4}\ln^3 a
\label{lo58}
\ee

Derivation of $\langle \phi^2 (x)\rangle $ at ${\cal O}(\beta^2)$ containing non-local contribution of the self energy using IR effective quantum field theory can  be seen in~\ref{nlo1}. 



\subsection{Two  loop ${\cal O}(\lambda^2)$ local contribution to $\langle \phi^2 (x) \rangle$}\label{lambdasquare}

There are three contributions here -- the snowman (the first of \ref{f2}), the sunset (the second of \ref{f2}) and the third of \ref{f2}. The snowman and sunset contributes partially locally to the self energy. All the non-local contributions of \ref{f2} is evaluated in \ref{nlo1}. We have (using $\eta \gtrsim \eta'$)
\begin{eqnarray}
&&\langle \phi^2(x)\rangle_{\lambda^2,\,{\rm sm}}=-\frac{\lambda^2}{4}\int a'^4 a''^4 d^4 x' d^4 x'' i\Delta (x'',x'') \left[i\Delta_{++}^2(x,x') (i\Delta_{++}^2(x',x'')-i\Delta_{+-}^2(x',x'')) \right. \nonumber\\ && \left.+ i\Delta_{+-}^2(x,x') (i\Delta_{--}^2(x',x'')-i\Delta_{-+}^2(x',x'')) \right]
 \label{nc8}
\end{eqnarray}
The local part corresponds to 
\begin{eqnarray}
-\frac{\lambda^2}{4}\int a'^4 a''^4 d^4 x' d^4 x'' i\Delta (x'',x'') \left[i\Delta_{-+}^2(x,x') i\Delta_{++}^2(x',x'') + i\Delta_{+-}^2(x,x') i\Delta_{--}^2(x',x'') \right]
 \label{nc9}
\end{eqnarray}
We substitute~\cite{Brunier:2004sb} 
$$i \Delta(x,x)=\frac{H^{2-\e}}{2^{2-\e} \pi^{2-\e/2}}\frac{\Gamma(2-\e)}{\Gamma(1-\e/2)}\left(\ln a +\frac{1}{\e} \right)$$
into the above. The divergence can be absorbed using the one loop bubble mass counterterm. We next use \ref{lo55}, \ref{lo56} into the above. The $\delta$-function takes $x''$ to $x'$, as required for the local self energy. The divergence can be absorbed by the quartic vertex counterterm~\cite{Brunier:2004sb}, to give the renormalised local contribution
\be
-\frac{i\lambda^2 H^2}{2^7\pi^4}\int d^4 x' a'^4 \ln^2 a' \left( i\Delta_{+-}^2(x,x')-i\Delta_{-+}^2(x,x')\right)
\label{nc10}
\ee
The leading late time contribution of the above can be derived as of \ref{lo1}, to yield
\begin{eqnarray}
&&\langle \phi^2(x)\rangle_{\lambda^2,\,{\rm sm, loc}}=\frac{\lambda^2 H^2}{2^{10} \times 3 \pi^6}\ln^4 a
\label{nc111}
\end{eqnarray}

Let us now evaluate the local contribution from the sunset diagram. We have
\begin{eqnarray}
&&\langle \phi^2(x)\rangle_{\lambda^2,\,{\rm ss}}=-\frac{\lambda^2 }{6}\int a'^d a''^d d^d x' d^d x''\left[i\Delta_{++}(x,x') i\Delta_{++}(x,x'') i\Delta_{++}^3(x',x'')- i\Delta_{++}(x,x') i\Delta_{+-}(x,x'') i\Delta_{+-}^3(x',x'')  \right. \nonumber\\ && \left. - i\Delta_{+-}(x,x') i\Delta_{++}(x,x'') i\Delta_{-+}^3(x',x'')+i\Delta_{+-}(x,x') i\Delta_{+-}(x,x'') i\Delta_{--}^3(x',x'')  \right] \nonumber\\
&&=-\frac{\lambda^2 }{6}\int (a' a'')^d d^d x' d^d x''\left[i\Delta_{-+}(x,x') i\Delta_{-+}(x,x'') i\Delta_{++}^3(x',x'')- i\Delta_{-+}(x,x') i\Delta_{+-}(x,x'') i\Delta_{+-}^3(x',x'')  \right. \nonumber\\ && \left. - i\Delta_{+-}(x,x') i\Delta_{-+}(x,x'') i\Delta_{-+}^3(x',x'')+i\Delta_{+-}(x,x') i\Delta_{+-}(x,x'') i\Delta_{--}^3(x',x'')  \right]
\label{nc12}
\end{eqnarray}
where we have used $\eta \lesssim \eta'$. The cube of the (anti)-Feynman propagators contain $\delta$-functions~\cite{Brunier:2004sb}, and hence the local part corresponds to
\begin{eqnarray}
-\frac{\lambda^2 }{6}\int a'^4 a''^4 d^4 x' d^4 x''\left[i\Delta_{-+}(x,x') i\Delta_{-+}(x,x'') i\Delta_{++}^3(x',x'')+i\Delta_{+-}(x,x') i\Delta_{+-}(x,x'') i\Delta_{--}^3(x',x'') \right]
\label{nc12'}
\end{eqnarray}
We have~\cite{Brunier:2004sb}
\be
(a'a'')^d i \Delta_{++}^3(x',x'')\vert_{\rm loc} = \frac{3iH^{2-\e}\mu^{-\e} a'^{d+\e}}{2^{6-\e}\pi^{4-\e}} \left[ \frac{(1-\e/2)(1-\e/4)\Gamma^2(1-\e/2)}{(1-3\e/2)\e^2} \left(\frac{H}{2\mu} \right)^{\e} a'^{\e}- \frac{2\Gamma(1-\e)}{\e} \ln a' - \frac{2\Gamma(1-\e)}{\e^2}\right]\delta^d(x'-x'')
\label{ss1}
\ee
which we substitute into \ref{nc12}. The divergences can be cancelled by the mass renormalisation as well as by some other diagrams.  The rest of the diagram is ultraviolet finite and hence can be evaluated as of \ref{lo1}. It turns out that the contribution exactly equals \ref{nc111}. Thus we have the total local contribution 
\begin{eqnarray}
&&\langle \phi^2(x)\rangle_{\lambda^2,\,{\rm loc}}=\frac{\lambda^2 H^2}{2^{9} \times 3 \pi^6}\ln^4 a
\label{ss2}
\end{eqnarray}

Derivation of $\langle \phi^2 (x)\rangle $ at ${\cal O}(\lambda^2)$ containing non-local contribution of the self energy using IR effective quantum field theory can  be seen in~\ref{nlo2}. 

\subsection{Two loop  ${\cal O}(\lambda \beta^2)$ local contribution to $\langle \phi^2 (x) \rangle$}\label{lo4}
%
There are total eight diagrams associated with the ${\cal O}(\lambda \beta^2)$ correction to the two point correlator, as shown in \ref{f3}. We wish to evaluate below the part of the first diagram that contributes partially locally to the self energy. None of the other diagrams, including the 1PI second one, of \ref{f3} contains any local contribution. While such non-contribution seems obvious for the 1PR diagrams, we note that the $\delta$-function required for a local self energy must come from the square or cube of the (anti-)Feynman propagators, \ref{ss1}, \ref{lo55}. It is easy to see simply by observing the second of \ref{f3} that no such scenario is possible for it.
The non-local part of all these diagrams  has been evaluated in \ref{nlo3}. 

For the local part, we have for the first of \ref{f3},
\begin{eqnarray}
&&\frac{i\lambda\beta^2}{4} \int (a' a'' a''')^d d^d x' d^d x'' d^d x'''  \left[i\Delta_{-+}^2(x,x''')\left\{i\Delta_{++}(x''',x'') i\Delta_{++}(x''',x') i\Delta_{++}^2(x',x'')\right.\right. \nonumber\\ &&\left.\left.- i\Delta_{++}(x''',x'') i\Delta_{+-}(x''',x') i\Delta_{-+}^2(x',x'')+i\Delta_{+-}(x''',x'') i\Delta_{+-}(x''',x') i\Delta_{--}^2(x',x'')) \right.\right.\nonumber\\&&\left.\left. -i\Delta_{+-}(x''',x'') i\Delta_{++}(x''',x') i\Delta_{+-}^2(x',x'')\right\}-{\rm c.c.} \right] 
\label{nc25}
\end{eqnarray}
where `c.c.' stands for complex conjugation. The local contribution from the above expression should only correspond to 
\begin{eqnarray}
&&\frac{i\lambda\beta^2}{4} \int (a' a'' a''')^d d^d x' d^d x'' d^d x'''  \left[i\Delta_{-+}^2(x,x''')i\Delta_{++}(x''',x'') i\Delta_{++}(x''',x') i\Delta_{++}^2(x',x'')\right. \nonumber\\ &&\left.-i\Delta_{+-}^2(x,x''')i\Delta_{--}(x''',x'') i\Delta_{--}(x''',x') i\Delta_{--}^2(x',x'')\right] 
\label{nc25'}
\end{eqnarray}
We first use \ref{lo55} for $i\Delta_{++}^2(x',x'')$ and $i\Delta_{--}^2(x',x'')$ above. The divergence can be absorbed using the counterterm diagram containing the one loop mass counterterm given below \ref{lo56}. Performing the $\delta$-function integration over $x''$ we use once again \ref{lo55} for the resulting $i\Delta^2_{++}(x''',x')$ and $i\Delta^2_{--}(x''',x')$. The resulting ${\cal O}(\e^{-1})$ and ${\cal O}(\e^{-2})$ are absorbed respectively by the quartic vertex and mass counterterm diagrams respectively with 
\begin{eqnarray}
\delta \lambda = -\left(\frac{\mu}{H}\right)^{-\e}\frac{\lambda \beta^2 \Gamma^2(1-\e/2)}{2^{5+\e}\pi^2 H^2\Gamma(2-\e)(1-\e)\e},  \qquad \delta m^2= \frac{\mu^{-\e}\lambda \beta^2 \Gamma(1-\e/2)}{2^8\pi^{4-\e/2}\e^2(1-\e)}
\label{nc25''}
\end{eqnarray}
The remaining integral is ultraviolet finite and can be evaluated using the IR effective tool of \ref{lo1}, to give 
\begin{eqnarray}
 \langle \phi^2(x)\rangle_{\lambda \beta^2,\,{\rm loc.} } = - \frac{\lambda\beta^2}{2^{11} \times 3\pi^6} \ln^4 a
\label{nc26n}
\end{eqnarray}
%

\section{Resummation of $\langle \phi^2\rangle_{\rm local}$ and dynamical mass generation  via quantum field theory}\label{resum}

In this section, we wish to once again try the formalism proposed in~\cite{Kamenshchik:2020yyn, Kamenshchik:2021tjh} outlined at the beginning of \ref{phiresum}, to resum  $\langle \phi^2 \rangle_{\rm loc}$, computed in the preceding sections up to two loop.   Combining \ref{free}, \ref{nc7}, \ref{lo58}, \ref{ss2} and \ref{nc26n}, we have the perturbative expansion in terms of the e-foldings, ${\cal N}= Ht$,
\begin{eqnarray}
f\equiv \langle\bar \phi^2 \rangle_{\rm loc}= \alpha_1 {\cal N} -\alpha_2\left(\lambda-\frac{\overline{\beta}^2}{2}\right) {\cal N}^3  + \alpha_3 {\lambda^2}{\cal N}^4 -\alpha_4 \lambda \beta^2 {\cal N}^4
\label{nc54}
\end{eqnarray}
where we have abbreviated for the sake of brevity
\begin{eqnarray}
\alpha_1=\frac{1}{4\pi^2},\qquad  \alpha_2=\frac{1}{2^4\times 9 \pi^4},\qquad \alpha_3=\frac{1}{2^9\times 3 \pi^6}, \qquad \alpha_4 = \frac{1}{2^{11}\times 3 \pi^6}
\label{n53}
\end{eqnarray}
Comparing \ref{nc54} with \ref{cc17}, we observe that the leading secular behaviour corresponding to the local part is sub-leading compared to that of the non-local part. In fact the power counting of $\ln a$ in any non-local contribution equals the total number of propagators involved in the corresponding diagram, whereas the non-local contribution is subleading compared to that. 

Differentiating now \ref{nc54} once with respect to ${\cal N}$, and promoting it now to non-perturbative level as earlier, we have
\begin{eqnarray}
 \frac{d f}{d {\cal N}}=\alpha_1  -  \frac{3\alpha_2}{\alpha_1^2}\left(\lambda-\frac{\overline{\beta}^2}{2}\right) f^2  +  \frac{4\alpha_3 \lambda^2}{\alpha_1^3} f^3-\frac{4\alpha_4\lambda \beta^2 }{\alpha_1^3} f^3
\label{nc58}
\end{eqnarray}
The above equation too, yields a complex $\langle \phi^2 \rangle_{\rm loc}$ as of the earlier case, \ref{phiresum}. However, if we drop the ${\cal O}(\lambda \beta^2)$, term, we indeed obtain a real and positive   $\langle \phi^2 \rangle_{\rm loc}$. The possible reason behind this could be the fact that the ${\cal O}(\lambda \beta^2)$ term gets overlapping contribution from both cubic and quartic self interactions. And hence in order to take into account the effects of such overlap, we need to include further such diagrams from higher order such as ${\cal O}(\lambda^2 \beta^2)$. 

\begin{figure}[h!]
\begin{center}
 \includegraphics[width=6.5cm]{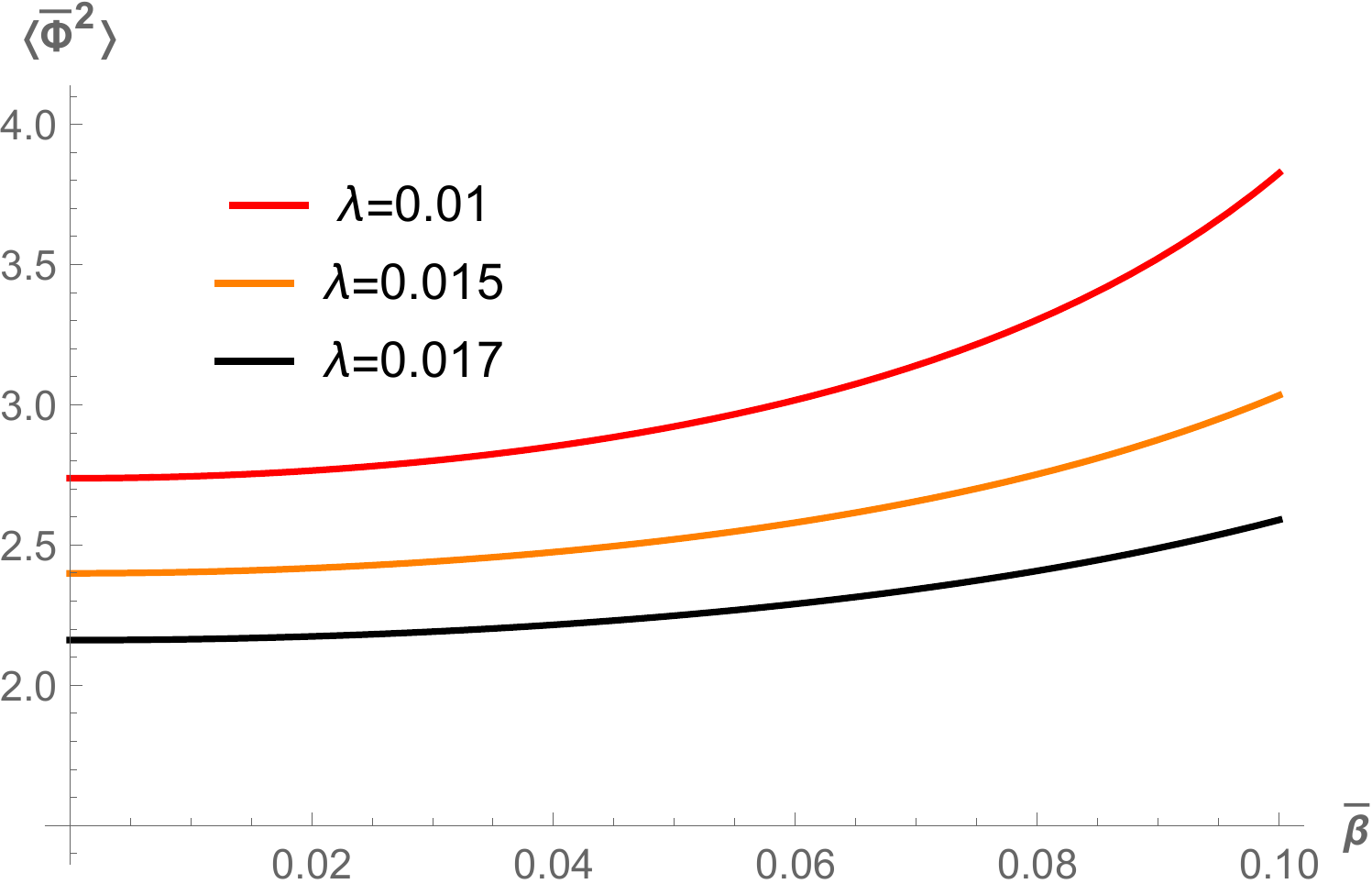}
 \includegraphics[width=7.5cm]{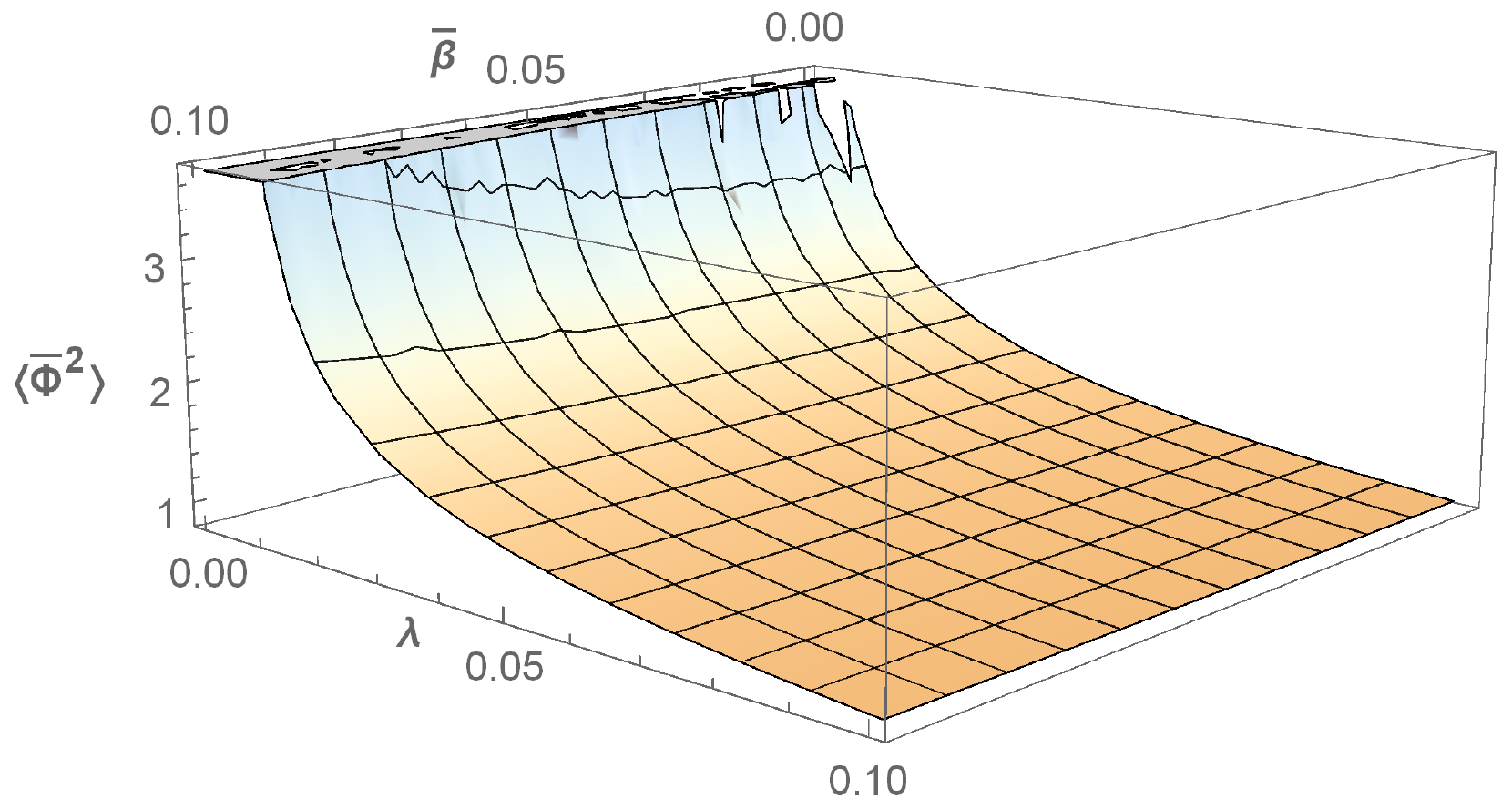}
  \caption{\small Variation of the resummed $\langle\bar\phi^2\rangle_{\rm loc}$ found via quantum field theory,  with respect to the coupling parameters. As $\lambda$ decreases, $\langle\bar\phi^2\rangle_{\rm loc}$ increases, whereas it also increases with increasing  $\bar\beta$. Although the overall qualitative behaviour seems to be similar to that of \ref{phis}, the numerical values disagree with increasing $\bar \beta$. }
  \label{fig1}
\end{center}
\end{figure}

\begin{figure}[h!]
\begin{center}
 \includegraphics[width=6.5cm]{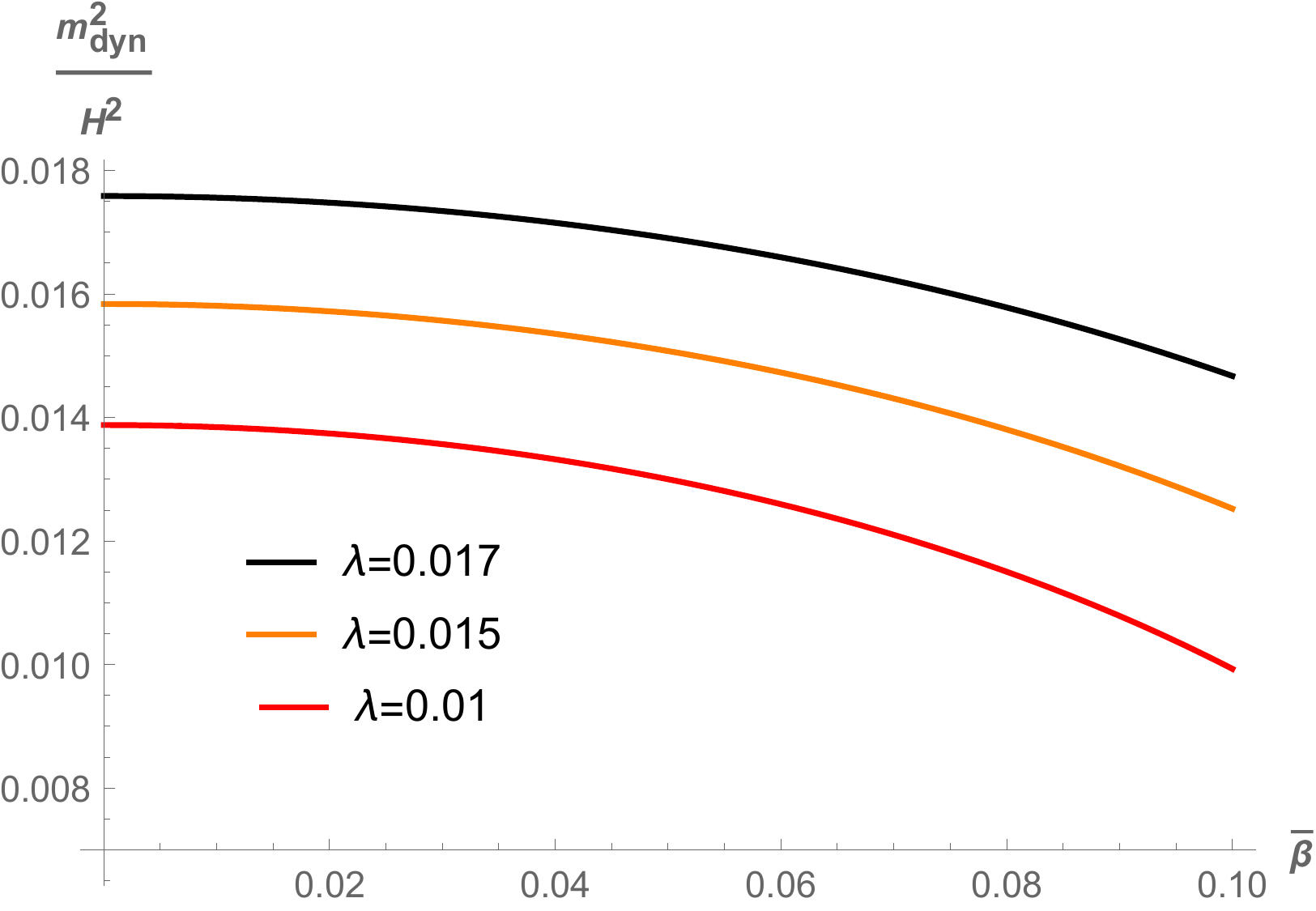}
 \includegraphics[width=7.0cm]{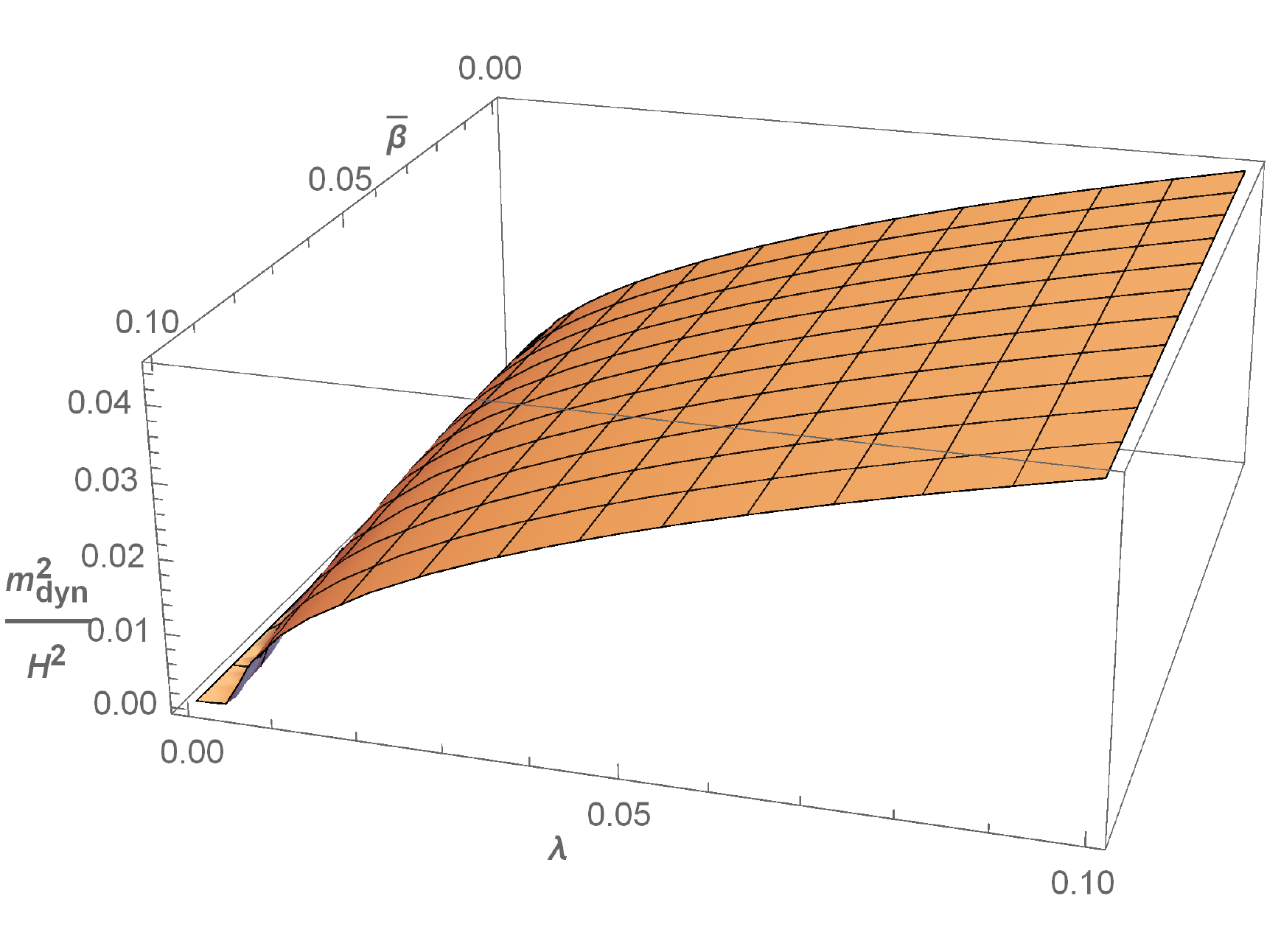}
  \caption{\small Variation of the dynamically generated mass using quantum field theory.  Although this behaviour seems to be as whole similar to the stochastically computed dynamical mass, \ref{dy'}, the numerical values differ with increasing $\bar \beta$. See main text for discussion.}
  \label{dy}
\end{center}
\end{figure}

The resulting $\langle \phi^2 \rangle_{\rm loc}$ found numerically and the dynamical mass found via \ref{dynamical}, have been plotted respectively in \ref{fig1} and \ref{dy}. We note that  qualitative behaviour of these quantities are similar to that of the stochastic method, \ref{stoch}.
However, we note also from \ref{phis}, \ref{dy'} that their numerical values differ and as we increase  $\bar\beta$, the disagreement grows. 

Let us also compare   $\langle \bar\phi \rangle$ computed stochastically in \ref{ssa} and using the same resummation method in quantum field theory in~\cite{Bhattacharya:2022aqi}. The latter simply reads
 \begin{eqnarray}
 \langle \bar\phi \rangle= -\frac{0.4781 \bar{\beta }}{ \lambda }
 \label{sou}
\end{eqnarray}
It seems hopeless to compare the above with \ref{ssa}. Although the above also shows monotonic behaviour with respect to the couplings similar to \ref{phi}, the numerical values differ, with the difference getting larger with increasing $\beta$, as earlier. Note also that the stochastic formalism predicts that $\langle \bar \phi \rangle$ approaches the classical value, $-3\bar{\beta}/\lambda$, as indicated by  \ref{phi}, for large $\bar \beta$. This is a feature qualitatively different from the quantum field theory result of \ref{sou}.

\begin{figure}[h!]
\begin{center}
 \includegraphics[width=7.0cm]{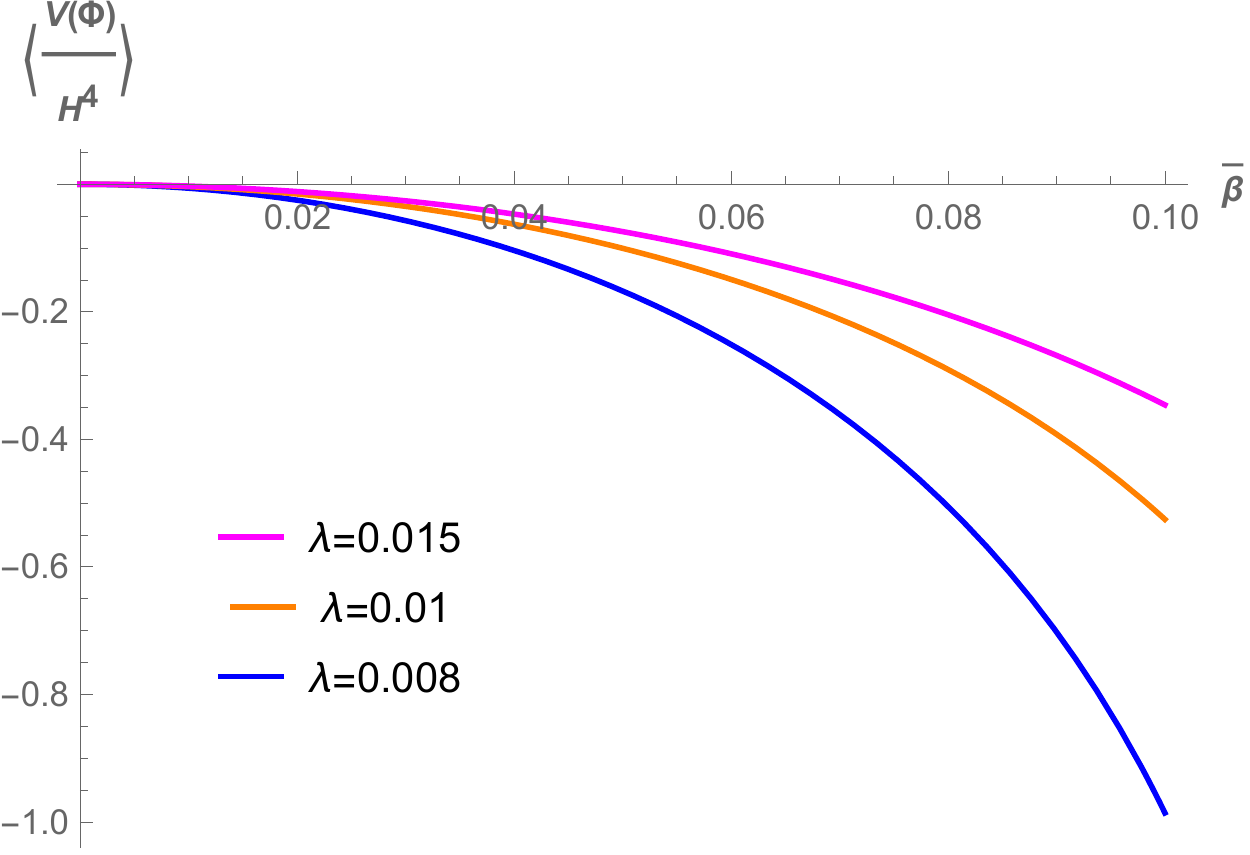}
  \caption{\small The variation of $\langle V(\phi)/H^4\rangle_{\rm{QFT}}$  with respect to $\lambda$ and $\bar \beta$.  Note that as $\lambda$ decreases, the expectation value   decreases rapidly and unboundedly.  See main text for discussion.}
  \label{qv}
\end{center}
\end{figure}

 We expect similar quantitative disagreement for $\langle V(\phi)\rangle$ as well, which we wish to point out very briefly before we end.  Let us estimate  the derivative of   $V(\phi)$ as
\begin{eqnarray}\label{vprime}
\left\langle\frac{ V'(\phi)}{H^3}\right\rangle \sim \frac12\frac{i\Delta(x,x)}{H^2}\left( \frac{\lambda \bar\phi}{3}+\bar\beta\right) 
\end{eqnarray}
so that $V(\phi)$ can be compared with an effective mass term $m_{\rm eff}^2\phi^2/2$, with 
\begin{eqnarray*}\label{masss}
m_{\rm eff}^2 \sim \left(\frac{\lambda\phi^2}{6}+\beta\phi\right)
\end{eqnarray*}
Using now $i\Delta(x,x)=\langle \phi^2 \rangle$, we estimate late time $\langle V^{\prime}(\phi)\rangle$, by replacing $\phi$ with $\langle\phi\rangle$. We substitute the numerical values of $\langle\bar\phi^2\rangle_{\rm{QFT}}$, \ref{fig1}, and also \ref{sou} into \ref{vprime}, and finally integrate it w.r.t to $\langle\bar\phi\rangle$ to generate \ref{qv}. If we compare this with the stochastic result of \ref{V}, we see that the qualitative behaviour are exactly the same although the stochastic result is more negative compared to that of quantum field theory and the difference increases with increasing $\bar\beta$ as for the earlier cases. In other words, quantum field theory predicts less late time backreaction compared to the stochastic method. We shall not go into any further detail of this here. 

Finally, we would like to emphasise  in particular that equation like  \ref{sou}, along with the other quantum field theoretic results found in this paper, manifestly show the strong  non-classical effects in the inflationary de Sitter background for a massless minimal scalar field. As we mentioned earlier, as of flat spacetime quantum field theory, we could  {\it a priori} have naively expected the field to simply rest around the classical minimum $\phi_{\rm cl, min} =-3\beta/\lambda$, of $V(\phi)$ at late times, \ref{poten}. In this case we would expect the field mass to be simply given by $V''(\phi_{\rm cl, min})=3\beta^2/2\lambda$, whose behaviour is obviously qualitatively very different from  \ref{sou} or   \ref{dy}.  For the stochastic formalism as well, strong non-classical effect is manifest, at least when the cubic coupling is not too strong, as demonstrated in  \ref{stoch}.

\section{Discussion}\label{discussion}
In this paper we have computed some non-perturbative vacuum expectation values in a massless minimally coupled quantum scalar field theory with $V(\phi)=\lambda \phi^4/4!+\beta \phi^3/3!$ ($\lambda >0$) in the inflationary de Sitter background. It is well known that the perturbation theory for such a scalar in de Sitter background breaks down at late times, necessitating some non-perturbative method to extract any sensible result.   We have used both stochastic formalism~\cite{Starobinsky:1986fx, Starobinsky:1994bd} as well as quantum field theoretic resummation technique proposed recently in~\cite{Kamenshchik:2020yyn, Kamenshchik:2021tjh}  in this paper. 

We have computed the late time $\langle \phi (x)\rangle$, $\langle \phi^2(x) \rangle$ and $\langle V(\phi) \rangle$ using the IR effective, super Hubble, non-perturbative  stochastic formalism in \ref{stoch}. Variations of these quantities with respect to the coupling parameters have been investigated numerically. We also have computed the dynamically generated mass of the scalar field, \ref{dynamical}, an important physical quantity associated with the non-perturbative $\langle \phi^2(x) \rangle$. As we have emphasised, the chief physical motivation behind this study comes from the fact that firstly, $V(\phi)$, \ref{poten}, as well as  $\phi$ can generate negative vacuum expectation values, and hence may dynamically lead to some screening of the inflationary $\Lambda$ at late cosmological times. Secondly, since $V(\phi)$ is bounded from below for any value of $\beta$, we expect an equilibrium state at late times. We have estimated the backreactions due to them in~\ref{backreaction}. We have shown that any significant screening requires high $\beta/H\lambda$ ratio, ${\cal O}(10^2)$, or more. However, whether such values could actually be allowed needs to be determined by analysing the observable correlation functions, which we reserve for a future work.  

While a resummed, non-perturbative $\langle \phi \rangle$ with the same $V(\phi)$ was found in~\cite{Bhattacharya:2022aqi} using the  formalism of~\cite{Kamenshchik:2020yyn, Kamenshchik:2021tjh}, we next compute a non-perturbative $\langle \phi^2 \rangle$  using the same  in \ref{phiresum} and \ref{resum}.  We have shown that perhaps the formalism of~\cite{Kamenshchik:2020yyn, Kamenshchik:2021tjh} may not be very useful to resum a perturbation series for $\langle \phi^2\rangle$ containing contributions from both local and non-local self energies. Accordingly, we have resummed a part of $\langle \phi^2 \rangle$ containing contributions from only the local self energy, \ref{fig1}. Using this, we have computed the dynamically generated mass in \ref{dy}. We have emphasised  strong non-classical effects,  manifested by both the formalism  (cf., the discussion below \ref{sou}).

However as we also have seen,  although the quantities found via these two different methods show similar qualitative behaviour, they differ numerically. We may speculate two possible reasons, either or both of which could be true, behind this mismatch. First, it could be possible that in order to apply the resummation technique in a theory with more than two coupling parameters such as ours, we have to go to further higher order of perturbative results, as we mentioned earlier. However, perhaps more importantly, keeping in mind the mismatch increases with increasing $\bar \beta$ and also the minimum of $
V(\phi)$  deepens with this increase,  perhaps it could also be speculated for our asymmetric $V(\phi)$, that the stochastic formalism does not {\it sufficiently} take into account the contribution while the field is away from the minimum and is in the relatively flat region of the potential, \ref{poten}.  There can be some non-trivial effects coming from the random behaviour of the field in this flat region also. Note that as discussed at the beginning of \ref{phiresum}, in order that the perturbation theory holds at initial times with respect to the Bunch-Davies vacuum,  we must have $\phi$ to be located around the flat valley around $\phi \sim 0$, initially.  In other words, it might be possible that the stochastic formalism does not sufficiently take into account the journey of the field and its random behaviour while away from the minima, starting from its initial Bunch-Davies state ($\phi\sim 0$).  Anything  concrete beyond this, however, remains elusive to us so far and it certainly seems to warrant further attention.

It seems to be an important task to further investigate this model using other non-perturbative quantum field theoretic techniques. For example, computation of the non-perturbative $\langle \phi^2\rangle$ and the dynamical mass via the Schwinger-Dyson resummation technique seems to be an important task. For the quartic self interaction, this can be seen in~\cite{Youssef:2013by}. Also computation of higher correlators using the non-perturbative effective action techniques seems to be both challenging and important. We hope to return  to these issues in our future publications.

\section*{Acknowledgements}
SB acknowledges partial support from the startup grant (S-3/122/22) of Jadavpur University, Kolkata, India. The authors would like to sincerely acknowledge anonymous referee for useful comments on an earlier version of this manuscript.

\bigskip
\appendix
\labelformat{section}{Appendix #1} 
\section{The in-in formalism}\label{A}
The standard in-out formalism for the S-matrix in quantum field theory may not be very useful in dynamical scenario, where the initial vacuum state evolves into another state and hence particle pair creation occurs. The cosmological spacetime we are interested in is one such example. In such backgrounds, one needs the Schwinger-Keldysh or the in-in formalism to compute the expectation value of any operator meaningfully, which we wish to review very briefly~\cite{Chou, Calzetta, Hu, Weinberg, Adshead}.

The functional integral representation of the standard in-out matrix elements in terms of the field basis reads
\begin{eqnarray}\label{sw}
\langle\phi|T(O[\phi])| \psi\rangle=\int \mathcal{D} \phi e^{i \int_{t_{i}}^{t_{f}} \sqrt{-g} d^{d} x \mathcal{L}[\phi]} \Phi^{\star}\left[\phi\left(t_{f}\right)\right] O[\phi] \Psi\left[\phi\left(t_{i}\right)\right]
\end{eqnarray}
where $T$ stands for time ordering, $O[\phi]$ is some observable and $\Phi$, $\Psi$ are the wave functionals with respect to the field kets $|\phi\rangle$ and $|\psi\rangle$ respectively. We can also have the functional representation for anti-time ordering as 
\begin{eqnarray}\label{ky}
\langle\psi|\bar{T}(P[\phi])| \phi\rangle=\int \mathcal{D} \phi e^{-i \int_{t_{i}}^{t_{f}} \sqrt{-g} d^{d} x \mathcal{L}[\phi]} \Phi\left[\phi\left(t_{i}\right)\right] P[\phi] \Psi^{\star}\left[\phi\left(t_{f}\right)\right]
\end{eqnarray}
 Combining the above with \ref{sw}, we have the in-in matrix representation
\begin{eqnarray}\label{ex}
\langle\psi|\bar{T}(P[\phi]) T(O[\phi])| \psi\rangle=\int \mathcal{D} \phi_{+} \mathcal{D} \phi_{-} \delta\left(\phi_{+}\left(t_{f}\right)-\phi_{-}\left(t_{f}\right)\right) e^{i \int_{t_{i}}^{t_{f}} \sqrt{-g} d^{d} x\left(\mathcal{L}\left[\phi_{+}\right]-\mathcal{L}\left[\phi_{-}\right]\right)} \Psi^{\star}\left[\phi_{-}\left(t_{i}\right)\right] P\left[\phi_{-}\right] O\left[\phi_{+}\right] \Psi\left[\phi_{+}\left(t_{i}\right)\right]
\end{eqnarray}
 Here $\phi_{+}$ evolves the system forward in time and $\phi_{-}$ evolves it backward in time. $\phi_{+}$  and $\phi_{+}$ are two different scalar field but they are coincident on the final hypersurface. We have also used  the completeness relation, 
 $$ \int \mathcal{D} \phi \Phi[\phi_{-}\left(t_{f}\right)] \Phi^{\star}[\phi_{+}\left(t_{f}\right)]=\delta(\phi_{+}(t_{f})-\phi_{-}(t_{f}))$$
on the final hypersurface  at $t=t_{f}$. 

\section{Tree level infrared correlators via quantum field theory }\label{corr1}
In this Appendix we shall review some basic ingredients to do late time, infrared (IR) effective quantum field theory in the de Sitter spacetime. The following framework, along with the preceding Appendix on the in-in formalism provides us the tool to compute the late time leading secular effect. The IR effective field theory corresponds to the redshifted super-Hubble modes at late cosmological times, and certainly is devoid of any ultraviolet divergences. Such field theory corresponds to IR truncated modes, written as, e.g.~\cite{Onemli:2015pma},
\begin{eqnarray}
\phi(\eta, \vec{x}) \vert_{\rm IR} = \int \frac{d^3 \vec{k}}{(2\pi)^{3/2}} \theta (Ha-k) \left[a_{\vec k}\, u(\vec{k},\eta)\vert_{\rm IR} e^{-i\vec{k}\cdot \vec{x}}+a^{\dagger}_{\vec k}\, u^{\star}(\vec{k},\eta)\vert_{\rm IR} e^{i\vec{k}\cdot \vec{x}}\right]
\label{c1}
\end{eqnarray}
where the operators satisfy the canonical commutation retaion $[a_{\vec k}, a^{\dagger}_{\vec k'}]= \delta^3(\vec{k}-\vec{k'})$. The mode functions $u(\vec{k},\eta)$ for a massless and minimally coupled scalar in the de Sitter background corresponding to the Bunch-Davies vacuum is given by \ref{hankel}. The late time ($\eta \to 0^-$), super-Hubble IR modes $(k \lesssim aH)$ can be read off as, 
\begin{eqnarray}
u(\vec{k},\eta) \vert_{\rm IR} \approx \frac{H}{\sqrt{2} k^{3/2}}\left[1+\frac12 \left(\frac{k}{Ha} \right)^2 + \frac{i}{3}\left(\frac{k}{Ha} \right)^3 +\,{\rm subleading~terms } \right]
\label{c2}
\end{eqnarray}
Thus the temporal part of the Bunch-Davies modes becomes nearly a constant in this limit. Substituting this into \ref{c1} and dropping  the suffix `IR' without any loss of generality, we have
\begin{eqnarray}
\phi(\eta, \vec{x})  =\frac{H}{\sqrt{2}} \int \frac{d^3 \vec{k}}{(2\pi)^{3/2}} \frac{\theta (Ha-k)}{k^{3/2}} \left[a_{\vec k}\,  e^{-i\vec{k}\cdot \vec{x}}+a^{\dagger}_{\vec k}\,  e^{i\vec{k}\cdot \vec{x}}\right]+\,{\rm subleading~terms}
\label{c3}
\end{eqnarray}
The step function appearing above ensures that we are essentially dealing with long wavelength modes. We also note that for $\eta' \leq \eta$,  we may take
$\theta (Ha-k)\theta (Ha'-k)\equiv \theta (Ha'-k) $. With this we have~\cite{Onemli:2015pma},
\begin{eqnarray}
\langle 0| \phi(\eta,\vec{x})\phi(\eta',\vec{x'})| 0\rangle=  \frac{H^2}{4\pi^2} \int_{0}^{\infty} \frac{d k}{k^2 r}\sin kr\, \theta(Ha'-k) 
=\frac{H^2}{4\pi^2} \left[{\rm Ci}(Ha' r)-\frac{\sin Ha'r}{Ha'r} -{\rm Ci}(H r)+ \frac{\sin Hr}{Hr}\right]
\label{c4}
\end{eqnarray}
where we have defined $r=|\vec{x}-\vec{x'}|$ and the Ci is the cosine integral function \cite{GR, abr},
$${\rm Ci}(\zeta)= \gamma + \ln \zeta +\sum_{n=1}^{\infty} \frac{(-1)^n \zeta^{2n}}{2n (2n)!}$$
where $\gamma$ is the Euler constant. Using the above and Taylor expanding the $\sin$ in \ref{c4} and also taking the coincidence limit, we have
\be
\langle 0| \phi^2(x)| 0\rangle=\frac{H^2}{4\pi^2} \ln a 
\label{free}
\ee
Note that since we are essentially working with the infrared sector of the field, \ref{c1}, the above expression is free of any ultraviolet divergence in the coincidence limit. Let us also compute 
\begin{eqnarray}
 \dot{\phi}(\eta, \vec{x})= H^2 \int \frac{d^3 \vec{k}}{(2\pi)^{3/2}} \frac{\delta(k-Ha)}{\sqrt{2k}}\left[a_{\vec k}\,  e^{-i\vec{k}\cdot \vec{x}}+a^{\dagger}_{\vec k}\,  e^{i\vec{k}\cdot \vec{x}}\right]
\label{c7}
\end{eqnarray}
where the dot denotes differentiation once with respect to the cosmological 
time. If we consider $\dot{\phi}$ to be the canonical momentum conjugate to $\phi$, it is easy to see from  \ref{c3} that their commutator vanishes. This is a manifestation of the fact that the infrared field is no longer quantum, but is stochastic.   \\

\noindent
Let us now look into the propagators necessary for the in-in computations, \ref{ex}. We have respectively the Feynman and anti-Feynman propagators, 
\begin{eqnarray}
&&i\Delta_{++}(x,x')= \theta(t-t') i\Delta_{-+}(x,x')+ \theta(t'-t) i\Delta_{+-}(x,x')\nonumber\\
&&i\Delta_{--}(x,x')= \theta(t-t') i\Delta_{+-}(x,x')+ \theta(t'-t) i\Delta_{-+}(x,x')
\label{nc1}
\end{eqnarray}
where $i\Delta_{+-}(x,x')$ and $i\Delta_{-+}(x,x')$ are the Wightman functions, given by
\begin{eqnarray}
i\Delta_{-+}(x,x')&&=\frac{H^2}{2} \int \frac{d^3 {\vec k}}{(2\pi)^3 k^3} e^{i{\vec k}\cdot({\vec x}-{\vec y})}\theta(Ha-k)\theta(Ha'-k)(1+ik\eta)(1-ik\eta')e^{-ik(\eta -\eta')}\nonumber\\ &&= \int \frac{d^3 {\vec k}}{(2\pi)^3} e^{i{\vec k}\cdot({\vec x}-{\vec y})}i \Delta_{-+}(k,\eta,\eta')\nonumber\\
i\Delta_{+-}(x,x') &&=\frac{H^2}{2} \int \frac{d^3 \vec{k}}{(2\pi)^3 k^3} e^{i\vec{k}\cdot(\vec{x}-\vec{y})}\theta(Ha-k)\theta(Ha'-k)(1-ik\eta)(1+ik\eta')e^{ik(\eta -\eta')}\nonumber\\&&= \int \frac{d^3 {\vec k}}{(2\pi)^3} e^{i{\vec k}\cdot({\vec x}-{\vec y})}i \Delta_{+-}(k,\eta,\eta') 
\label{nc2}
\end{eqnarray}
where $k=|\vec{k}|$.
Note that $(i\Delta_{-+}(x,x'))^{\star} =i\Delta_{+-}(x,x')$. Thus we also have $i\Delta_{--}(x,x') = (i\Delta_{++}(x,x'))^{\star}$. Using now \ref{c2}, we compute at the leading order,
\begin{eqnarray}
 i \Delta_{+-} (k,\eta',\eta'')&&\approx  \frac{H^2 \theta(Ha'-k)\theta(Ha''-k)}{2k^3} \left(1+\frac{ik^3}{3H^3 a''^3} \right)\qquad (\eta' \gtrsim \eta'')\nonumber\\
&& \approx \frac{H^2 \theta(Ha'-k)\theta(Ha''-k)}{2k^3} \left(1-\frac{ik^3}{3H^3 a'^3} \right)\qquad (\eta'' \gtrsim \eta')\label{nc4'}
\end{eqnarray}
We also have for our purpose
\begin{eqnarray}
&& i \Delta_{+-} (k,\eta,\eta') - i \Delta_{-+} (k,\eta,\eta')\approx \nonumber\\
&& = \frac{i \theta(Ha-k)\theta(Ha'-k)}{H a'^3}\left[ \frac{a'}{a} \left(1-\frac{a'}{a}\right) +\frac{1}{3}\left( 1- \frac{a'}{a}\right)^3\right]=-\frac{i \theta(Ha-k)\theta(Ha'-k)}{H a^3}\left[ \frac{a}{a'} \left(1-\frac{a}{a'}\right) +\frac{1}{3}\left( 1- \frac{a}{a'}\right)^3\right]\nonumber\\
&&i \Delta_{+-} (k,\eta,\eta') + i \Delta_{-+} (k,\eta,\eta')\approx \frac{H^2}{k^3}\theta(Ha-k)\theta(Ha'-k)
\label{nc4}
\end{eqnarray}

The upper limit of the loop momentum will naturally be decided by the step functions. The lower limit will be taken to be $H$, consistent with the IR effective description we are interested in~\cite{Onemli:2015pma}.  Also the time argument   ($\eta$) of the scalar field appearing in the external legs of a diagram will be taken to be greater than or at most equal to any $\eta$ appearing inside a loop integral. With all these equipments, in the next section we are ready to calculate the loop corrections.

\section{Two loop ${\cal O}(\lambda^2)$ non-local contribution to the 2-point correlator}\label{nlo1}

There are three diagrams here, given by the first three of \ref{f2}, of which the local part of the first and second  has been  computed in \ref{lambdasquare}. We wish to compute below the non-local parts of \ref{f2}. For the first or the snowman diagram, we have
\begin{eqnarray}
&&\langle \phi^2(x)\rangle_{\lambda^2,\,{\rm sm}}=-\frac{\lambda^2}{4}\int a'^4 a''^4 d^4 x' d^4 x'' i\Delta (x'',x'') \left[i\Delta_{++}^2(x,x') (i\Delta_{++}^2(x',x'')-i\Delta_{+-}^2(x',x'')) \right. \nonumber\\ && \left.+ i\Delta_{+-}^2(x,x') (i\Delta_{--}^2(x',x'')-i\Delta_{-+}^2(x',x'')) \right]
\nonumber\\&& = -\frac{\lambda^2}{4} \int d\eta' a'^4 d\eta'' a''^4 d^3 {\vec x'}d^3 {\vec x''} (i\Delta_{++}(x,x')+i\Delta_{+-}(x,x'))(i\Delta_{++}(x,x')-i\Delta_{+-}(x,x'))
\nonumber\\&& \times (i\Delta_{++}(x',x'')+i\Delta_{+-}(x',x''))(i\Delta_{++}(x',x'')-i\Delta_{+-}(x',x'')) i \Delta (x'',x'')
 \label{nc8}
\end{eqnarray}
where we have used $(i\Delta_{+-})^{\star}= i \Delta_{-+}$ and $(i\Delta_{++})^{\star}= i \Delta_{--}$ and also the fact that $(i\Delta^2_{++}-i\Delta^2_{+-})=(i\Delta_{++}+i\Delta_{+-})(i\Delta_{++}-i\Delta_{+-})$ is purely imaginary, via \ref{nc1}.
Converting now into the 3-momentum space, the above integral becomes
\begin{eqnarray}
&&-\frac{\lambda^2}{4} \int d\eta' a'^4 d\eta'' a''^4 \int \frac{d^3{\vec k_1} d^3 {\vec k_2}}{(2\pi)^6}(i\Delta_{++}(k_1,\eta,\eta')+i\Delta_{+-}(k_1,\eta,\eta'))(i\Delta_{++}(k_1,\eta,\eta')-i\Delta_{+-}(k_1,\eta,\eta'))
\nonumber\\&& \times (i\Delta_{++}(k_2,\eta',\eta'')+i\Delta_{+-}(k_2,\eta',\eta''))(i\Delta_{++}(k_2,\eta',\eta'')-i\Delta_{+-}(k_2,\eta',\eta'')) \int \frac{d^3{\vec k_3}}{(2\pi)^3} i \Delta (k_3,\eta'',\eta'')
\label{nc9}
\end{eqnarray}
Note from the first of \ref{nc1} that the above integral is non-vanishing only with the temporal hierarchy : $\eta'' \lesssim \eta' \lesssim \eta$. Thus we have 
\begin{eqnarray}
&&-\frac{\lambda^2}{2^5 \pi^6} \int d\eta' a'^4 d\eta'' a''^4 \int k_1^2 k_2^2\,dk_1 dk_2(i\Delta_{-+}(k_1,\eta,\eta')+i\Delta_{+-}(k_1,\eta,\eta'))(i\Delta_{-+}(k_1,\eta,\eta')-i\Delta_{+-}(k_1,\eta,\eta'))
\nonumber\\&& \times (i\Delta_{-+}(k_2,\eta',\eta'')+i\Delta_{+-}(k_2,\eta',\eta''))(i\Delta_{-+}(k_2,\eta',\eta'')-i\Delta_{+-}(k_2,\eta',\eta'')) \int k_3^2 dk_3 \,i \Delta (k_3,\eta'',\eta'')
\label{nc10}
\end{eqnarray}
Substituting from \ref{nc4}, using the aforementioned temporal hierarchy in order to fix the limits of various integrals, we find
\begin{eqnarray}
&&\langle \phi^2(x)\rangle_{\lambda^2,\,{\rm sm}}=\frac{\lambda^2 H^2}{2^6 \times 135 \pi^6}\ln^5 a
\label{nc11}
\end{eqnarray}

The contribution for the the second or the sunset diagram is given by \ref{nc8}, which is non-vanishing for only two temporal hierarchies, $\eta\gtrsim \eta' \gtrsim \eta''$ and  $\eta\gtrsim \eta''\gtrsim \eta'$. For the first, \ref{nc8} becomes
\begin{eqnarray}
&&\langle \phi^2(x)\rangle_{\lambda^2,\,{\rm ss}}\big \vert_{\eta\gtrsim\eta'\gtrsim \eta''}=-\frac{\lambda^2 }{6}\int a'^4 a''^4 d^4 x' d^4 x''(i\Delta_{+-}(x,x')-i\Delta_{-+}(x,x')) (i\Delta_{+-}(x,x'')i\Delta^3_{+-}(x',x'')-{\rm c.c.})\nonumber\\
&&=-\frac{\lambda^2 }{6}\int a'^4 a''^4 d \eta' d \eta'' \frac{d^3{\vec k_1} d^3{\vec k_2}d^3{\vec k_3}}{(2\pi)^9}(i\Delta_{+-}(k_1,\eta,\eta')-i\Delta_{-+}(k_1,\eta,\eta')) \nonumber\\
&& \times \left[i\Delta_{+-}(k_1,\eta,\eta'') i\Delta_{+-}(k_2,\eta',\eta'')i\Delta_{+-}(k_3,\eta',\eta'')i\Delta_{+-}(|\vec{k_1}-{\vec k_2}-{\vec k_3}|,\eta',\eta'')-{\rm c.c.}\right]\nonumber\\
&&= \frac{\lambda^2 H^2}{2^4 \times 27 \times (2\pi)^9} \int_1^a  \frac{da'}{a'}\int_1^{a'}\frac{da''}{a''}  d^3{\vec k_1} d^3{\vec k_2}d^3{\vec k_3} \theta(Ha'' -k_1)\theta(Ha'' -k_2)\theta(Ha'' -k_3)\theta(Ha'' -|{\vec k_1}-{\vec k_2}-{\vec k_3}|) \nonumber\\
&& \times \left[\frac{1}{k_1^3 k_2^3 k_3^3}+ \frac{1}{k_1^3 k_2^3 |{\vec k_1}-{\vec k_2}-{\vec k_3}|^3}+\frac{1}{k_1^3 k_3^3 |{\vec k_1}-{\vec k_2}-{\vec k_3}|^3}+\frac{1}{k_2^3 k_3^3 |{\vec k_1}-{\vec k_2}-{\vec k_3}|^3}\right]
\label{nc13}
\end{eqnarray}
Let us consider the first momentum integral. The step function containing three spatial momenta indicates some kind of orientation between them. The maximum contribution to this integral comes when we take in the upper limits $k_1,\,k_2,\,k_3\sim Ha''$. The fourth step function then indicates the late time orientation $|\hat{k}_1-\hat{k}_2-\hat{k}_3| \sim 1$, where the hatted quantities are the unit vectors along the spatial momenta. We may take lesser final values of $k_1,\,k_2,\,k_3$ in this integral, but it is easy to see that their contribution is less than that of the above. The other momentum integrals can be performed by redefining the dummy momentum variables. Each of the integrals then yields the same contribution and we have 
\begin{eqnarray}
\langle \phi^2(x)\rangle_{\lambda^2,\,{\rm ss}}\big \vert_{\eta \gtrsim\eta' \gtrsim \eta''} = \frac{\lambda^2 H^2}{2^7   \times 135 \pi^6} \ln^5 a
\label{nc14}
\end{eqnarray}

Let us now come to the other case, $\eta' \lesssim \eta''$, which we expect to be the same as above due to the symmetry of the digram under the interchange of $x'$ and $x''$. Indeed, we have 
\begin{eqnarray}
&&\langle \phi^2(x)\rangle_{\lambda^2,\,{\rm ss}}\big\vert_{\eta \gtrsim\eta'' \gtrsim \eta'} = -\frac{\lambda^2}{6} \int (a' a'')^4 d^4 x' d^4 x'' (i\Delta_{+-}(x,x'')-i\Delta_{-+}(x,x''))(i\Delta_{+-}(x,x') i\Delta^3_{+-}(x'',x')-{\rm c.c.})\nonumber\\
&&=-\frac{\lambda^2 }{6}\int a'^4 a''^4 d \eta' d \eta'' \frac{d^3{\vec k_1} d^3{\vec k_2}d^3{\vec k_3}}{(2\pi)^9}(i\Delta_{+-}(k_1,\eta,\eta'')-i\Delta_{-+}(k_1,\eta,\eta'')) \nonumber\\
&& \times \left[i\Delta_{+-}(k_1,\eta,\eta') i\Delta_{+-}(k_2,\eta'',\eta')i\Delta_{+-}(k_3,\eta'',\eta')i\Delta_{+-}(|\vec{k_1}+{\vec k_2}+{\vec k_3}|,\eta'',\eta')-{\rm c.c.}\right]\nonumber\\
&&=\frac{\lambda^2H^2}{2^5\times 27 \pi^6}\int_{1}^{a}\frac{da''}{a''} \int_1^{a''} \frac{da'}{a'}\ln^3 a' =\frac{\lambda^2H^2}{2^7\times 135 \pi^6}\ln^5a,
\label{nc15}
\end{eqnarray}
exactly the same as \ref{nc14}. We have also used $i\Delta_{-+}(x',x'')=i\Delta_{+-}(x'',x')$ in the above computation. The full expression for the two loop two point sunset correlator  at late times is found by adding the above result with \ref{nc14},
\begin{eqnarray}
\langle \phi^2(x)\rangle_{\lambda^2,\,{\rm ss}} = \frac{\lambda^2 H^2}{2^6  \times 135 \pi^6} \ln^5 a
\label{nc16}
\end{eqnarray}
Note that the above is dominant compared to the local part, which is ${\cal O}(\ln^4 a)$, derived in \ref{lambdasquare}.

Finally, we wish to compute the non-1PI process, given by the third of \ref{f2},
\begin{eqnarray}
&&-\frac{\lambda^2}{4} \int (a'a'')^4 d^4 x' d^4 x'' \left[ i\Delta_{-+}(x,x') i\Delta_{-+}(x,x'') i\Delta_{++}(x',x'') - i\Delta_{-+}(x,x') i\Delta_{+-}(x,x'') i\Delta_{+-}(x',x'') \right. \nonumber\\ &&\left. - i\Delta_{+-}(x,x') i\Delta_{-+}(x,x'') i\Delta_{-+}(x',x'') + i\Delta_{+-}(x,x') i\Delta_{+-}(x,x'') i\Delta_{--}(x',x'') \right] i\Delta(x',x') i\Delta(x'',x'') \nonumber\\
\label{nc17a1}
\end{eqnarray}

For $\eta \gtrsim \eta' \gtrsim \eta''$, the above integral becomes,
\begin{eqnarray}
&&-\frac{\lambda^2}{4} \int (a'a'')^4 d^4 x' d^4 x'' (i\Delta_{+-}(x,x')-i\Delta_{-+}(x,x')) (i\Delta_{+-}(x,x'')i\Delta_{+-}(x',x'') -{\rm c.c.}) i\Delta(x',x')i\Delta(x'',x'')\nonumber\\
&&=\frac{\lambda^2H^2}{2^7 \times 135 \pi^6}\ln^5 a
\label{nc17a2}
\end{eqnarray}
On the other hand for $\eta \gtrsim \eta'' \gtrsim \eta'$,  \ref{nc17a1} gives the same expression as above
\begin{eqnarray}
&&-\frac{\lambda^2}{4} \int (a'a'')^4 d^4 x' d^4 x'' (i\Delta_{+-}(x,x'')-i\Delta_{-+}(x,x'')) (i\Delta_{+-}(x,x')i\Delta_{+-}(x'',x') -{\rm c.c.}) i\Delta(x',x')i\Delta(x'',x'')\nonumber\\
&&= \frac{\lambda^2H^2}{2^7\times 135 \pi^6}\ln^5 a,
\label{nc17a3}
\end{eqnarray}
as expected. Combining now  \ref{nc11}, \ref{nc16}, \ref{nc17a2} and \ref{nc17a3} we obtain the full two loop expression at late times corresponding to the quartic sector, 
\begin{eqnarray}
\langle \phi^2(x)\rangle_{\lambda^2} = \frac{\lambda^2 H^2}{2^6  \times 45 \pi^6} \ln^5 a
\label{nc17}
\end{eqnarray}
%

\section{One loop ${\cal O}(\beta^2)$  non-local contribution to $\langle \phi^2(x) \rangle$ }\label{nlo2}
There are two diagrams which contribute to $\langle \phi^2 \rangle$ at ${\cal O}(\beta^2)$, given by the fourth and fifth of \ref{f2}. The local contribution for the fourth has been computed in \ref{lo3}. We wish to compute below the non-local contribution for the same, as well as the contribution from the fifth diagram, which is entirely non-local.

The fourth diagram represent the loop integral \ref{lo52}. As of the sunset diagram computed above, we shall break the above integral into two temporal higherarhies : $\eta \gtrsim\eta' \gtrsim \eta''$ and $\eta \gtrsim\eta'' \gtrsim \eta'$. For the first, \ref{lo52} becomes, 
\begin{eqnarray}
&&\langle \phi^2(x)\rangle_{\beta^2}\big\vert_{\eta\gtrsim \eta'\gtrsim \eta''} = -\frac{\beta^2}{2} \int (a'a'')^4 d^4 x' d^4 x'' (i\Delta_{+-}(x,x')-i\Delta_{-+}(x,x')) (i\Delta_{+-}(x,x'')i\Delta^2_{+-}(x',x'')-{\rm c.c.})\nonumber\\
&&= \frac{\beta^2}{2^5 \times 3\pi^4} \int_{1}^{a} \frac{da'}{a'} \int_{1}^{a'} \frac{da''}{a''}\ln^2 a'' =\frac{\beta^2}{2^7\times 9\pi^4}\ln^4 a
\label{nc19}
\end{eqnarray}
Likewise for $\eta\gtrsim \eta'' \gtrsim \eta'$, we have
\begin{eqnarray}
&&\langle \phi^2(x)\rangle_{\beta^2}\big\vert_{\eta\gtrsim \eta''\gtrsim \eta'} = -\frac{\beta^2}{2} \int (a'a'')^4 d^4 x' d^4 x'' (i\Delta_{+-}(x,x'')-i\Delta_{-+}(x,x''))(i\Delta_{+-}(x,x') i\Delta^2_{+-}(x'',x')-{\rm c.c.})\nonumber\\
&&=\frac{\beta^2}{2^5 \times 3\pi^4} \int_{1}^{a}\frac{da''}{a''} \int_{1}^{a''} \frac{da'}{a'}\ln^2 a'  =\frac{\beta^2}{2^7\times 9\pi^4}\ln^4 a
\label{nc20}
\end{eqnarray}
which is the same as \ref{nc19}, as expected from the symmetry of the diagram. Combining the above with \ref{nc19}, we have 
\begin{eqnarray}
&&\langle \phi^2(x)\rangle=\frac{\beta^2}{2^6\times 9\pi^4}\ln^4 a
\label{nc21}
\end{eqnarray}
Note that as of the sunset diagram, the above non-local contribution is also dominant compared to the local contribution, derived in \ref{lo3}.

Let us now compute the tadpole contribution to the two point function i.e., the fifth of \ref{f2}. This cannot have any local contribution, as has been discussed in the main body of the paper.  The corresponding integral becomes
\begin{eqnarray}
&&-\frac{\beta^2}{2} \int (a' a'')^4 d^4 x' d^4 x'' \left[i\Delta_{-+}^2(x,x') (i\Delta_{++}(x',x'')-i\Delta_{+-}(x',x'')) \right. \nonumber\\ && \left. -i\Delta_{+-}^2(x,x') (i\Delta_{-+}(x',x'')-i\Delta_{--}(x',x'')) \right]i\Delta(x'',x''),
\label{nc22}
\end{eqnarray}
which is non-vanishing only for $\eta \gtrsim \eta' \gtrsim \eta''$. Thus the above integral becomes
\begin{eqnarray}
&&-\frac{\beta^2}{2} \int (a' a'')^4 d^4 x' d^4 x''  (i\Delta_{+-}(x',x'')-i\Delta_{-+}(x',x'')) (i\Delta^2_{+-}(x,x')-i\Delta^2_{-+}(x,x'))i\Delta(x'',x'')\nonumber\\
&& = \frac{\beta^2}{2^6\times 9 \pi^4} \ln^4 a
\label{nc23}
\end{eqnarray}
Combining the above with \ref{nc21}, we find the one loop two point correllator for the cubic sector,
\begin{eqnarray}
&&\langle \phi^2(x)\rangle_{\beta^2}=\frac{\beta^2}{2^5\times 9\pi^4}\ln^4 a
\label{nc24}
\end{eqnarray}
%

\section{Two loop ${\cal O}(\lambda \beta^2)$ non-local contribution to $\langle \phi^2(x) \rangle$}\label{nlo3}
%
There are total eight diagrams associated with the ${\cal O}(\lambda \beta^2)$ correction to the two point correlator, as shown in \ref{f3}. 
The first of \ref{f3} is given by \ref{nc25}, whose local part has been computd in \ref{lo4}. We now wish to compute its non-local contribution below, using the IR effective tool outlined in \ref{corr1}. 

Keeping in mind $\eta \gtrsim \eta', \eta'', \eta'''$, let us check for the six possible temporal hierarchies in \ref{nc25},
\be
\eta''' \gtrsim \eta' \gtrsim \eta'',\quad \eta''' \gtrsim \eta'' \gtrsim \eta',\quad \eta'' \gtrsim \eta''' \gtrsim \eta',\quad \eta'' \gtrsim \eta' \gtrsim \eta''',\quad  \eta' \gtrsim \eta'' \gtrsim \eta''',\quad \eta' \gtrsim \eta''' \gtrsim \eta''
\label{nc25'}
\ee
It is easy to see using \ref{nc1} that the contribution from the last four hierarchies vanish, leaving us only the first two to deal with. For $\eta''' \gtrsim \eta' \gtrsim \eta''$, the above integral becomes
\begin{eqnarray}
&&\frac{i\lambda\beta^2}{4} \int (a' a'' a''')^4 d^4 x' d^4 x'' d^4 x''' \left[i\Delta_{-+}^2(x,x''')(i\Delta_{+-}(x''',x')-i\Delta_{-+}(x''',x'))(i\Delta_{+-}(x''',x'')i\Delta_{+-}^2(x',x'')-{\rm c.c.})\right. \nonumber\\ &&\left.-{\rm c.c.} \right]\nonumber\\
&&=\frac{i\lambda\beta^2}{4} \int (a' a'' a''')^4 d\eta' d\eta'' d\eta''' \frac{d^3 {\vec k_1} d^3 {\vec k_2}d^3 {\vec k_3}}{(2\pi)^9}\left[i\Delta_{-+}^2(k_1,\eta,\eta''') (i\Delta_{+-}(k_2,\eta''',\eta')-i\Delta_{-+}(k_2,\eta''',\eta')) \right. \nonumber\\ &&\left.
\times (i\Delta_{+-}(k_2,\eta''',\eta'')i\Delta_{+-}(k_3,\eta',\eta'') i\Delta_{+-}(|{\vec k_2+{\vec k_3}}|,\eta',\eta'') -{\rm c.c.}) -{\rm c.c.} \right]
\nonumber\\ 
&&= -\frac{\lambda \beta^2}{2^7\times 9\pi^6} \int_1^{a}\frac{da'''}{a'''}\ln a''' \int_1^{a'''}\frac{da'}{a'} \int_1^{a'} \frac{da''}{a''}\ln^2a''=-\frac{\lambda \beta^2}{2^7\times 9 \pi^6} \frac{\ln^6 a}{72}
\label{nc26}
\end{eqnarray}

For $\eta''' \gtrsim \eta'' \gtrsim \eta'$, \ref{nc25} becomes,
\begin{eqnarray}
&&\frac{i\lambda\beta^2}{4} \int (a' a'' a''')^4 d^4 x' d^4 x'' d^4 x''' \left[i\Delta_{-+}^2(x,x''')(i\Delta_{+-}(x''',x'')-i\Delta_{-+}(x''',x''))(i\Delta_{+-}(x''',x')i\Delta_{-+}^2(x',x'')-{\rm c.c.})\right. \nonumber\\ &&\left.-{\rm c.c.} \right]\nonumber\\
&&=\frac{i\lambda\beta^2}{4} \int (a' a'' a''')^4 d\eta' d\eta'' d\eta''' \frac{d^3 {\vec k_1} d^3 {\vec k_2}d^3 {\vec k_3}}{(2\pi)^9}\left[i\Delta_{-+}^2(k_1,\eta,\eta''') (i\Delta_{+-}(|{\vec k_2}+{\vec k_3}|,\eta''',\eta'')-i\Delta_{-+}(|{\vec k_2}+{\vec k_3}|,\eta''',\eta'')) \right. \nonumber\\ &&\left.
\times (i\Delta_{+-}(|{\vec k_2}+{\vec k_3}|,\eta''',\eta')i\Delta_{-+}(k_2,\eta',\eta'') i\Delta_{-+}(k_3,\eta',\eta'') -{\rm c.c.}) -{\rm c.c.} \right] \nonumber\\
&&=-\frac{\lambda \beta^2}{2^7 \times 9\pi^6}\int_1^{a}\frac{da'''}{a'''}\ln a''' \int_{1}^{a'''}\frac{da''}{a''} \int_{1}^{a''}\frac{da'}{a'}\ln^2 a'= -\frac{\lambda \beta^2}{2^7\times 9 \pi^6} \frac{\ln^6 a}{72},
\label{nc27}
\end{eqnarray}
which is exactly the same as \ref{nc26}, expected from the symmetry of the diagram (first of \ref{f3}). The full expression is obtained by combining  these two results,
\be
-\frac{\lambda \beta^2}{2^7\times 9 \pi^6} \frac{\ln^6 a}{36}
\label{nc28}
\ee

Let us next consider the second diagram of \ref{f3}, which reads
\begin{eqnarray}
&&\frac{i\lambda \beta^2}{4}\int (a'a''a''')^4 d^4 x' d^4 x'' d^4 x''' \left[\left\{i\Delta_{-+}(x,x') i\Delta_{-+}(x,x'')i\Delta_{++}(x',x'')i\Delta_{++}(x',x''')i\Delta_{++}(x'',x''')\right.\right. \nonumber\\&& \left.\left.
-i\Delta_{-+}(x,x') i\Delta_{-+}(x,x'')i\Delta_{++}(x',x'')i\Delta_{+-}(x',x''')i\Delta_{+-}(x'',x''')\right. \right.\nonumber\\&& \left. \left.
-i\Delta_{-+}(x,x') i\Delta_{+-}(x,x'')i\Delta_{+-}(x',x'')i\Delta_{++}(x',x''')i\Delta_{-+}(x'',x''')\right.\right. \nonumber\\&& \left.\left.
+i\Delta_{-+}(x,x') i\Delta_{+-}(x,x'')i\Delta_{+-}(x',x'')i\Delta_{+-}(x',x''')i\Delta_{--}(x'',x''')\right\}-{\rm c.c.}
\right]\times  i\Delta(x''',x''')
\nonumber\\&&
=\frac{i\lambda \beta^2}{4}\int (a'a'' a''')^4 d\eta'd \eta'' d\eta''' \frac{d^3{\vec k_1} d^3{\vec k_2}d^3{\vec k_3}}{(2\pi)^9} i\Delta(k_3,\eta''',\eta''')\nonumber\\&& \times\left[\left\{ i\Delta_{-+}(k_1,\eta,\eta')i\Delta_{-+}(k_1,\eta,\eta'')i\Delta_{++}(k_2,\eta',\eta'')i\Delta_{++}(|{\vec k_1}-{\vec k_2}|,\eta',\eta''')i\Delta_{++}(|{\vec k_1}-{\vec k_2}|,\eta'',\eta''')\right.\right. \nonumber\\&& \left.\left.
- i\Delta_{-+}(k_1,\eta,\eta')i\Delta_{-+}(k_1,\eta,\eta'')i\Delta_{++}(k_2,\eta',\eta'')i\Delta_{+-}(|{\vec k_1}-{\vec k_2}|,\eta',\eta''')i\Delta_{+-}(|{\vec k_1}-{\vec k_2}|,\eta'',\eta''')\right. \right.\nonumber\\&& \left. \left.
- i\Delta_{-+}(k_1,\eta,\eta')i\Delta_{+-}(k_1,\eta,\eta'')i\Delta_{+-}(k_2,\eta',\eta'')i\Delta_{++}(|{\vec k_1}-{\vec k_2}|,\eta',\eta''')i\Delta_{-+}(|{\vec k_1}-{\vec k_2}|,\eta'',\eta''')\right.\right. \nonumber\\&& \left.\left.
+i\Delta_{-+}(k_1,\eta,\eta') i\Delta_{+-}(k_1,\eta,\eta'')i\Delta_{+-}(k_2,\eta',\eta'')i\Delta_{+-}(|{\vec k_1}-{\vec k_2}|,\eta',\eta''')i\Delta_{--}(|{\vec k_1}-{\vec k_2}|,\eta'',\eta''')\right\}-{\rm c.c.}
\right] \hskip .3cm
\label{nc29}
\end{eqnarray}
We argued in \ref{lo3} that this diagram contains only non-local contribution to the self-energy.

It is easy to see that the first two temporal hierarchies of \ref{nc25'} yields vanishing contribution to the above integral, leaving us only with the remaining four hierarchies to deal with. For $\eta'\gtrsim \eta'' \gtrsim \eta'''$, the above integral becomes,
\begin{eqnarray}
&&\frac{i\lambda \beta^2}{4}\int (a'a'' a''')^4 d\eta'd \eta'' d\eta''' \frac{d^3{\vec k_1} d^3{\vec k_2}d^3{\vec k_3}}{(2\pi)^9} i\Delta(k_3,\eta''',\eta''')\nonumber\\
&&\times\left[ \left\{i\Delta_{-+}(k_1,\eta,\eta') (i\Delta_{-+}(k_1,\eta,\eta'')i\Delta_{-+}(k_2,\eta',\eta'')-{\rm c.c.})\times (i\Delta_{-+}(|{\vec k_1}-{\vec k_2}|,\eta',\eta''')i\Delta_{-+}(|{\vec k_1}-{\vec k_2}|,\eta'',\eta''')-{\rm c.c.})\right\}\right. \nonumber\\&&\left.-{\rm c.c.}\right]\nonumber\\
&&= -\frac{\lambda \beta^2}{2^6 \times 27 \pi^6}\int_1^{a}\frac{da'}{a'}\int_1^{a'}\frac{da''}{a''}\ln a'' \int_1^{a''} \frac{da'''}{a'''}\ln^2 a'''= - \frac{\lambda \beta^2}{2^7 \times 27 \pi^6}\frac{\ln^6 a}{45}
\label{nc30}
\end{eqnarray}
For  $\eta''\gtrsim \eta' \gtrsim \eta'''$,
the contribution is exactly the same as above. For $\eta'\gtrsim \eta''' \gtrsim \eta''$ and  $\eta''\gtrsim \eta''' \gtrsim \eta'$, each of the contributions equals,
$$-\frac{\lambda \beta^2}{2^7 \times 27\pi^6} \frac{\ln^6a}{60}$$
Combining the above with \ref{nc30}, we find the total contribution, 
\begin{eqnarray}
- 2\times\frac{\lambda \beta^2}{2^7 \times 27 \pi^6}\frac{\ln^6 a}{45}-2\times\frac{\lambda \beta^2}{2^7 \times 27\pi^6} \frac{\ln^6a}{60}= - \frac{\lambda \beta^2}{2^7 \times 9 \pi^6}\frac{7 \ln^6 a}{270}
\label{nc31}
\end{eqnarray}

Let us now compute the third process of \ref{f3}.  It equals
\begin{eqnarray}
&&\frac{i\lambda\beta^2}{4} \int (a' a'' a''')^4 d^4 x' d^4 x'' d^4 x''' \left[i\Delta^2_{-+}(x,x''') i\Delta^2_{++}(x''',x'')i\Delta_{++}(x'',x')\right. \nonumber\\&& \left.+i\Delta^2_{+-}(x,x''') i\Delta^2_{--}(x''',x'')i\Delta_{-+}(x'',x')+i\Delta^2_{+-}(x,x''') i\Delta^2_{-+}(x''',x'')i\Delta_{+-}(x'',x')\right. \nonumber\\ && \left.
+i\Delta^2_{-+}(x,x''') i\Delta^2_{+-}(x''',x'')i\Delta_{--}(x'',x')-{\rm c.c.}
\right] \times i\Delta(x',x')
\label{nc32}
\end{eqnarray}
The above integral gives non-vanishing contribution only with $\eta'''\gtrsim \eta'' \gtrsim \eta'$, 
\begin{eqnarray}
&&\frac{i\lambda\beta^2}{4} \int (a' a'' a''')^4 d^4 x' d^4 x'' d^4 x''' \left[i\Delta^2_{-+}(x,x''') i\Delta^2_{-+}(x''',x'')i\Delta_{-+}(x'',x')\right. \nonumber\\&& \left.+i\Delta^2_{+-}(x,x''') i\Delta^2_{+-}(x''',x'')i\Delta_{-+}(x'',x')+i\Delta^2_{+-}(x,x''') i\Delta^2_{-+}(x''',x'')i\Delta_{+-}(x'',x')\right. \nonumber\\ && \left.
+i\Delta^2_{-+}(x,x''') i\Delta^2_{+-}(x''',x'')i\Delta_{+-}(x'',x')-{\rm c.c.}
\right] \times i\Delta(x',x')\nonumber\\
&&= \frac{i\lambda \beta^2}{4} \int (a' a'' a''')^4 d\eta' d\eta'' d\eta''' \frac{d^3{\vec k_1} d^3{\vec k_2}d^3{\vec k_3} }{(2\pi)^9} (i\Delta_{-+}^2(k_1,\eta,\eta''')-i\Delta_{+-}^2(k_1,\eta,\eta''')) \nonumber\\ &&\times (i\Delta_{-+}^2(k_2,\eta''',\eta'')-i\Delta_{+-}^2(k_2,\eta''',\eta''))(i\Delta_{-+}(0,\eta'',\eta')-i\Delta_{+-}(0,\eta'',\eta'))  i\Delta(k_3, \eta',\eta')\nonumber\\
&&= -\frac{\lambda \beta^2}{2^6 \times 27 \pi^6}\int_1^a \frac{da'''}{a'''}\ln a''' \int_1^{a'''} \frac{da''}{a''}\ln a'' \int_1^{a''} \frac{da'}{a'}\ln a'=-\frac{\lambda \beta^2}{2^7 \times 9 \pi^6}\frac{\ln^6 a}{72}
\label{nc33}
\end{eqnarray}

Let us now come to the fourth process of \ref{f3}, which equals,
\begin{eqnarray}
&&\frac{i\lambda\beta^2}{4} \int (a' a'' a''')^4 d^4 x' d^4 x'' d^4 x''' \left[i\Delta^2_{-+}(x,x') i\Delta_{++}(x',x'')i\Delta^2_{++}(x'',x''')\right. \nonumber\\&& \left.+i\Delta^2_{-+}(x,x') i\Delta_{+-}(x',x'')i\Delta^2_{--}(x'',x''')+i\Delta^2_{+-}(x,x') i\Delta_{-+}(x',x'')i\Delta^2_{+-}(x'',x''')\right. \nonumber\\ && \left.
+i\Delta^2_{+-}(x,x') i\Delta_{--}(x',x'')i\Delta^2_{-+}(x'',x''')-{\rm c.c.}
\right] \times i\Delta(x''',x''')
\label{nc34}
\end{eqnarray}
The only non-vanishing contribution comes from $\eta'\gtrsim \eta'' \gtrsim \eta'''$, yielding
\begin{eqnarray}
&&\frac{i\lambda\beta^2}{4} \int (a' a'' a''')^4 d^4 x' d^4 x'' d^4 x''' i\Delta(x''',x''') (i\Delta_{+-}(x,x')+i\Delta_{-+}(x,x'))(i\Delta_{+-}(x'',x''')+i\Delta_{-+}(x'',x'''))\nonumber\\
&& \times (i\Delta_{-+}(x,x')-i\Delta_{+-}(x,x'))(i\Delta_{-+}(x',x'')-i\Delta_{+-}(x',x'')) (i\Delta_{-+}(x'',x''')-i\Delta_{+-}(x'',x'''))
\nonumber\\
&&=\frac{i\lambda\beta^2}{4} \int (a' a'' a''')^4 d \eta' d \eta'' d \eta'''  \frac{d^3{\vec k_1} d^3{\vec k_2}d^3{\vec k_3} }{(2\pi)^9}(i\Delta_{+-}(k_1,\eta,\eta')+i\Delta_{-+}(k_1,\eta,\eta'))(i\Delta_{+-}(k_2,\eta'',\eta''')+i\Delta_{-+}(k_2,\eta'',\eta'''))\nonumber\\
&& \times (i\Delta_{-+}(k_1,\eta,\eta')-i\Delta_{+-}(k_1,\eta,\eta'))(i\Delta_{-+}(0,\eta',\eta'')-i\Delta_{+-}(0,\eta',\eta'')) \nonumber\\
&& \times (i\Delta_{-+}(k_2,\eta'',\eta''')-i\Delta_{+-}(k_2,\eta'',\eta'''))i\Delta(k_3,\eta''',\eta''')\nonumber\\&&
=-\frac{\lambda \beta^2}{2^6 \times 27\pi^6}\int_1^a\frac{da'}{a'}\ln a' \int_1^{a''} \frac{da''}{a''} \int_1^{a''}\frac{da'''}{a'''} \ln^2 a'''=-\frac{\lambda \beta^2}{2^7 \times 9\pi^6}\frac{\ln^6 a}{108}
\label{nc35}
\end{eqnarray}

The fifth process of \ref{f3} yields,
\begin{eqnarray}
&&\frac{i\lambda\beta^2}{4} \int (a' a'' a''')^4 d^4 x' d^4 x'' d^4 x''' i\Delta(x'',x'') i\Delta(x''',x''')\left[i\Delta_{-+}^2(x,x')i\Delta_{++}(x',x'')i\Delta_{++}(x'',x''') \right. \nonumber\\ &&\left. 
+i\Delta_{-+}^2(x,x')i\Delta_{+-}(x',x'')i\Delta_{--}(x'',x''')+
i\Delta_{+-}^2(x,x')i\Delta_{--}(x',x'')i\Delta_{-+}(x'',x''')\right. \nonumber\\ &&\left.+i\Delta^2_{+-}(x,x')i\Delta_{-+}(x',x'')i\Delta_{+-}(x'',x''') -{\rm c.c.}
\right]
\label{nc36}
\end{eqnarray}
The only non-vanishing contribution comes from $\eta' \gtrsim \eta'' \gtrsim \eta'''$,
\begin{eqnarray}
&&\frac{i\lambda\beta^2}{4} \int (a' a'' a''')^4 d^4 x' d^4 x'' d^4 x''' i\Delta(x'',x'') i\Delta(x''',x''')(i\Delta_{-+}(x,x')+i\Delta_{+-}(x,x'))\nonumber\\&&
(i\Delta_{-+}(x,x')-i\Delta_{+-}(x,x'))(i\Delta_{-+}(x'',x''')-i\Delta_{+-}(x'',x'''))(i\Delta_{-+}(x',x'')-i\Delta_{+-}(x',x''))\nonumber\\&&
=\frac{i\lambda\beta^2}{4} \int (a' a'' a''')^4 d \eta' d \eta'' d \eta''' \frac{d^3{\vec k_1} d^3{\vec k_2}d^3{\vec k_3} }{(2\pi)^9}i\Delta(k_2,\eta'',\eta'') i\Delta(k_3,\eta''',\eta''')(i\Delta_{-+}(k_1,\eta,\eta')+i\Delta_{+-}(k_1,\eta,\eta'))\nonumber\\&&
\times(i\Delta_{-+}(k_1,\eta,\eta')-i\Delta_{+-}(k_1,\eta,\eta'))(i\Delta_{-+}(0,\eta'',\eta''')-i\Delta_{+-}(0,\eta'',\eta'''))(i\Delta_{-+}(0,\eta',\eta'')-i\Delta_{+-}(0,\eta',\eta''))\nonumber\\
&&= -\frac{\lambda \beta^2}{2^7 \times 27 \pi^6} \int_1^a \frac{da'}{a'}\ln a' \int_1^{a'}\frac{da''}{a''}\ln a''  \int_1^{a''}\frac{da'''}{a'''}\ln a'''= -\frac{\lambda \beta^2}{2^7 \times 9 \pi^6}\frac{\ln^6 a}{144}
\label{nc37}
\end{eqnarray}

The sixth process of \ref{f3} reads,
\begin{eqnarray}
&&\frac{i\lambda\beta^2}{4} \int (a' a'' a''')^4 d^4 x' d^4 x'' d^4 x'''  i\Delta(x''',x''')\left[i\Delta_{-+}(x,x')i\Delta_{-+}(x,x''')i\Delta^2_{++}(x',x'') i\Delta_{++}(x'',x''') \right. \nonumber\\ &&\left. 
+i\Delta_{-+}(x,x')i\Delta_{+-}(x,x''')i\Delta^2_{+-}(x',x'') i\Delta_{--}(x'',x''')+
i\Delta_{+-}(x,x')i\Delta_{+-}(x,x''')i\Delta^2_{-+}(x',x'') i\Delta_{+-}(x'',x''')\right. \nonumber\\ &&\left.+i\Delta_{+-}(x,x')i\Delta_{-+}(x,x''')i\Delta^2_{--}(x',x'') i\Delta_{-+}(x'',x''') -{\rm c.c.}
\right]
\label{nc38}
\end{eqnarray}
There are four hierarchies which contribute to the above integral, $\eta'\gtrsim \eta''\gtrsim \eta'''$, $\eta'\gtrsim \eta'''\gtrsim \eta''$, $\eta'''\gtrsim \eta''\gtrsim \eta'$ and $\eta'''\gtrsim \eta'\gtrsim \eta''$.
For the first, the above integral becomes,
\begin{eqnarray}
&&\frac{i\lambda\beta^2}{4} \int (a' a'' a''')^4 d^4 x' d^4 x'' d^4 x'''  i\Delta(x''',x''')(i\Delta_{-+}(x,x')-i\Delta_{+-}(x,x'))(i\Delta_{-+}(x',x'')+i\Delta_{+-}(x',x''))\nonumber\\&&
\times (i\Delta_{-+}(x',x'')-i\Delta_{+-}(x',x''))(i\Delta_{-+}(x,x''')i\Delta_{-+}(x'',x''')-{\rm c.c.})\nonumber\\
&&=\frac{i\lambda\beta^2}{4} \int (a' a'' a''')^4 d \eta' d \eta'' d \eta'''  \frac{d^3{\vec k_1}d^3{\vec k_2}d^3{\vec k_3}}{(2\pi)^9} i\Delta(k_3,\eta''',\eta''')(i\Delta_{-+}(k_1,\eta,\eta')-i\Delta_{+-}(k_1,\eta,\eta'))
\nonumber\\&& \times (i\Delta_{-+}(k_2,\eta',\eta'')+i\Delta_{+-}(k_2,\eta',\eta''))
(i\Delta_{-+}(|{\vec k_1}-{\vec k_2}|,\eta',\eta'')-i\Delta_{+-}(|{\vec k_1}-{\vec k_2}|,\eta',\eta''))\nonumber\\&& \times (i\Delta_{-+}(k_1,\eta,\eta''')i\Delta_{-+}(k_1,\eta'',\eta''')-{\rm c.c.})
\nonumber\\
&&=-\frac{\lambda \beta^2}{2^5 \times 27 \pi^6} \int_1^a \frac{da'}{a'} \int_1^{a'} \frac{da''}{a''} \ln a'' \int_1^{a''} \frac{da'''}{a'''} \ln^2 a''' = -\frac{\lambda \beta^2}{2^6\times 27 \pi^6}\frac{\ln^6 a}{45}
\label{nc39}
\end{eqnarray}

For $\eta'\gtrsim \eta'''\gtrsim \eta''$, $\eta'''\gtrsim \eta''\gtrsim \eta'$ and $\eta'''\gtrsim \eta'\gtrsim \eta''$, the contributions equal
\begin{eqnarray}
-\frac{\lambda \beta^2}{2^8 \times 9\pi^6} \frac{\ln^6 a}{90}, \qquad -\frac{\lambda \beta^2}{2^8 \times 9\pi^6} \frac{\ln^6 a}{72}, \qquad -\frac{\lambda \beta^2}{2^8 \times 9\pi^6} \frac{\ln^6 a}{72}
\label{nc40}
\end{eqnarray}

Combining the above with \ref{nc39}, we have the total contribution for this process,
\begin{eqnarray}
-\frac{\lambda \beta^2}{2^7 \times 9 \pi^6} \frac{37 \ln^6 a}{1080}
\label{nc41}
\end{eqnarray}

The seventh process of \ref{f3} is given by
\begin{eqnarray}
&&\frac{i\lambda\beta^2}{6} \int (a' a'' a''')^4 d^4 x' d^4 x'' d^4 x''' \left[i\Delta_{-+}^2(x,x')i\Delta_{++}(x',x'') i\Delta_{++}^3(x'',x''')+ i\Delta_{-+}^2(x,x')i\Delta_{+-}(x',x'') i\Delta_{--}^3(x'',x''')\right. \nonumber\\ && \left. + i\Delta_{+-}^2(x,x')i\Delta_{--}(x',x'') i\Delta_{-+}^3(x'',x''')+ i\Delta_{+-}^2(x,x')i\Delta_{-+}(x',x'') i\Delta_{+-}^3(x'',x''')-{\rm c.c.}\right]
\label{nc42}
\end{eqnarray}
The only non-vanishing contribution comes from $\eta' \gtrsim \eta'' \gtrsim \eta'''$, giving
\begin{eqnarray}
&&\frac{i\lambda\beta^2}{6} \int (a' a'' a''')^4 d \eta' d \eta'' d \eta'''\frac{d^3{\vec k_1}d^3{\vec k_2}d^3{\vec k_3} }{(2\pi)^9} \left[(i\Delta_{-+}(0,\eta',\eta'')-i\Delta_{+-}(0,\eta',\eta''))(i\Delta_{-+}(k_1,\eta,\eta')-i\Delta_{+-}(k_1,\eta,\eta')) \right. \nonumber\\ && \left.
\times (i\Delta_{-+}(k_1,\eta,\eta')+i\Delta_{+-}(k_1,\eta,\eta'))(i\Delta_{-+}(k_2,\eta'',\eta''')-i\Delta_{+-}(k_2,\eta'',\eta''')) \right. \nonumber\\ && \left.
\times (i\Delta_{-+}(k_3,\eta'',\eta''')i\Delta_{-+}(|{\vec k_2}+{\vec k_3}|,\eta'',\eta''')+i\Delta_{+-}(k_3,\eta'',\eta''')i\Delta_{+-}(|{\vec k_2}+{\vec k_3}|,\eta'',\eta''')\right. \nonumber\\&&\left. +i\Delta_{+-}(k_3,\eta'',\eta''')i\Delta_{-+}(|{\vec k_2}+{\vec k_3}|,\eta'',\eta'''))\right]
\nonumber\\&&=-\frac{\lambda \beta^2}{2^6 \times 27 \pi^6} \int_1^a \frac{da'}{a'}\ln a' \int_1^{a'}\frac{da''}{a''} \int_1^{a''}\frac{da'''}{a'''}\ln^2 a'''  =-\frac{\lambda \beta^2}{2^7 \times 9 \pi^6} \frac{\ln^6 a}{108}
\label{nc43}
\end{eqnarray}

Finally, the last process gives,
\begin{eqnarray}
&&\frac{i\lambda\beta^2}{8} \int (a' a'' a''')^4 d^4 x' d^4 x'' d^4 x''' i\Delta(x'',x'') i\Delta(x''',x''') \left[ i\Delta_{-+}^2(x,x')i\Delta_{++}(x',x'')i\Delta_{++}(x',x''') \right. \nonumber\\ &&\left. 
+i\Delta_{-+}^2(x,x')i\Delta_{+-}(x',x'')i\Delta_{+-}(x',x''')+i\Delta_{+-}^2(x,x')i\Delta_{-+}(x',x'')i\Delta_{--}(x',x''')\right. \nonumber\\ &&\left. +i\Delta_{+-}^2(x,x')i\Delta_{--}(x',x'')i\Delta_{-+}(x',x''') -{\rm c.c.}
\right]
\label{nc44}
\end{eqnarray}

The non-vanishing contribution to the above integral comes from the hierarchies $\eta' \gtrsim \eta'' \gtrsim \eta'''$ and $\eta' \gtrsim \eta''' \gtrsim \eta''$. For the first the above becomes, 
\begin{eqnarray}
&&\frac{i\lambda\beta^2}{8} \int (a' a'' a''')^4 d^4 x' d^4 x'' d^4 x''' i\Delta(x'',x'') i\Delta(x''',x''') (i\Delta_{-+}(x,x')-i\Delta_{+-}(x,x'))\nonumber\\&& \times (i\Delta_{-+}(x,x')+i\Delta_{+-}(x,x'))(i\Delta_{-+}(x',x'')-i\Delta_{+-}(x',x''))(i\Delta_{-+}(x',x''')-i\Delta_{+-}(x',x''')) \nonumber\\
&&=\frac{i\lambda\beta^2}{8} \int (a' a'' a''')^4 d \eta' d \eta'' d \eta'''\frac{d^3{\vec k_1}d^3{\vec k_2}d^3{\vec k_3} }{(2\pi)^9} i\Delta(k_2, \eta'', \eta'')i\Delta(k_3, \eta''', \eta''')(i\Delta_{-+}(k_1,\eta, \eta')+i\Delta_{+-}(k_1,\eta, \eta')) \nonumber\\&&
\times (i\Delta_{-+}(k_1,\eta, \eta')-i\Delta_{+-}(k_1,\eta, \eta'))(i\Delta_{-+}(0,\eta', \eta'')-i\Delta_{+-}(0,\eta', \eta''))(i\Delta_{-+}(0,\eta', \eta''')-i\Delta_{+-}(0,\eta', \eta'''))\nonumber\\
&&=-\frac{\lambda \beta^2}{2^8 \times 27 \pi^6}\int_1^a \frac{da'}{a'}\ln a' \int_1^{a'} \frac{da''}{a''}\ln a'' \int_1^{a''} \frac{da'''}{a'''}\ln a''' = -\frac{\lambda \beta^2}{2^8 \times 27 \pi^6} \frac{\ln^6 a}{48},
\label{nc45}
\end{eqnarray}
whereas for the second we get exactly the same answer as above. Thus the total contribution for the last diagram of \ref{f3} reads,
\begin{eqnarray}
-\frac{\lambda \beta^2}{2^7 \times 9 \pi^6} \frac{\ln^6 a}{144}
\label{nc46}
\end{eqnarray}

Combining now \ref{nc28}, \ref{nc31}, \ref{nc33}, \ref{nc35}, \ref{nc37}, \ref{nc41}, \ref{nc43}, \ref{nc45}, we find the ${\cal O}(\lambda \beta^2)$ correction to the two point correlator at late times for a massless minimal scalar field,
\begin{eqnarray}
\langle \phi^2\rangle_{\lambda \beta^2}= -\frac{\lambda \beta^2}{2^{10}\times 9 \pi^6} \frac{29 \ln^6 a}{27}
\label{nc47}
\end{eqnarray}

\bigskip

\end{document}